\documentclass[aps,prd,showpacs,eqsecnum,floatfix]{revtex4}
\usepackage{amsmath,amssymb}
\usepackage{graphicx}

\begin{document}

\title{Longterm general relativistic simulation of binary neutron
stars \\ collapsing to a black hole}

\author{Kenta Kiuchi$^{1}$
\footnote{\affiliation\ kiuchi@gravity.phys.waseda.ac.jp}}
\author{Yuichiro Sekiguchi$^{2}$
\footnote{\affiliation\ sekig@th.nao.ac.jp}}
\author{Masaru Shibata$^{3}$}
\author{Keisuke Taniguchi$^{4}$}

\affiliation{$^{1}$Department of Physics, Waseda University, 3-4-1
 Okubo, Shinjuku-ku, Tokyo 169-8555, Japan~}
\affiliation{$^{2}$Division of Theoretical Astronomy/Center for
 Computational Astrophysics, National Astronomical Observatory of
 Japan, 2-21-1, Osawa, Mitaka, Tokyo, 181-8588, Japan~}
\affiliation{$^{3}$Yukawa Institute for Theoretical Physics, 
Kyoto University, Kyoto, 606-8502, Japan~} 
\affiliation{$^{4}$ 
 Department of Physics, University of Wisconsin-Milwaukee, P.O. Box 413,
 Milwaukee, Wisconsin 53201 USA}

\date{\today}

\begin{abstract}
General relativistic simulations for the merger of binary neutron
stars are performed as an extension of a previous work
\cite{Shibata:2006nm}. We prepare binary neutron stars with a large
initial orbital separation and employ the moving-puncture formulation,
which enables to follow merger and ringdown phases for a long time,
even after black hole formation.  For modeling inspiraling neutron
stars, which should be composed of cold neutron stars, the
Akmal-Pandharipande-Ravenhall (APR) equation of state (EOS) is
adopted.  After the onset of merger, the hybrid-type EOS is used;
i.e., the cold and thermal parts are given by the APR and $\Gamma$-law
EOSs, respectively. Three equal-mass binaries, each with mass
$1.4M_\odot$, $1.45M_\odot$, and $1.5M_\odot$, and two unequal-mass
binaries with mass, 1.3 vs $1.6M_\odot$ and 1.35 vs $1.65M_\odot$, are
prepared. We focus primarily on the black hole formation case, and
explore mass and spin of the black hole, mass of disks which surround
the black hole, and gravitational waves emitted during the black hole
formation.  We find that (i) the black hole is promptly formed if
total mass of the system initially satisfies $m_0$ $\gtrsim
2.9M_\odot$; (ii) for the systems of $m_0=2.9$--$3.0M_\odot$ and of
mass ratio $\approx 0.8$, the mass of disks which surround the formed
black hole is 0.006--$0.02M_{\odot}$; (iii) the spin of the formed
black hole is $0.78 \pm 0.02$ when a black hole is formed after the
merger in the dynamical time scale. This value depends weakly on the
total mass and mass ratio, and is about $0.1$ larger than that of a
black hole formed from nonspinning binary black holes; (iv) for the
black-hole formation case, Fourier spectrum shape of gravitational
waves emitted in the merger and ringdown phases has a universal
qualitative feature irrespective of the total mass and mass ratio, but
quantitatively, the spectrum reflects the parameters of the binary
neutron stars.
\end{abstract} 


\maketitle


\section{Introduction}\label{sec:Intro}

Coalescence of binary neutron stars is one of the most promising
sources for kilometer-size laser interferometric detectors such as the
LIGO~\cite{Barish:1999vh,Cutler:2002me}, GEO~\cite{Hild:2006},
VIRGO~\cite{Acernese:2002bw,Acernese:2006bj}, and
TAMA~\cite{Tsubono:1994sg}.  Latest statistical estimate indicates
that detection rate of gravitational waves from binary neutron stars
will be 1 event per $\sim 40$--300 years for the first-generation
interferometric detectors and $\sim 10$--100 events per year for the
advanced detectors~\cite{Kalogera:2003tn,Bel}.  This suggests that
gravitational waves from binary neutron stars will be detected within
the next decade.

The merger of the binary neutron stars has also been proposed as a 
likely candidate for the central engine of short $\gamma$-ray bursts
(GRBs)~\cite{Narayan:1992iy,Zhang:2003uk}.  The observational facts
that short GRBs have a cosmological origin (e.g.,
Ref.~\cite{Fox:2005kv}) indicate that the central engine supplies a
large amount of energy $\agt 10^{48}$ ergs in a very short time scale
$\sim 0.1$--1 s~\cite{Piran:2004ba}.  According to a standard scenario
based on the merger hypothesis, a stellar-mass black hole 
surrounded by a hot and massive disk (or torus) should be formed after
the merger. Possible relevant processes to extract the energy of 
this black hole-accretion disk system for launching a relativistic jet
are neutrino-anti neutrino annihilation and magnetically driven mechanisms. 

Recent semi-analytic calculations (e.g., Ref.~\cite{GRBdiskA}) show
that an accretion rate of $\dot{M}\agt 0.1M_{\odot}$/s is required for
neutrino annihilation models to achieve sufficiently high energy
efficiency.  Also, recent numerical studies (e.g.,
Ref.~\cite{GRBdiskN}) suggest that if the disk had a mass $\agt
0.01M_\odot$, it could supply the required energy by neutrino
radiation.  Recent general relativistic magnetohydrodynamic
simulations \cite{GRBMHD} of (Kerr) black hole-accretion disk system
find that the rotational energy of the black hole can be extracted via
magnetohydrodynamics processes as a form of the Poynting flux.  In
particular the results are in general agreement with the predictions
of Blandford and Znajek \cite{BZ}. In this case, the black hole is
required to be rapidly rotating for efficient energy extraction.
This fact motivates to calculate final mass and spin of the black hole
formed after the merger, and to clarify whether such a massive
disk can be formed, and to clarify what type of the progenitors are
necessary for formation of a system composed of a black hole and a
massive disk.

For theoretically studying the late inspiral, merger, and ringdown
phases of the binary neutron stars, numerical relativity is the unique
approach.  Until quite recently, there has been no longterm general
relativistic simulation that self-consistently clarifies the inspiral
and merger phases because of limitation of the computational resources
or difficulty in simulating a black-hole spacetime, although a number
of simulations have been done for qualitative studies 
\cite{Shibata:1999hn,Shibata:1999wm,Shibata:2002jb,Shibata:2003ga,
  Shibata:2005ss, Shibata:2006nm,Miller:2003vc,Marronetti:2003hx}.
The most crucial drawbacks in the previous works are summarized as
follows (but see Refs. \cite{Yamamoto:2008js,Baiotti:2008ra}); (i) the
simulations were not able to be continued for a long time after
formation of a black hole and/or (ii) the simulations were short-term
for the inspiral phase; the inspiral motion of the binary neutron
stars is followed only for $\sim 1$--2 orbits. The simulations were
usually started with a quasiequilibrium state which is obtained assuming
that approaching velocity between two neutron stars is zero, that is
not realistic. As a result of this treatment, non-zero approaching
velocity at the onset of merger is not correctly taken into
account, and moreover, effects of non-zero eccentricity could play an
unfavored role \cite{foot0}.
 
In the present work, we perform an improved simulation overcoming
these drawbacks; (i) we adopt a moving-puncture approach
\cite{Campanelli:2005dd,Brugmann:2008zz}, which enables us to evolve
black hole spacetimes for an arbitrarily long time; (ii) we prepare
binary neutron stars in quasiequilibrium states of a large separation
as the initial condition. In the chosen initial data, the binary
neutron stars spend $\sim 4$ orbits in the inspiral phase before the
onset of merger, and hence, approximately correct non-zero approaching
velocity and nearly zero eccentricity results. Furthermore, we add a
non-zero approaching velocity at $t=0$. The magnitude of the
approaching velocity can be estimated by a post-Newtonian (PN) analysis
(see Sec.~\ref{sec:form} in detail). Adding the approaching velocity
suppresses an artificial orbital eccentricity caused by an
incompleteness of the initial conditions \cite{Boyle:2007ft}.  In
addition, a non-uniform grid with a sufficiently large computational
domain
\cite{Campanelli:2005dd,Shibata:2006ks,Shibata:2007zm,Liu:2008xy} is
employed to perform longterm accurate simulations with a relatively
low computational cost.

A longterm simulation of binary neutron stars, in which the inspiral
phase is followed for 3--5 orbits, has very recently been performed by
Baiotti et al. using the polytropic and $\Gamma$-law EOSs for modeling
the neutron stars \cite{Baiotti:2008ra}. With these simple EOSs, they
self-consistently investigated the inspiral, merger, and ringdown
processes.  However, as is well known, such EOSs are oversimplified 
for modeling the neutron stars, and hence, the results are only
qualitative. In this paper, we perform a longterm simulation not with
such simple EOSs but by adopting a nuclear-theory-based EOS. We focus
in particular on quantitatively clarifying the formation process of a 
black hole for the case that it is formed promptly (i.e., in the 
dynamical time scale $\sim 1$--2 ms) after the onset of merger. More 
specifically, the primary purpose of this paper is (1) to determine
the condition for prompt black hole formation, (2) to determine the
final mass and spin of the black hole formed after the merger, (3) to
clarify quantitative features of gravitational waves emitted in the
merger and ringdown phases, and (4) to estimate the mass of disks
surrounding the formed black hole.

Here, some words of explanation are necessary about nuclear-theory-based
EOS (often referred to as realistic EOS).  There are two types of
nuclear-theory-based EOSs for modeling the neutron stars. One is
zero-temperature nuclear EOS such as the Akmal-Pandharipande-Ravenhall
(APR) EOS \cite{APR:1998}. This type of EOSs could be suitable for
modeling not so young neutron stars, such as neutron stars in binary
neutron stars just before the merger, for which the thermal energy per
nucleon is much smaller than the Fermi energy.  The other is
finite-temperature EOS such as the Shen's EOS \cite{SHEN}.  This type
of EOSs should be employed for studying high-energy phenomena such as
supernova core collapse and the merger phase of binary neutron stars.

One problem faced when choosing EOSs is that no one knows the truly
realistic EOS for high-density nuclear matter. This implies that for
studying the inspiral and merger phases of binary neutron stars by
numerical simulation, it is necessary to systematically perform many
simulations adopting a variety of nuclear-theory-based EOSs.  Indeed,
one of the important roles in gravitational-wave detection is to
constrain nuclear EOS from detected gravitational waves
\cite{Cutler:2002me}.  For this purpose, numerical simulation for a
variety of EOSs has to be performed for deriving all the possible
gravitational waveforms.  There are many cold EOSs that have been
proposed \cite{Lattimer:2000nx}, and a systematic theoretical survey
of the gravitational waveforms for clarifying their dependence on the
EOSs is possible.  By contrast, there are only a few
finite-temperature EOSs \cite{SHEN,LS}. This implies that a systematic
study is not possible when adopting the finite-temperature EOSs.
 
In the series of our papers
\cite{Shibata:2002jb,Shibata:2003ga,Shibata:2006nm}, we have found
that shock heating during the merger phase increases the thermal
energy of the neutron stars. This indicates that adopting
finite-temperature EOSs is desirable for a sophisticated study of the
merger process.  However, the shock heating is not very strong during 
the merger and the thermal energy per nucleon after the
shock heating is only $\sim 10\%$ of the Fermi energy of the nuclear
matter in the central region of the merged object (see Sec. IV). Thus,
the thermal energy is not essential at least for short-term evolution
of the main body of the merged object.  In particular, the thermal
energy plays a minor role in determining the criterion for prompt
black hole formation at the merger and its process. Because of this 
reason, in the previous two papers
\cite{Shibata:2005ss,Shibata:2006nm}, we have adopted a hybrid EOS in
which the pressure and thermal energy are divided into cold and hot
parts. The cold part is determined by the zero-temperature
nuclear-theory-based EOSs, and the hot part is simply modeled by a
$\Gamma$-law EOS. In this method, the main part (cold part) of the EOS is
appropriately modeled, although a minor part (finite-temperature part)
is modeled in an approximate manner. The merit of this EOS is that a
variety of the cold EOSs can be systematically employed for the
simulation.  The hybrid EOS may not be appropriate for studying
the longterm evolution of a hypermassive neutron star which is formed when
the total mass of the system is not large enough for prompt black hole
formation.  The reason for this is that in the hypermassive neutron
star, longterm shock heating and cooling via neutrino emission are
likely to play a role. However, for studying prompt black hole
formation, which is the main subject of this paper, the hybrid EOS
seems to be acceptable because the roles of the shock heating and
neutrino cooling do not seem to be essential.  Thus, following
previous papers \cite{Shibata:2005ss,Shibata:2006nm}, numerical
simulations are performed, employing the hybrid EOS.  Specifically, we
adopt the APR EOS for the cold part following
Ref.~\cite{Shibata:2006nm}.

This paper is organized as follows. In Sec.~\ref{sec:form}, we
describe the formalism for numerical solution of the Einstein and of
relativistic-hydrodynamics equations, the EOS adopted in this paper,
and the method for extraction of gravitational waves.  The initial
condition for binary neutron stars and grid structure for the
simulation are summarized in Sec.~\ref{sec:init}. In
Sec.~\ref{sec:result}, numerical results are presented, focusing in
particular on the case that a black hole is formed in the dynamical
time scale after the onset of merger. Sec.~\ref{sec:sum} is
devoted to a summary. Throughout this paper, unless otherwise stated, we
adopt the geometrical units in which $G=c=1$ where $G$ and $c$ are the
gravitational constant and the speed of light. Greek and Latin
indices denote the spacetime and space components, respectively. 

\section{Formulation}\label{sec:form}
\subsection{Numerical methods}

Our formulation for the fully general relativistic simulation  
is the same as in Refs.~\cite{Shibata:2003ga,Shibata:2006nm,Shibata:2007zm}, 
to which the reader may refer for details of the basic equations. 

For solving the Einstein evolution equations, we use the original
version of the Baumgarte-Shapiro-Shibata-Nakamura (BSSN) formalism~
\cite{Shibata:1995we}: We evolve the conformal factor,
$\phi=(\ln\gamma)/12$, the trace part of the extrinsic curvature, $K$,
the conformal three-metric,
$\tilde{\gamma}_{ij}\equiv\gamma^{-1/3}\gamma_{ij}$, the tracefree
extrinsic curvature, $\tilde{A}_{ij}\equiv
\gamma^{-1/3}(K_{ij}-K\gamma_{ij}/3)$, and a three-auxiliary variable,
$F_i\equiv\delta^{jk}\partial_j\tilde{\gamma}_{ik}$. Here
$\gamma_{ij}$ is the three-metric, $K_{ij}$ the extrinsic curvature,
$\gamma\equiv\text{det}(\gamma_{ij})$, and $K\equiv
K_{ij}\gamma^{ij}$.  As in Ref.~\cite{Shibata:2007zm}, we evolve the
conformal factor $\phi$, not the inverse of $\psi$, because the
cell-centered grid is adopted in our code, and hence, the coordinate
singularity at the puncture is avoided in moving puncture
frameworks \cite{Campanelli:2005dd,Brugmann:2008zz}.

For the conditions of the lapse, $\alpha$, and the shift vector,
$\beta^i$, we adopt a dynamical gauge condition in the following
forms,
\begin{eqnarray}
&&
(\partial_t-\beta^i\partial_i) \ln\alpha = - 2 K,\label{eq:gaug1}\\
&&
\partial_t \beta^i = 0.75\tilde{\gamma}^{ij}(F_j + \Delta t
\partial_t F_j),\label{eq:gaug2}
\end{eqnarray}
where $\Delta t$ denotes the time step in the numerical simulations,
and the second term on the right-hand side of Eq.~(\ref{eq:gaug2}) is
introduced for stabilizing the numerical computations.  The 
gauge condition (\ref{eq:gaug2}), which was originally proposed in
Ref.~\cite{Shibata:2003iw}, is slightly different from that usually
used in the moving puncture framework (see, e.g.,~\cite{Baiotti:2008ra}). 
We note that Ref.~\cite{Yamamoto:2008js} shows that this 
gauge is as suitable as popular one for simulating black hole spacetimes.  

The numerical scheme for solving the Einstein equations is essentially
the same as that in Ref.~\cite{Yamamoto:2008js}. We use the
fourth-order finite difference scheme in the spatial direction and a
fourth-order Runge-Kutta scheme in the time integration, where the
advection terms such as $\beta^i\partial_i\phi$ are evaluated by a
fourth-order upwind scheme, as proposed in Ref.~\cite{Brugmann:2008zz}. 
In a previous uni-grid simulation, we use a third-order scheme for
the time integration~\cite{Shibata:2007zm}.  We have found that the
fourth-order scheme can give more accurate results, and hence, updated 
the scheme for this work. 

The location and properties of the black hole, such as the area and the
circumferential radii, are determined by analyzing an apparent
horizon. Our method for finding the apparent horizon is described in 
Refs.~\cite{Shibata:1997nc,Shibata:2000nw}. From the area, and 
polar and equatorial circumferential proper lengths, we infer the 
mass and spin of the black hole. 

The numerical code for the hydrodynamics is the same as that in
Refs.~\cite{Shibata:2006ks,Shibata:2007zm}: As the variables to be
evolved, we adopt $\rho_* \equiv \rho \alpha u^t e^{6\phi}$, $\hat u_i
\equiv h u_i$, and $e_* \equiv h \alpha u^t -P/(\rho \alpha u^t)$,
where $\rho$ is the rest-mass density, $u_i$ is the three-component of
the four velocity, $u^t$ is the time component of the four velocity,
$P$ is the pressure, $h$ is the specific enthalpy defined by $h
\equiv 1 + \varepsilon + P/\rho$, and $\varepsilon$ is the specific
internal energy. To handle advection terms in the hydrodynamic
equations, a high-resolution central scheme \cite{KT} is adopted with
a third-order piecewise parabolic interpolation and with a steep
min-mod limiter. In the present work, the limiter parameter, $b$, is
set to be $2.5$ (see Ref. \cite{Shibata:2003iy} for detail about our 
interpolation scheme and the parameter $b$). 

\subsection{Equations of state}

Following Refs.~\cite{Shibata:2003ga,Shibata:2006nm}, we adopt a
hybrid EOS for modeling the EOS of neutron stars. In this EOS, the pressure and
the specific internal energy are written in the form 
\begin{eqnarray}
&&
P = P_{\rm cold} + P_{\rm th},\label{eq:EOS1}\\
&&
\varepsilon = \varepsilon_{\rm cold} + \varepsilon_{\rm th},\label{eq:EOS2}
\end{eqnarray}
where $P_{\rm cold}$ and $\varepsilon_{\rm cold}$ are the cold
(zero-temperature) parts, and are written as functions of rest-mass
density $\rho$.  In general, any nuclear-theory-based EOS for zero-temperature
nuclear matter can be employed for assigning $P_{\rm cold}$ and
$\varepsilon_{\rm cold}$.  In this paper, we adopt the APR
EOS~\cite{APR:1998}, for which $P$ and $\varepsilon$ are tabulated as
functions of the baryon rest-mass density for a wide density range.
We use a fitting formula for the data in the range, $10^{10}~{\rm
g/cm^3}\le\rho\lesssim 10^{16}~{\rm g/cm^3}$.  The method of the
fitting was first developed in Ref.~\cite{Haensel:2004nu} and slightly
modified in Ref.~\cite{Shibata:2006nm,Shibata:2005ss}. We adopt the
fitting parameters listed in Table I of Ref.~\cite{Shibata:2006nm}.

$P_{\rm th}$ and $\varepsilon_{\rm th}$ in Eqs.~(\ref{eq:EOS1}) and
(\ref{eq:EOS2}) denote thermal (finite-temperature) parts which are
zero in the absence of shocks, but become non-zero if shocks are
formed. Specifically, they have finite values after the merger sets
in.  During the simulation, $\rho$ and $\varepsilon$ are determined
from the evolved variables $\rho_*$, $e_*$, and the normalization
relation of the four-velocity. Then, $\varepsilon_{\rm th}$ is
determined by $\varepsilon-\varepsilon_{\rm cold}(\rho)$, and
subsequently, the thermal part of the pressure $P_{\rm th}$ is related
to the specific thermal energy $\varepsilon_{\rm th}\equiv
\varepsilon-\varepsilon_{\rm cold}$ as
\begin{eqnarray}
P_{\rm th} = (\Gamma_{\rm th} - 1 )\rho \varepsilon_{\rm th},\label{eq:EOS3}
\end{eqnarray}
where $\Gamma_{\rm th}$ is an adiabatic constant for which we set 
$\Gamma_{\rm th}=2$ taking into account the fact that the EOSs for 
high-density nuclear matter are stiff. 

\subsection{Extracting Gravitational Waves}\label{sec:GW}

To extract gravitational waves from numerical data, we compute the
outgoing component of the Newman-Penrose quantity $\Psi_4$ (e.g.,
Refs.~\cite{Brugmann:2008zz,Buonanno:2006ui,Yamamoto:2008js} for
detail). From $\Psi_4$, loss rates of energy and angular momentum
carried by gravitational waves are computed by
\begin{eqnarray}
&&
\frac{dE}{dt} = \lim_{r\to\infty}\left[\frac{r^2}{16\pi}
\oint_S d(\cos\theta)d\varphi
\left|
\int^t \Psi_4 dt'
\right|^2
\right],\label{eq:dE}\\
&&
\frac{dJ}{dt} = \lim_{r\to\infty}{\rm Re}\left[\frac{r^2}{16\pi}
\oint_S d(\cos\theta)d\varphi
\left(\int^t \partial_\varphi \Psi_4 dt'
\right)
\left(\int^t\int^{t'}\bar{\Psi}_4 dt'dt''
\right)
\right],\label{eq:dJ}
\end{eqnarray}
where $\oint d(\cos\theta)d\varphi$ denotes an integral on two surface
of a constant coordinate radius and $\bar{\Psi}_4$ is the complex
conjugate of $\Psi_4$. In numerical simulation, the surface integral
is performed for several radii near the outer boundaries, and we check
that the resulting gravitational waveforms depend only weakly on the
extracted radii.  The outer boundaries along each axis are located 
at $r=L \sim \lambda_0$ where $\lambda_0$ is wave length of 
gravitational waves emitted by inspiraling binary neutron stars at $t=0$  
and gravitational waves are extracted for $r \approx 0.7$--0.95$L$  
(cf. Table~\ref{tab:numset}; maximum extraction radii are denotes by  
$r_{\rm ex}$ in this table). 

\begin{figure*}
  \begin{center}
  \vspace*{40pt}
    \begin{tabular}{c}
      \begin{minipage}{0.5\hsize}
      \includegraphics[width=9.0cm]{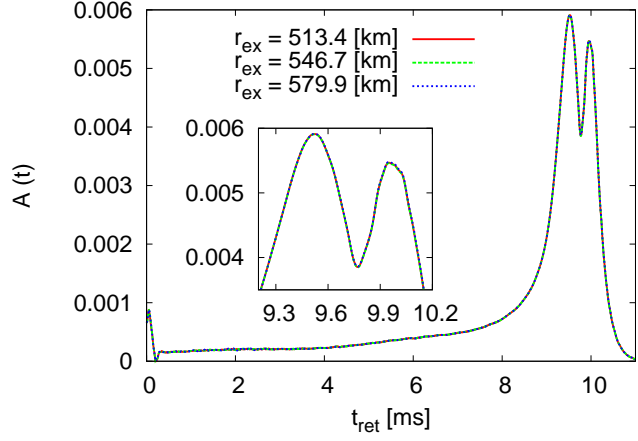}
      \end{minipage}
    \end{tabular}
    \caption{\label{fig:ext} The gravitational wave amplitude as a
      function of retarded time for run APR145145H.  The solid,
      dashed, and dotted curves denote the amplitudes extracted at
      $513.4$, $546.7$, and $579.9$ km, respectively.  }
  \end{center}
\end{figure*}

The amplitude and phase of gravitational waves extracted at several
radii should be extrapolated to obtain those at infinity for 
deriving precise waveforms, as often done in the simulation for 
binary black holes~\cite{Hannam:2007ik,Boyle:2007ft,Scheel:2008rj,Boyle:2009vi,Hannam:2009rd}. 
However, numerical waveforms computed in hydrodynamic simulation are not as 
precise as those in the simulations for vacuum spacetime, because 
the order of the accuracy is reduced at shocks, discontinuities, 
and places where the gradient of hydrodynamic quantities is large 
in the standard numerical hydrodynamics. This implies that 
numerical error is primarily determined by such sources, and hence, 
the finiteness of the extraction radius does not become the main source 
of the numerical error. To illustrate this fact, we generate 
Fig.~\ref{fig:ext} which shows the time evolution of gravitational 
wave amplitude for a typical model in this paper. 
The amplitude $A(t)$ is defined from $\Psi_4$ as 
\begin{eqnarray}
r\Psi_4 = A(t) {\rm e}^{i\phi(t)},
\end{eqnarray}
where $r$ is the extraction radius and $\phi$ the phase.  This plot
shows that the amplitude depends very weakly on the 
extraction radius; relative difference among three results 
is much smaller than 1\%, typical error size induced by the poor resolution 
of hydrodynamics mentioned above. Thus, we do not use extrapolation 
of amplitude and wave phase in this work. 

For calculating the two polarization modes $h_{+,\times}$ from
$\Psi_4$, we perform the time integration of $\Psi_4$ twice with
appropriate choice of integration constants and subtraction of
unphysical drift, perhaps associated primarily with the drift of the
mass center of the system, that often occurs in hydrodynamic
simulation due to accumulation of numerical error.  Specifically,
whenever the time integration is performed, we subtract a function of
the form $a_2 t^2 + a_1 t + a_0$ where $a_0$--$a_2$ denote constants
which are determined by the least-square fitting to the original
numerical data. 

From a time sequence of $dE/dt$ and $dJ/dt$, we compute 
total radiated energy and angular momentum by time integration as
\begin{eqnarray}
&& \Delta E = \int dt {dE \over dt},\\
&& \Delta J = \int dt {dJ \over dt}. 
\end{eqnarray}

\subsection{Comparison with PN approximation}\label{sec:PN}

Gravitational waveforms in the inspiral phase computed in numerical
simulation should be compared with those by the PN 
approximation, because it provides an accurate waveform if the orbital
velocity $v$ is not extremely relativistic (i.e., $v \alt 0.3c$ ).
Assuming that binary components are point masses and their
quasicircular orbits evolve adiabatically, orbital evolution of the
binary stars can be analytically determined with 3PN accuracy and
gravitational waveforms are derived with the 3.5PN accuracy
\cite{Blanchet:2006mz}.  Recently, several phenomenological
prescription for improving the PN results, which is applicable even to
highly relativistic orbits of $v \sim 0.3c$, have been proposed. Among
them, high-accuracy simulations for equal-mass and nonspinning binary
black holes~\cite{Boyle:2007ft} have proven that the so-called Taylor
T4 formula provides the orbital evolution and gravitational waveforms
with a high accuracy at least up to about one orbit before the onset
of the merger.  The Taylor T4 formula appears to be a good
approximation also for unequal-mass and nonspinning binaries~\cite{Ajith}
 (but see Ref.~\cite{Hannam:2007wf} for a depression
of the accuracy for the Taylor T4 in spinning cases).  Because we
focus only on nonspinning binaries, it is safe to assume that the
Taylor T4 formula would be a good approximate formula, and thus, we
explore a match between the numerical and Taylor-T4's waveforms for the
inspiral phase. 

One issue in comparison in the present context is that the 
effect of tidal deformation of neutron stars, which is not 
taken into account in the Taylor T4 formula, plays an important 
role for close orbits. This implies that the numerical waveforms 
should not agree with the waveforms by the Taylor T4 formula for 
such orbits. Thus, we compare two waveforms for $m_0 \Omega \alt 0.04$ 
(see Sec. III A for definition of $m_0$). 

More specifically, in comparison, two waveforms have to appropriately
align as done in Refs.~\cite{Boyle:2007ft,Hannam:2007ik,Boyle:2008ge}.
Our procedure in this work is as follows: First we calculate the
orbital angular velocity by $\Omega\equiv d\Phi/dt$.  Then, we
determine the reference time at which two waveforms give the same
value of $\Omega$. Recent studies in the context of the binary black
holes have shown that the reference value of $\Omega$ should be $\alt
0.1/m_0$ because beyond this value the PN approximation breaks down~
\cite{Buonanno:2006ui,Baker:2006ha,Hannam:2007ik,Boyle:2007ft}.  For
the binary neutron stars, the value of $0.1/m_0$ is even too large
because the compactness of neutron stars is much smaller than that of
black holes and hence the merger already started at such a high value.
Also, the effect of tidal deformation of neutron stars plays an
important role for $m_0\Omega \agt 0.04$. Due to this reason, the
reference value of $\Omega$ is chosen to be $m_0\Omega \approx 0.04$
for all the runs in this paper. 

\section{Initial model and simulation setting}\label{sec:init}
\subsection{initial model}

Except for the orbit just before the merger, binary neutron stars are
in a quasicircular orbit because the time scale of gravitational
radiation reaction at Newtonian order $\sim (5/64)\Omega^{-1} (M
\Omega)^{5/3}$ (see, e.g., \cite{shapiro}) is several times longer
than the orbital period.  Hence, following our previous works
\cite{Shibata:2000nw,Shibata:2003ga,Shibata:2006ks,Shibata:2006nm,
Shibata:2007zm,Yamamoto:2008js}, we adopt binary neutron stars in
quasiequilibrium states as initial conditions. The quasiequilibrium
states are computed in the so-called conformal-flatness formalism for
the Einstein equations~\cite{Wilson:1995uh}.  The irrotational 
velocity field is assumed because it is considered to be a realistic
velocity field for coalescing binary neutron stars in nature
\cite{Kochanek:1992wk,Bildsten:1992my}.  We employ numerical solutions
computed by a code in the LORENE library
\cite{Gourgoulhon:2000nn,Taniguchi:2002ns,Taniguchi:2003hx,LoreneURL}.
Table~\ref{tab:model} lists the several key quantities for the models
adopted in this paper.  We select three equal-mass (APR1414,
APR145145, and APR1515) and two unequal-mass models (APR1316 and
APR135165). Note that the numbers in the model name denote the ADM
mass of the two neutron stars in isolation (e.g., for APR1316, mass of
two neutron stars in isolation is $m_1=1.3M_{\odot}$ and
$m_2=1.6M_{\odot}$). In the following, we use $m_0$, $M_0$, and $M_*$
as the sum of masses of two neutron stars in isolation
($m_0=m_1+m_2$), initial total ADM mass, and total rest mass of the
system, respectively.

\subsection{grid setting}

In the simulation, the cell-centered Cartesian, $(x,y,z)$, grid is
adopted. In these coordinates, we can avoid the situation that the
location of the puncture (which always stays on the $z=0$ plane)
coincides with one of the grid points. Equatorial plane symmetry
is also assumed. The computational domain of $-L\le x \le L$, $-L\le y
\le L$, and $0\le z \le L$ is covered by the grid size $(2N,2N,N)$ for
$(x, y, z)$, where $L$ and $N$ are constants. Following
Refs.~\cite{Shibata:2006ks,Shibata:2007zm}, we adopt a nonuniform grid
as follows; an inner domain is covered with a uniform grid of spacing
$\Delta x$ and with the grid size, $(2N_0,2N_0,N_0)$. Outside this
inner domain, the grid spacing is increased according to the relation,
$\xi\tanh[(i-N_0)/\Delta i]\Delta x$, where $i$ denotes the $i$-th
grid point in each positive direction, and $N_0$, $\Delta i$, and
$\xi$ are constants.  Then, the location of $i$-th grid, $x^k(i)$, in
each direction is
\begin{eqnarray}
x^k(i)=\left\{
\begin{array}{ll}
(i+1/2)\Delta x & 0 \leq i \leq N_0 \\ 
(i+1/2)\Delta x + \xi \Delta i
\Delta x \log [ \cosh \{(i-N_0)/\Delta i \}] & i > N_0\\
\end{array}
\right.
\end{eqnarray}
and $x^k(-i-1)=-x^k(i)$, where $i=0,1,\cdots N$ for $x^k=x,y,$ and $z$. The
chosen parameters of the grid structure for each simulation are listed in
Table~\ref{tab:numset}.  

For investigating convergence of numerical results, we perform
simulations with three grid resolutions for all the models; labels 
``L'', ``M'', and ``H'' denote the low, medium, and high grid resolutions.
We note that for all the runs, the
grid resolution around the neutron stars is much better than that in
the previous work \cite{Shibata:2006nm}, in which the major diameter
of the neutron stars is covered only by 45 grid points. Since it is
covered by $\approx 60$--80 grid points, we expect that numerical results
in this paper are much more accurate than those in the previous work.
In physical units, the grid spacing around the neutron stars are about 
200 meter for the highest grid resolution. 
On the other hand, the grid spacing in a wave zone, where 
gravitational waves are extracted, is $\approx 4$ km. 
Radius of apparent horizon of black hole finally formed is 
covered by about ten grid points. 

For high-resolution run (run ``H''), about 200 GBytes computational
memory is necessary and 240 CPU hours was spent using 512 processors on
Cray XT4 system at the Center for Computational Astrophysics (CfCA) in
the National Astronomical Observatory of Japan (NAOJ).  For run ``L'',
the computational time is $\sim 100$ CPU hours using the same processors. 

\subsection{approaching velocity}\label{sec:approach}

In computing quasiequilibrium states in the conformal-flatness formalism,
approaching velocity, which should be present in reality due to
gravitational radiation reaction, is not taken into account.  Lack of
the approaching velocity induces a non-zero orbital eccentricity and
resulting modulation in gravitational waveforms (e.g.,
Ref.~\cite{Boyle:2007ft}). If a simulation is started from an initial
condition in which the initial orbital separation is sufficiently large,
the value of the approaching velocity is negligible, and hence, its lack
is not a serious problem; in such case, the approaching velocity
settles to a correct one and eccentricity approaches zero in a few
orbits, because of gravitational radiation reaction~\cite{Peters:1963}. 
However, in the present choice of the
initial condition, the initial orbital separation is not sufficiently
large. Thus, we initially add an approaching velocity to improve 
the quality of the initial condition. 
For calculating an approximate value of the approaching velocity, 
we assume that the two neutron stars may be approximated by 
point particles and the so-called Taylor T4 
formula~\cite{Buonanno:2006ui,Boyle:2007ft} is used for predicting their 
orbital motion. In the following, we describe our method 
(see relevant works \cite{Buonanno:2006ui,Pfeiffer:2007yz,Husa:2007rh} 
in simulations of binary black holes): Assuming that the
density maxima of the neutron stars are located along $x$-axis at
$t=0$, the approaching velocity $v^x$ may be calculated by 
\begin{eqnarray}
v^x = \frac{dr_{12}}{dt} = -\frac{G
  m_0}{\gamma_{12}^2c^2}\frac{d\gamma_{12}}{dt},
\label{eq:PN1}
\end{eqnarray}
where $\gamma_{12}\equiv G m_0/r_{12}c^2$ and $r_{12}$ is a coordinate
orbital separation in the harmonic gauge.  $m_0$ is the total mass
defined by the sum of masses of two neutron stars at a state when they
are in isolation.  Here, we recover $G$ and $c$ to explicitly clarify
that $\gamma_{12}$ is dimensionless.  $\gamma_{12}$ is written by a
gauge-invariant quantity, $x\equiv(G m_0\Omega/c^3)^{2/3}$, at 3PN
order as
\begin{eqnarray}
\gamma_{12}=x\left[1+\left(1-\frac{\nu}{3}\right)x
+\left(1-\frac{65}{12}\nu\right)x^2
+\left(1+\left[-\frac{2203}{2520}-\frac{41}{192}\pi^2-\frac{22}{3}\ln
\left(\frac{r_{12}}{r_0'}\right)\right]\nu
+\frac{229}{36}\nu^2+\frac{1}{81}\nu^3\right)x^3\right],\label{eq:PN2}
\end{eqnarray}
where $\Omega$ denotes the orbital frequency, $\nu=m_1m_2/m_0^2$ is
the ratio of the reduced mass to the total mass with $m_1(m_2)$ being 
the mass of the star 1 (2) in isolation, and $r_0'$ is ``logarithmic
barycenter''~\cite{Blanchet:2006mz}.  In the Taylor T4
formula~\cite{Boyle:2007ft}, the evolution equation of $x$ is derived
in an adiabatic approximation as 
\begin{eqnarray}
&& \frac{dx}{dt}=\frac{16c^3}{5G m_0}x^5\Bigg\{ 1-\frac{487}{168}x+4\pi
x^{3/2} + \frac{274229}{72576}x^2-\frac{254}{21}\pi x^{5/2}\nonumber\\
&&
+\left[\frac{178384023737}{3353011200}+\frac{1475}{192}\pi^2-\frac{1712}{105}\gamma_E-\frac{856}{105}\ln(16x)\right]x^3
+\frac{3310}{189}\pi x^{7/2} \Bigg\},\label{eq:PN3}
\end{eqnarray}
where $\gamma_E$ is Euler's constant. From 
Eqs.~(\ref{eq:PN1})--(\ref{eq:PN3}), the approaching velocity is derived.

We note that this approaching velocity is not gauge-invariant and has
a strict meaning only in the harmonic gauge condition. Because the
initial condition is not obtained in this gauge condition, it would be
necessary to carry out a coordinate transformation to obtain the
approaching velocity in our chosen gauge condition. However, the
initial condition is obtained in the conformal flatness formalism, and
thus, the gauge condition is not clear, because the metric components
are oversimplified in this formalism. Thus, in this work, we use the
approaching velocity defined in Eq. (\ref{eq:PN1}) with no
modification.  After calculating $v^x$, the four velocity is modified
from its quasiequilibrium value $u_i^{(\rm eq)}$ by adding a
correction $u_x^{\rm corr} = w {\rm e}^{4\phi}v^x / \alpha$ with $w
\equiv \alpha u^t = (1+\gamma^{ij}u_i^{(\rm eq)}u_j^{(\rm eq)})^{1/2}$
where we drop a correction in $\alpha$ and $\beta^i$ due to the change
of $u_i$, because they only give higher-order corrections.  Then, we
uniformly add the specific momentum $h u_x^{\rm corr}$ to each neutron
star and reimpose the Hamiltonian and momentum constraints.  Note that
it is nontrivial to relate velocity to momentum in general relativity
(but see Ref.~\cite{Brugmann:2007zj} for discussion about this issue).

\begin{figure*}
  \begin{center}
  \vspace*{40pt}
    \begin{tabular}{c}
      \begin{minipage}{0.5\hsize}
      \includegraphics[width=9.0cm]{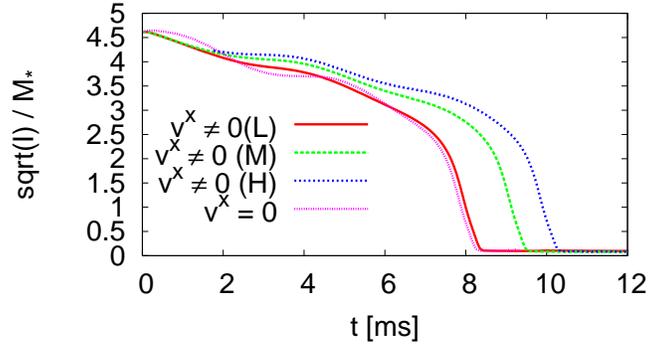}
      \end{minipage}
    \end{tabular}
    \caption{\label{fig:appr} The evolution of a coordinate separation
      defined by Eq. (\ref{coorsep}) for model APR1515. The solid,
      dashed, and short-dashed curves show the results for runs APR1515L, M,
      and H with approaching velocity, respectively. The dotted curve
      does for run APR1515L without approaching velocity.}
  \end{center}
\end{figure*}

Figure \ref{fig:appr} plots the evolution of a coordinate orbital
separation $I^{1/2}$ with/without approaching velocity for model
APR1515. Here, $I$ is defined by
\begin{eqnarray}
I = \frac{I_{xx}+I_{yy}}{M_*}, \label{coorsep}
\end{eqnarray}
where $I_{ij}\equiv\int \rho \alpha u^t x^i x^j \sqrt{\gamma}d^3x$
with $u^t$ being a time component of the four velocity.  In the
absence of the approaching velocity, $I$ does not decrease
monotonically but oscillates with time (the dotted curve in
Fig. \ref{fig:appr}).  In the presence of the approaching velocity,
this oscillation is suppressed (the solid curve in
Fig. \ref{fig:appr}).  This figure illustrates the advantage of our
treatment.  It should be noted that the orbital eccentricity is not
completely suppressed even in this method.  As discussed in
Sec.~\ref{sec:gw0}, indeed, gravitational waveforms in an early phase
slightly disagree with that calculated in the Taylor T4 framework
(e.g., angular velocity computed by gravitational waves does not
increase monotonically but has a modulation; see Sec.~\ref{sec:gw0}).
However, the radiation reaction circularizes the binary orbit and
gravitational waves in the late inspiral phase agree approximately
with the prediction by the Taylor T4 formula in a better manner 
(cf. Fig. 13 (b)).

\begin{table*}
\centering
\begin{minipage}{140mm}
\caption{\label{tab:model} List of several key quantities for the
  initial data of binary neutron stars in quasiequilibrium state.  The
  ADM mass of each star when they are in isolation ($m_1$ and $m_2$),
  the maximum baryon rest-mass density for each star, the baryon mass
  ratio $Q_M \equiv M_{*2}/M_{*1}$, the total baryon rest mass $M_*$,
  the total ADM mass $M_0$, nondimensional spin parameter $J_0/M_0^2$,
  orbital period $P_0$, and the orbital angular velocity in units of
  $M_0^{-1}$, $M_0\Omega_0$, where $\Omega_0$ denotes initial orbital
  angular velocity. }
\begin{tabular}{lcccccccc}
\hline\hline
Model                               &
$m_1$, $m_2 (M_\odot)$              &
$\rho~(10^{15}~{\rm g}/{\rm cm^3})$ &
$Q_M$                               &
$M_*(M_\odot)$                      & 
$M_0(M_\odot)$                      &
~$J_0/M_0^2$~                       &
~~$P_0(\rm ms)$~~                   &
~~$M_0\Omega_0$~~ \\
\hline
APR1414   & 1.40,~1.40 & 0.887,~0.887 & 1.00   & 3.106 & 2.771 & 0.983 & 3.185 & 0.0269\\
APR145145 & 1.45,~1.45 & 0.935,~0.935 & 1.00   & 3.232 & 2.870 & 0.983 & 3.299 & 0.0269\\
APR1515   & 1.50,~1.50 & 0.961,~0.961 & 1.00   & 3.359 & 2.969 & 0.983 & 3.412 & 0.0269\\
APR1316   & 1.30,~1.60 & 0.864,~1.015 & 0.7943 & 3.238 & 2.870 & 0.976 & 3.299 & 0.0269\\
APR135165 & 1.35,~1.65 & 0.887,~1.045 & 0.7992 & 3.365 & 2.970 & 0.978 & 3.412 & 0.0269\\
\hline\hline
\end{tabular}
\end{minipage}
\end{table*}

\begin{table*}
\centering
\begin{minipage}{140mm}
\caption{\label{tab:numset} Parameters for the grid structure employed
  in the numerical simulation. The grid number for covering one
  positive direction $(N)$, that for the inner uniform grid zone
  $(N_0)$, the parameters for nonuniform-grid domain $(\Delta i,\xi)$,
  the approximate grid number for covering the major diameter of
  massive neutron star $(L_{\rm NS})$, the ratio of the outer grid
  spacing to the wavelength of fundamental quasinormal mode of the
  formed black hole $(\lambda_{\rm QNM})$, the ratio of the location
  of outer boundaries along each axis to the initial gravitational
  wave length $(\lambda_0=\pi/\Omega_0)$ and maximum radius of
  gravitational wave extraction, $r_{\rm ex}$, in units of kilometer
  (and in units of $L$).  The last column shows the final outcome; BH
  and NS denote a black hole and neutron star, respectively. }
\begin{tabular}{lccccccccc}
\hline\hline
Model                                    &
~~$N$~~                                  &
~~$N_0$~~                                &
~~$\Delta i$~~                           &
~~$\xi$~~                                & 
$L_{\rm NS}/\Delta x$                    &
$L/\lambda_0$                            &
$\lambda_{\rm QNM}/\Delta x$             &
$r_{\rm ex}~(r_{\rm ex}/L)$              &
Outcome                                 \\
\hline
APR1414L   & 224 & 114 & 30 & 15   & 60 & 0.97 & --   &4.15E+2~(0.90)& NS \\
APR1414M   & 234 & 120 & 30 & 17.5 & 70 & 0.98 & --   &4.35E+2~(0.94)& NS \\
APR1414H   & 253 & 139 & 30 & 20   & 80 & 0.97 & --   &4.28E+2~(0.93)& NS \\
APR145145L & 224 & 110 & 30 & 20   & 60 & 1.24 & 7.0  &4.41E+2~(0.72)& BH \\
APR145145M & 252 & 123 & 30 & 20.5 & 70 & 1.24 & 8.3  &5.67E+2~(0.92)& BH \\
APR145145H & 282 & 141 & 30 & 21   & 80 & 1.24 & 9.2  &5.80E+2~(0.95)& BH \\
APR1515L   & 224 & 114 & 30 & 19.5 & 60 & 1.10 & 7.3  &5.06E+2~(0.90)& BH \\
APR1515M   & 250 & 128 & 30 & 20   & 70 & 1.09 & 8.3  &5.24E+2~(0.94)& BH \\
APR1515H   & 280 & 145 & 30 & 21   & 80 & 1.10 & 9.6  &5.42E+2~(0.94)& BH \\
APR1316L   & 239 & 130 & 30 & 19   & 60 & 1.11 & 7.7  &5.05E+2~(0.92)& BH \\
APR1316M   & 278 & 160 & 30 & 20   & 70 & 1.10 & 8.6  &4.93E+2~(0.91)& BH \\
APR1316H   & 300 & 170 & 30 & 21   & 80 & 1.11 & 9.5  &5.13E+2~(0.94)& BH \\
APR135165L & 240 & 130 & 30 & 20   & 60 & 1.11 & 7.8  &5.15E+2~(0.90)& BH \\
APR135165M & 278 & 155 & 30 & 20.5 & 70 & 1.11 & 8.9  &5.14E+2~(0.90)& BH \\
APR135165H & 310 & 175 & 30 & 21   & 80 & 1.10 & 9.9  &5.19E+2~(0.92)& BH \\
\hline\hline
\end{tabular}
\end{minipage}
\end{table*}

\section{Result}\label{sec:result}

\subsection{General feature for merger process} \label{sec:dyn}

We have already performed simulations for binary neutron stars
employing nuclear-theory-based EOSs 
\cite{Shibata:2005ss,Shibata:2006nm}. Although the qualitative feature
for the merger process found in the present work is the same as in the
previous works, we here summarize generic feature of the merger again.

Figure \ref{fig:alpc-rhoc} plots the evolution of the minimum value of
the lapse function, $\alpha_{\rm min}$, and maximum baryon rest-mass
density, $\rho_{\rm max}$, for all the models studied in this paper.
For models APR145145, APR1515, APR1316, and APR135165 for which a
black hole is formed in the dynamical time scale ($\sim \rho^{-1/2}$),
$\alpha_{\rm min}$ ($\rho_{\rm max}$) decreases (increases)
monotonically after the onset of merger.  In the high-resolution runs
for these models, two neutron stars come into the first contact at
$t\sim 9~{\rm ms}$. This merger time is slightly underestimated
because of the effect of finite grid resolution, but the numerical
results in the chosen grid resolution is in a convergent regime as
discussed in Sec.~\ref{sec:gw0}.
For all the cases, an apparent horizon is formed when $\alpha_{\rm
min}$ reaches $\sim0.03$.  For model APR1414, $\alpha_{\rm min}$
($\rho_{\rm max}$) steeply decreases (increases) after the onset of
merger, but then, they start oscillating and eventually settle down to
relaxed values. This indicates that the outcomes is a hypermassive
neutron star, for which the baryon rest mass is $\sim 30\%$ larger
than the maximum allowed mass of the spherical neutron stars
\cite{BSS}.

Figures~\ref{fig:14-eq}--\ref{fig:1316-eq} display the evolution of
density contour curves for the rest mass in the equatorial plane of
the inspiral, merger, and ringdown phases for runs APR1414H, APR1515H,
and APR1316H, respectively.  For the case of the equal-mass binaries,
two neutron stars are tidally deformed in a noticeable manner only
just before the onset of merger (see Figs.~\ref{fig:14-eq} and
\ref{fig:15-eq}).  If the total mass is not large enough to form a
black hole in the dynamical time scale after the onset of the merger,
a hypermassive neutron star of nonaxisymmetric structure is formed in
the central region. In this case, spiral arms are formed in the outer
region (see Fig. \ref{fig:14-eq}), and subsequently, they wind around
the formed hypermassive neutron star, generating shocks in its outer
region. Because of angular momentum dissipation by gravitational
radiation, hydrodynamic interaction, and hydrodynamic transport
process of angular momentum, the density and specific angular momentum
of the hypermassive neutron star are redistributed during the
subsequent evolution and eventually it relaxes to a moderately
ellipsoidal configuration. Here, the ellipticity remains because it is
rapidly rotating and also the EOS is stiff enough as mentioned in
Refs.~\cite{Shibata:2005ss,Shibata:2006nm}. As a result, the
hypermassive neutron star emits gravitational waves subsequently, and
it secularly evolves due to gravitational radiation reaction.

For the case when the total mass is large enough, the merged object
collapses to a black hole in the dynamical time scale after the onset
of merger (see Fig. \ref{fig:15-eq}).  In the equal-mass case, nearly
all the material are swallowed by the black hole in $\sim 1$ ms. 
The resulting final outcome is a black hole surrounded by a tiny disk 
of mass $< 10^{-3} M_{\odot}$. 

In the unequal-mass case, the less massive neutron star is tidally
deformed $\sim 1$ orbit before the onset of merger (see the second
panel of Fig. \ref{fig:1316-eq}).  Then, mass shedding occurs, and as
a result, the material of the less massive neutron star accrete onto
the massive companion.  During the merger, it is highly tidally
deformed, and thus, an efficient angular momentum transport occurs.
Due to this, the material in the outer region of the less massive
neutron star spread outward to form a spiral arm. This process helps
formation of accretion disks around the formed black hole. The disk
will survive for a time scale much longer than the dynamical time
scale as shown in Sec.~\ref{sec:BHmass}.

Figure \ref{fig:med} displays the density contour curves for the
rest-mass density and the local velocity field on the $x=0$ plane for
runs APR1414H, APR145145H, APR1316H, and APR135165H. These also show
that (i) a hypermassive neutron star of ellipsoidal shape is the
outcome for model APR1414 (panel (a)), (ii) a black hole with tiny
surrounding material is the outcome for model APR145145 (panel (b)),
and (iii) a black hole surrounded by disks is the outcome for models
APR1316 and APR135165 (panels (c) and (d)).

All these features are qualitatively the same as those reported in
Ref.~\cite{Shibata:2006nm}. However, in the previous work, we employed
initial conditions for which the initial separation is much smaller
than that in the present work, as $m_0\Omega_0 \sim 0.05$ where
$\Omega_0$ denotes initial orbital angular velocity. In the present
work, $m_0\Omega_0 \approx 0.027$, and this modification changes the
results quantitatively. Furthermore, we performed the simulations for
different total masses, and this leads to a new quantitative finding.
In the remaining part of this subsection, we summarize the
updated results.

As reported in the previous works
\cite{Shibata:2005ss,Shibata:2006nm}, the final outcome formed after
the merger (a black hole or neutron star) is primarily determined by a
relation between the initial total mass $m_0$ and the threshold mass
$M_{\rm thr}$ and by the mass ratio of the binary for a given EOS: The
binary neutron stars of $m_0 > M_{\rm thr}$ collapse to a black hole
after the onset of merger in dynamical time scale $\sim 1$ ms. On
the other hand, a hypermassive neutron star is formed for $m_0 <
M_{\rm thr}$, at least for a time much longer than the dynamical time
scale. The threshold mass depends weakly on the mass ratio of the
binary neutron stars.

The threshold mass, $M_{\rm thr}$, depends strongly on the EOSs (on
the stiffness of the EOS), and stiffer EOSs give larger values of
$M_{\rm thr}$. We reported in the previous work that for the APR EOS,
$2.8M_\odot < M_{\rm thr} < 3.0M_\odot$. In the present work, we find
that the value of $M_{\rm thr}$ is in a narrower range between
$2.8M_\odot$ and $2.9M_\odot$ via the study for models APR1414,
APR145145, and APR1316.  (Note that for models APR1414, APR145145, and
APR1515, $m_0=2.8M_{\odot}$, $2.9M_{\odot}$, and $3.0M_{\odot}$, 
respectively.)

In the previous paper \cite{Shibata:2006nm}, we reported that the disk mass
around the black hole is $\sim 4 \times 10^{-4}M_{\odot}$ for the
equal-mass binary neutron star of total mass $m_0=3M_{\odot}$ and
$\sim 0.003 M_{\odot}$ for the binary neutron star of mass 1.35 and
1.65$M_{\odot}$.  We find that the mass of the disk at $t-t_{\rm AH}=3~{\rm
  ms}$ is $\approx 7 \times 10^{-5}M_{\odot}$ for model APR1515H and
$\approx 6 \times 10^{-3} M_{\odot}$ for model APR135165 (see
Fig. \ref{fig:AH-disk}).  Here, $t_{\rm AH}$ is the time when the apparent
horizon is first formed.  Thus, the disk mass is corrected by a factor
of several, although the order of magnitude does not change. 
Also, we did not clarify the dependence of the disk
mass on the total mass $m_0$ in the previous work. For model APR1316,
we find that the disk mass at $t-t_{\rm AH}=3~{\rm ms}$ is $2.4 \times
10^{-2} M_{\odot}$, and thus, the disk mass depends strongly on the
total mass of the system. More details about the merger process, final
outcome, and disk mass will be discussed in subsequent subsections for
each model separately. 

\subsubsection{APR1414}\label{subsec1414}

\begin{figure*}
  \begin{center}
  \vspace*{40pt}
    \begin{tabular}{cc}
      \begin{minipage}{0.5\hsize}
      \includegraphics[width=8.0cm]{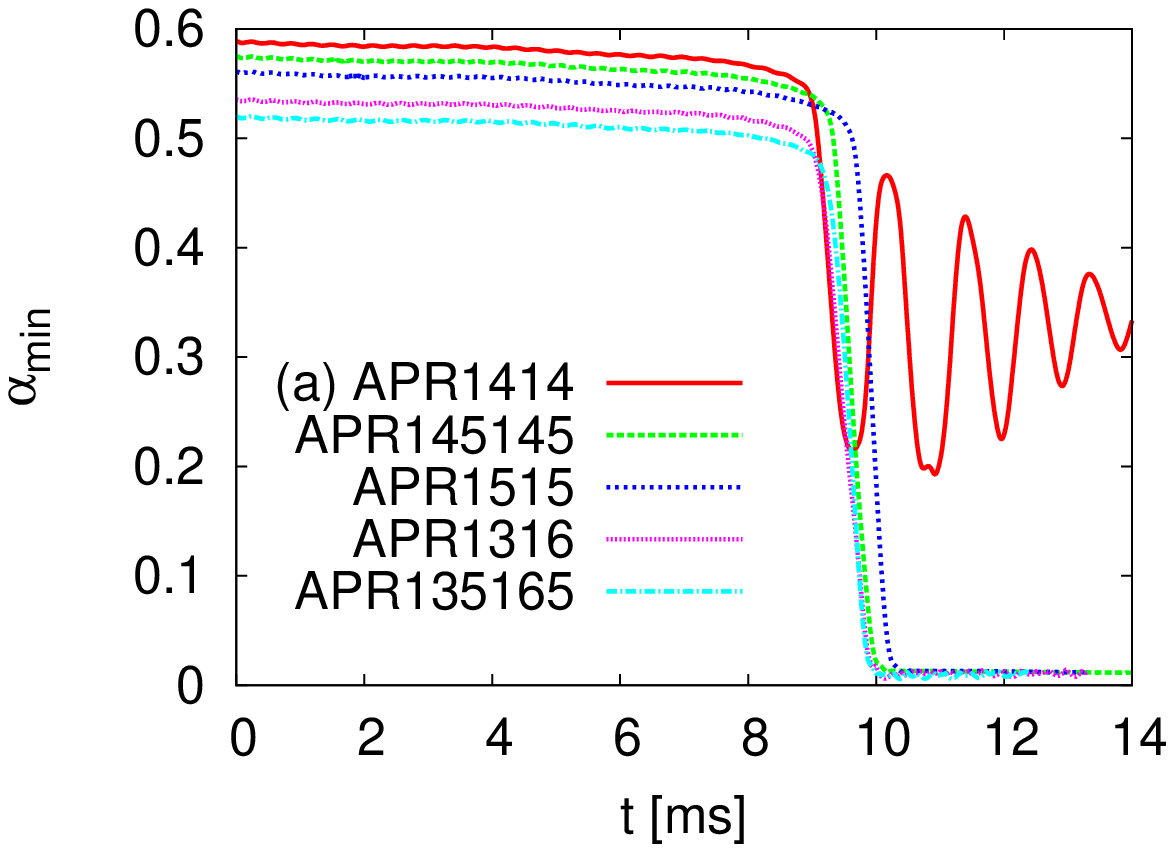}
      \end{minipage}
      \hspace{-1.0cm}
      \begin{minipage}{0.5\hsize}
      \includegraphics[width=8.0cm]{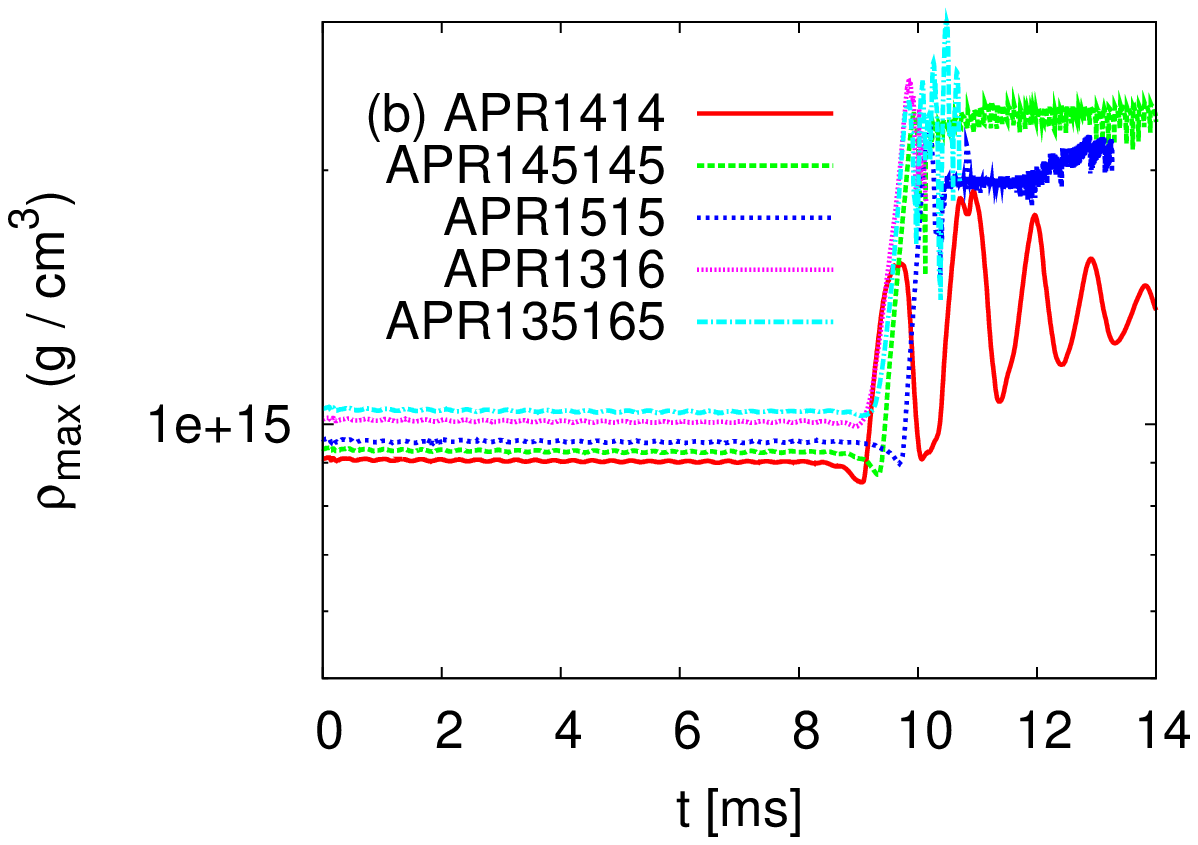}
      \end{minipage}
    \end{tabular}
    \caption{\label{fig:alpc-rhoc} The evolution of the minimum value
      of the lapse function, $\alpha_{\rm min}$, and the maximum
      rest-mass density $\rho_{\rm max}$ for runs APR1414H,
      APR145145H, APR1515H, APR1316H, and APR135165H.  }
  \end{center}
\end{figure*}

\begin{figure*}
  \begin{center}
  \vspace*{40pt}
    \begin{tabular}{ccc}
      \begin{minipage}{0.5\hsize}
      \includegraphics[width=8.0cm]{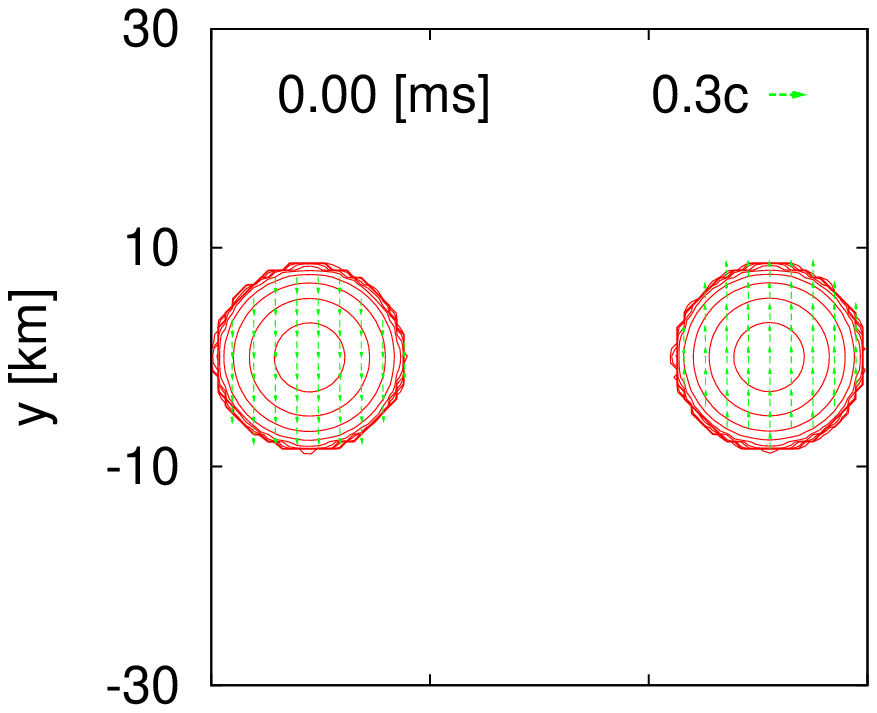}
      \end{minipage}
      \hspace{-4.95cm}
      \begin{minipage}{0.5\hsize}
      \includegraphics[width=8.0cm]{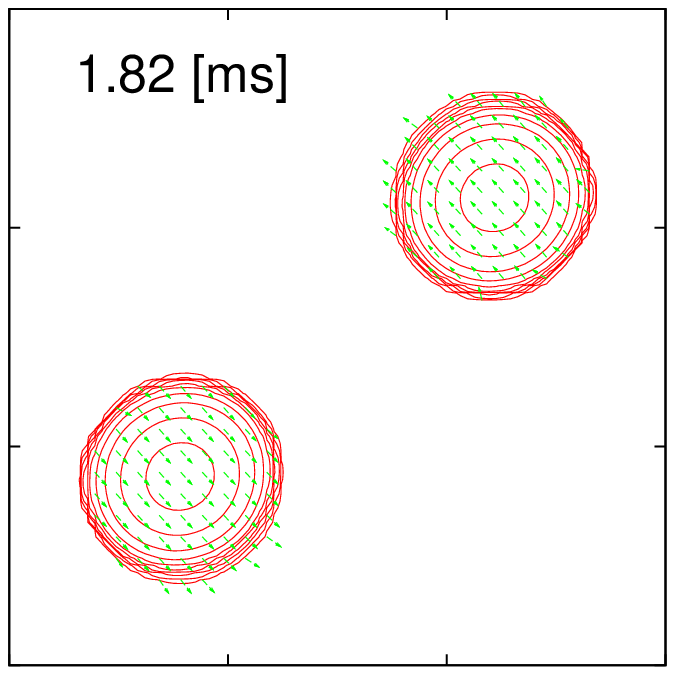}
      \end{minipage}
      \hspace{-4.92cm}
      \begin{minipage}{0.5\hsize}
      \includegraphics[width=8.0cm]{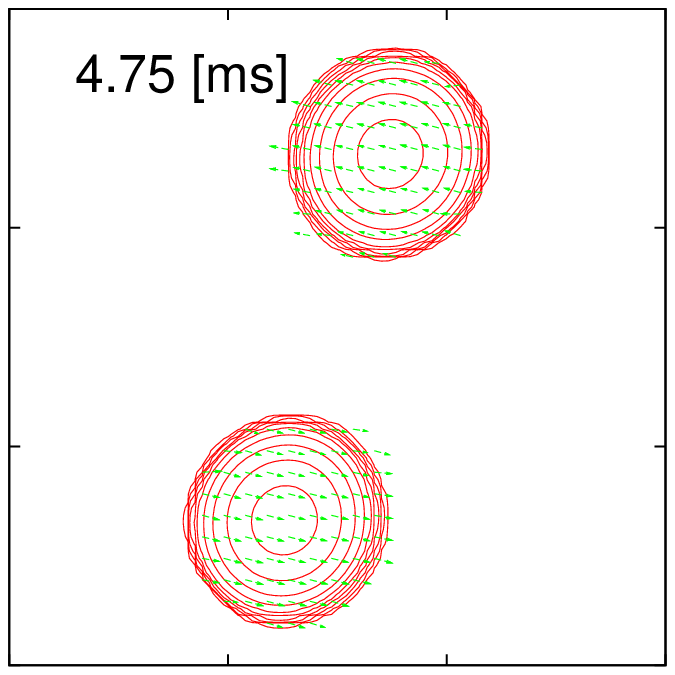}
      \end{minipage}
      \vspace{-1.39cm}
      \\
      \begin{minipage}{0.5\hsize}
      \includegraphics[width=8.0cm]{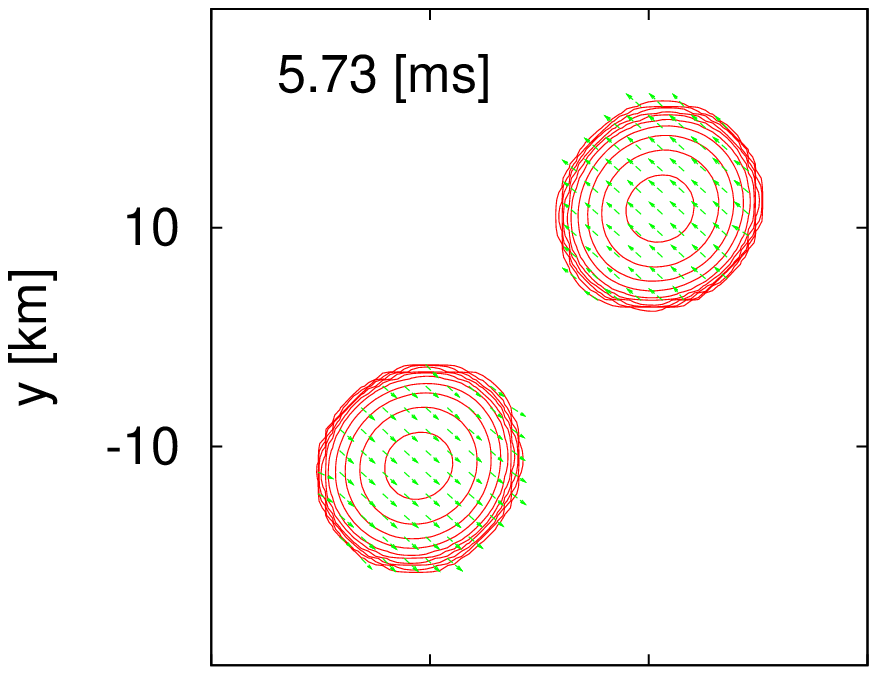}
      \end{minipage}
      \hspace{-4.95cm}
      \begin{minipage}{0.5\hsize}
      \includegraphics[width=8.0cm]{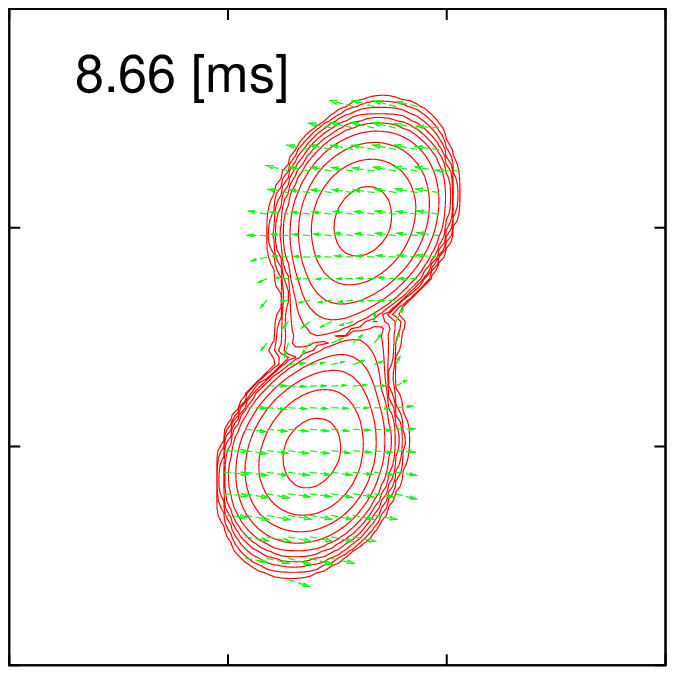}
      \end{minipage}
      \hspace{-4.92cm}
      \begin{minipage}{0.5\hsize}
      \includegraphics[width=8.0cm]{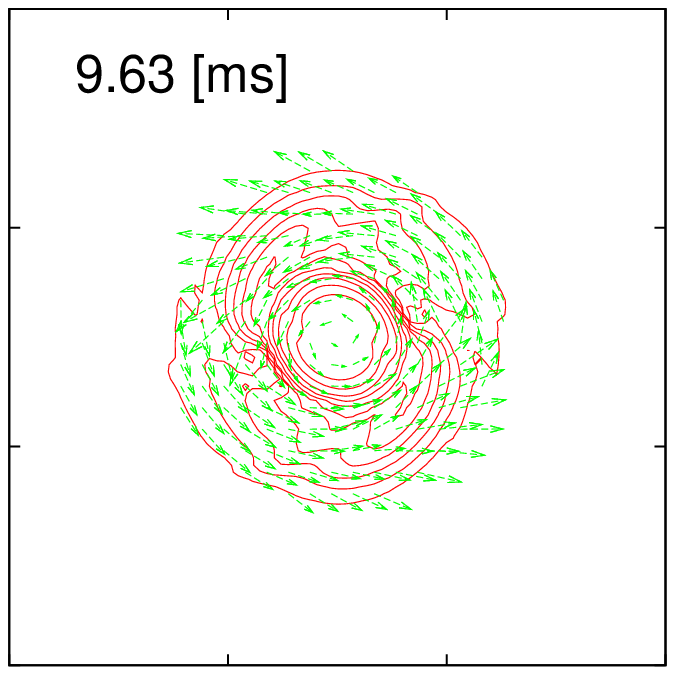}
      \end{minipage}
      \vspace{-0.75cm}
      \\
      \begin{minipage}{0.5\hsize}
      \includegraphics[width=8.0cm]{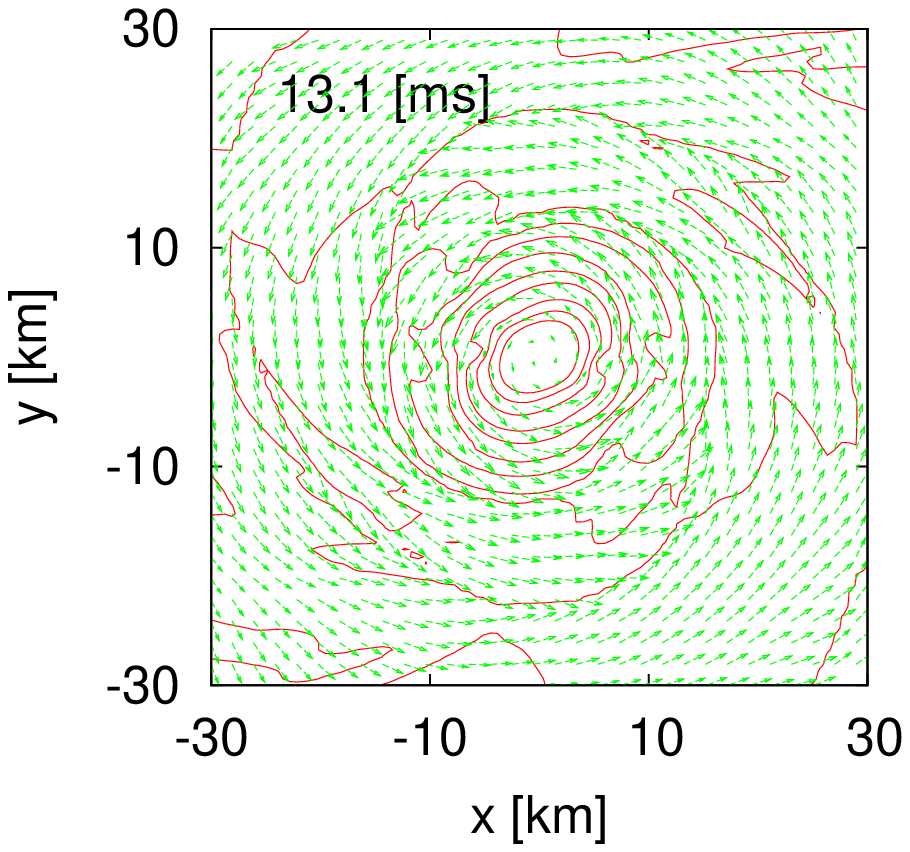}
      \end{minipage}
      \hspace{-5.24cm}
      \begin{minipage}{0.5\hsize}
      \includegraphics[width=8.0cm]{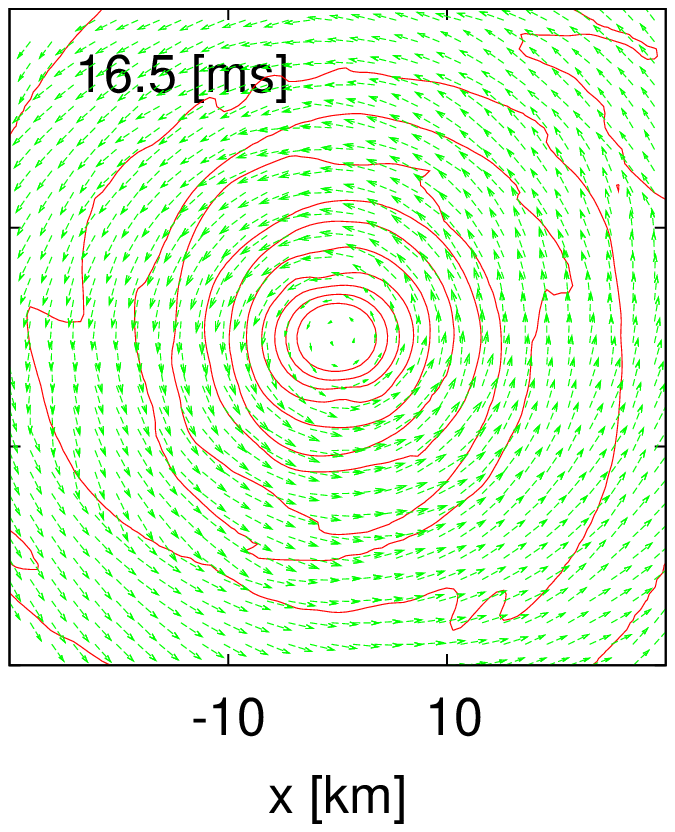}
      \end{minipage}
      \hspace{-4.94cm}
      \begin{minipage}{0.5\hsize}
      \includegraphics[width=8.0cm]{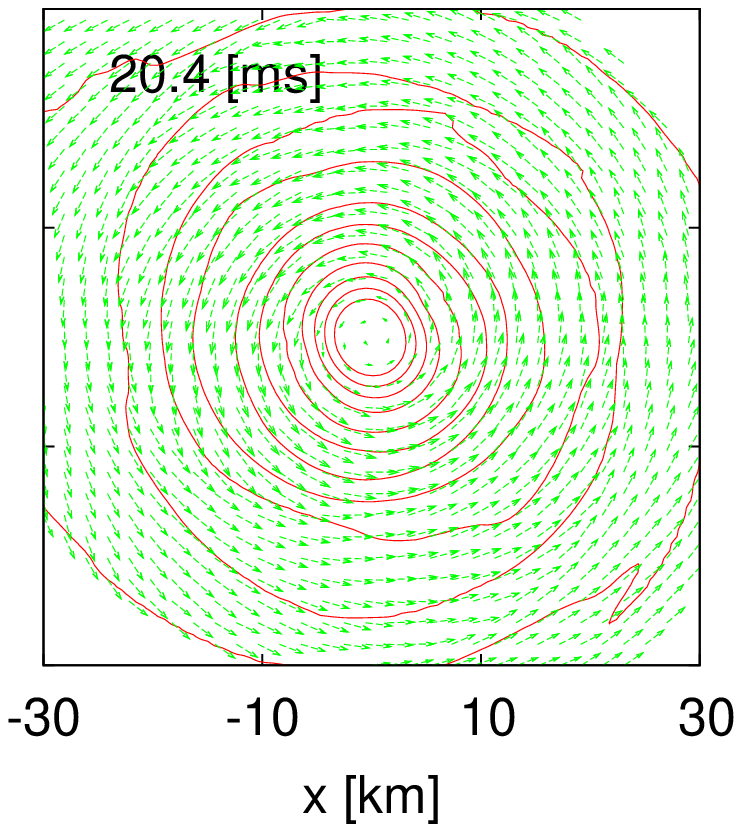}
      \end{minipage}
      \hspace{0.1cm}
      \\
    \end{tabular}
    \caption{\label{fig:14-eq} Snapshots of the density contour curves
      for $\rho$ and velocity field $(v^x,v^y)$ on the equatorial
      plane for run APR1414H. The solid contour curves are drawn for
      $\rho=i\times2\times10^{14}{\rm g/cm^3}(i=1,2,\cdots)$ and for
      $1\times10^{14-0.5i}{\rm g/cm^3}(i=1\sim 6)$. The time shown in
      the upper-left side denotes the elapsed time from the beginning
      of the simulation. }
  \end{center}
\end{figure*}

\begin{figure*}
  \begin{center}
  \vspace*{40pt}
    \begin{tabular}{ccc}
      \begin{minipage}{0.5\hsize}
      \includegraphics[width=8.0cm]{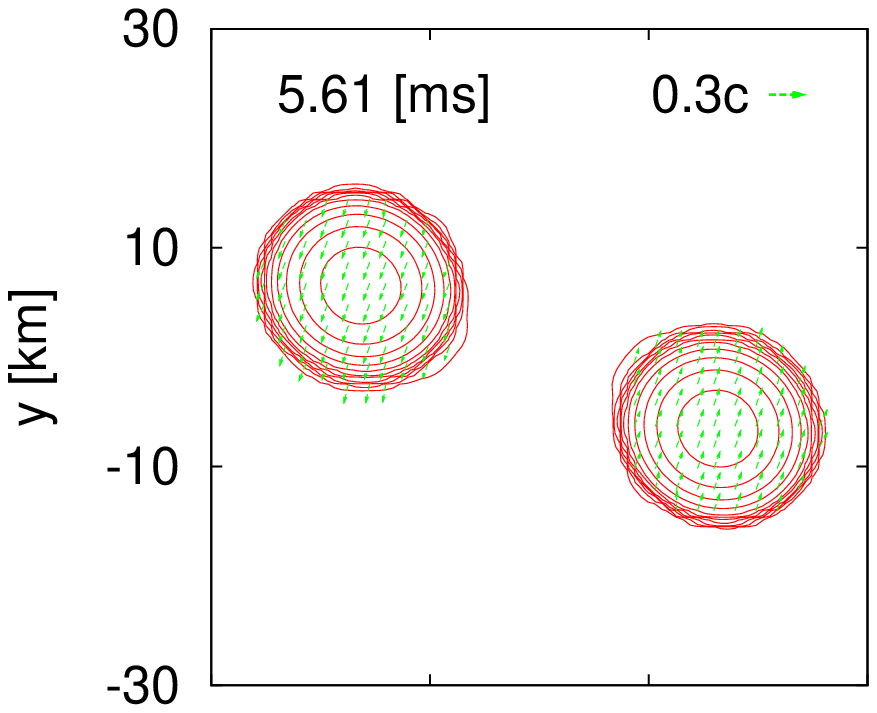}
      \end{minipage}
      \hspace{-4.95cm}
      \begin{minipage}{0.5\hsize}
      \includegraphics[width=8.0cm]{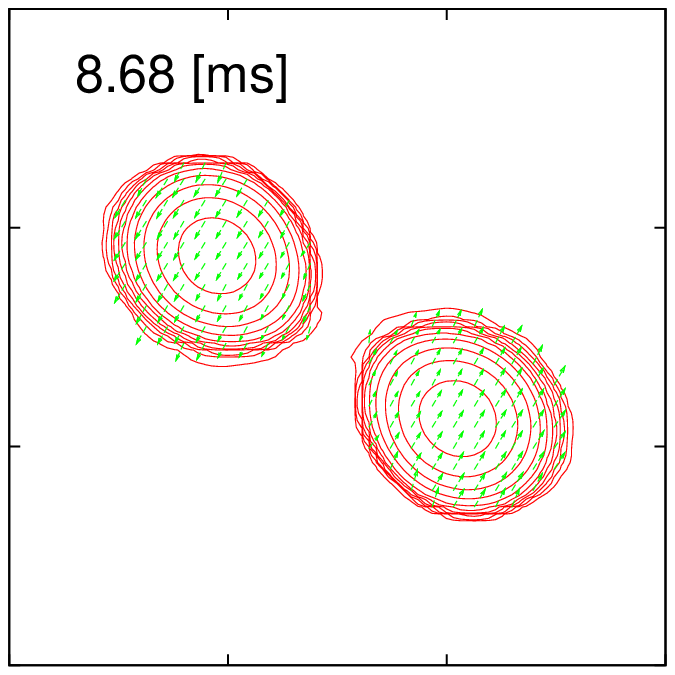}
      \end{minipage}
      \hspace{-4.92cm}
      \begin{minipage}{0.5\hsize}
      \includegraphics[width=8.0cm]{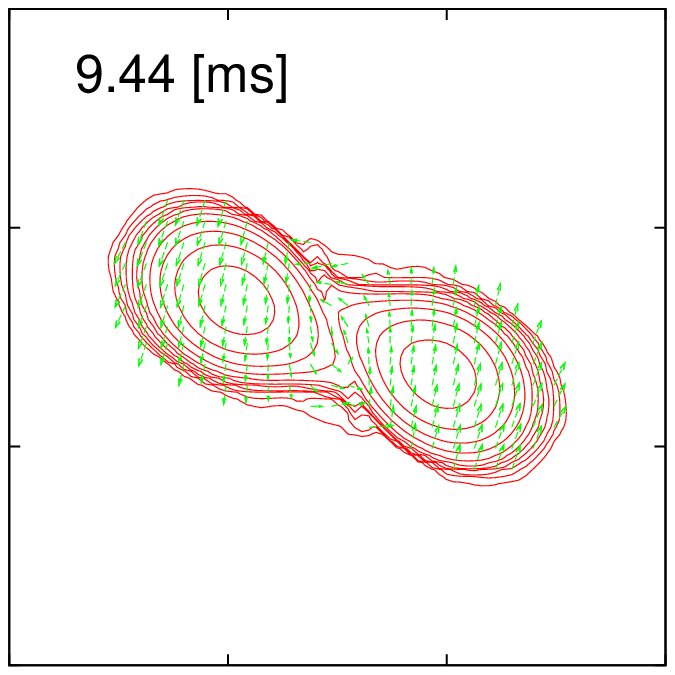}
      \end{minipage}
      \vspace{-0.75cm}
      \\
      \begin{minipage}{0.5\hsize}
      \includegraphics[width=8.0cm]{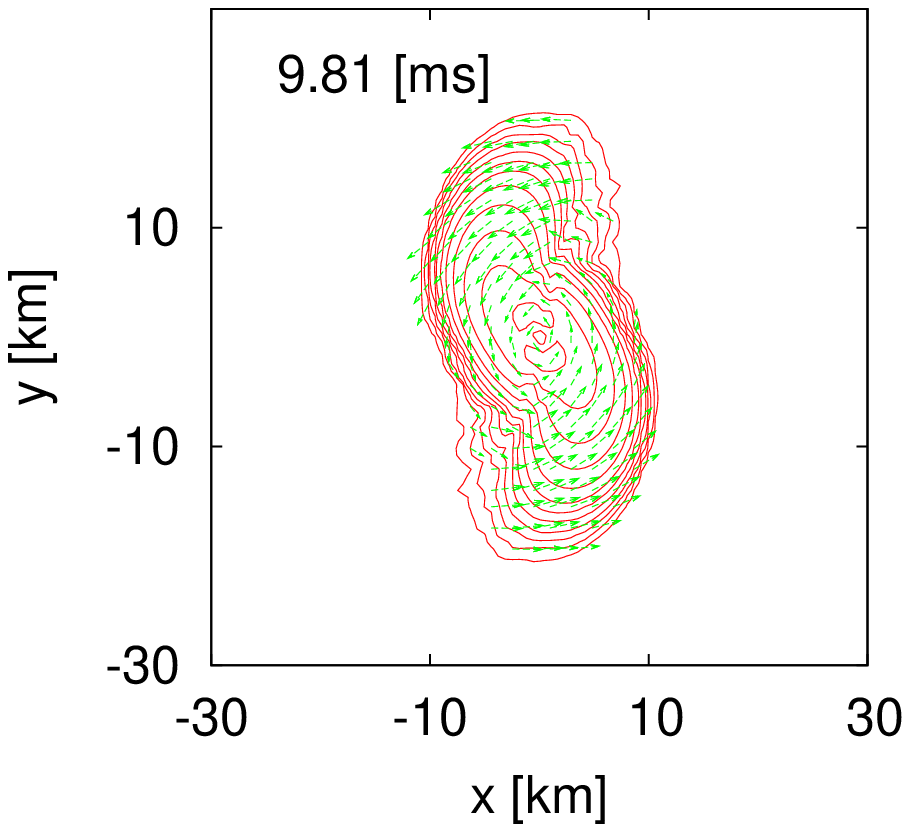}
      \end{minipage}
      \hspace{-5.24cm}
      \begin{minipage}{0.5\hsize}
      \includegraphics[width=8.0cm]{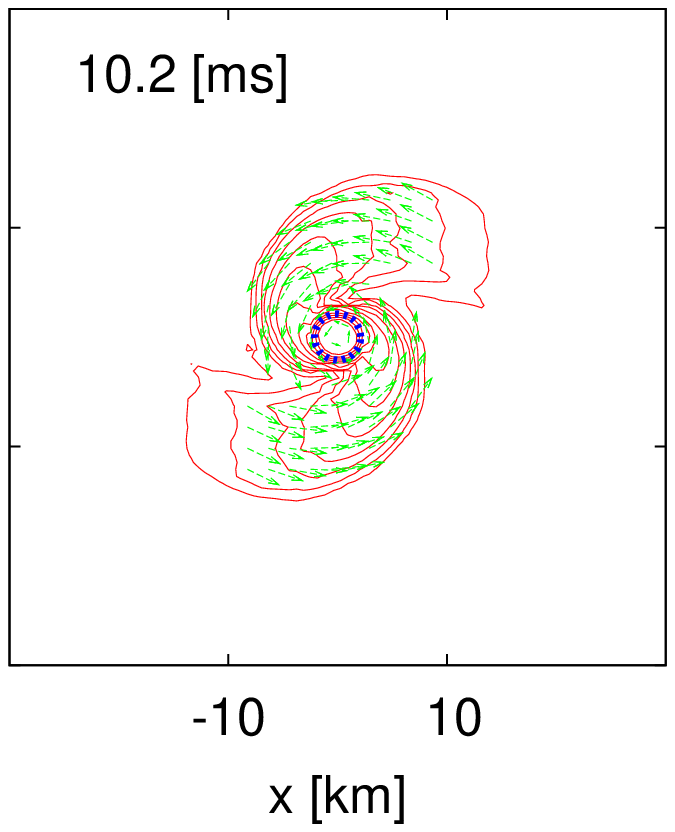}
      \end{minipage}
      \hspace{-4.94cm}
      \begin{minipage}{0.5\hsize}
      \includegraphics[width=8.0cm]{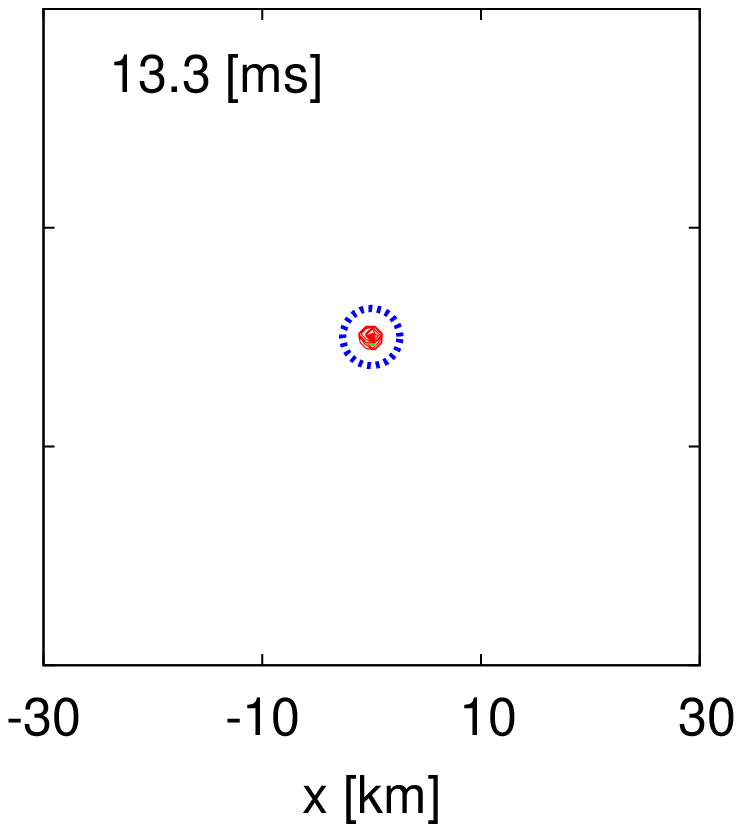}
      \end{minipage}
      \hspace{0.1cm}
      \\
    \end{tabular}
    \caption{\label{fig:15-eq} The same as Fig. \ref{fig:14-eq} but
      for run APR1515H.  The short-dashed circles around the origin in
      the last two panels denote the location of the apparent horizon.
    }
  \end{center}
\end{figure*}

\begin{figure*}
  \begin{center}
  \vspace*{40pt}
    \begin{tabular}{ccc}
      \begin{minipage}{0.5\hsize}
      \includegraphics[width=8.0cm]{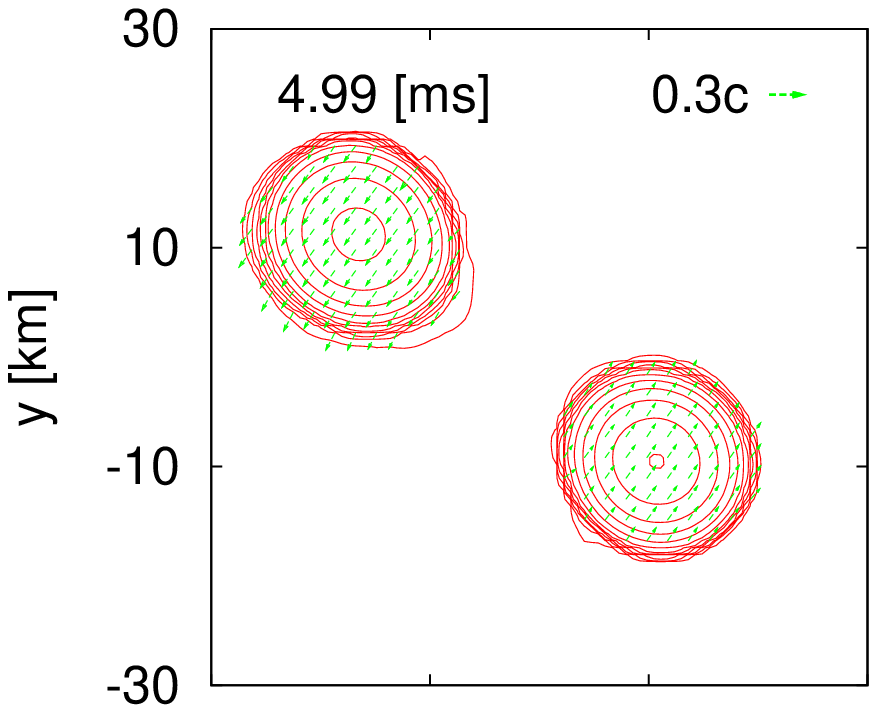}
      \end{minipage}
      \hspace{-4.95cm}
      \begin{minipage}{0.5\hsize}
      \includegraphics[width=8.0cm]{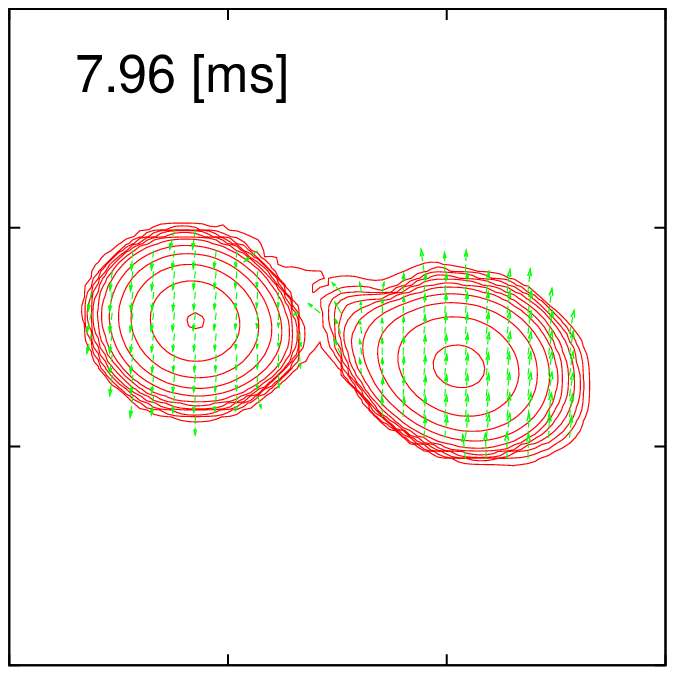}
      \end{minipage}
      \hspace{-4.92cm}
      \begin{minipage}{0.5\hsize}
      \includegraphics[width=8.0cm]{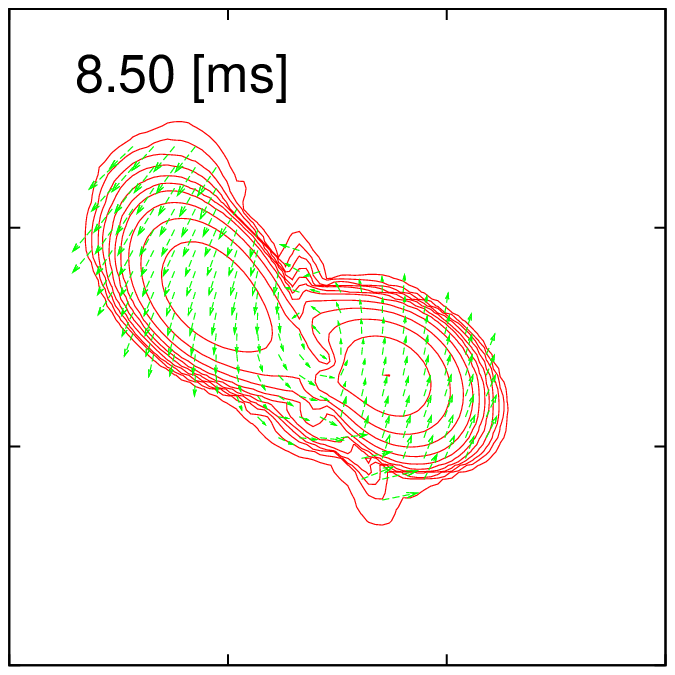}
      \end{minipage}
      \vspace{-0.75cm}
      \\
      \begin{minipage}{0.5\hsize}
      \includegraphics[width=8.0cm]{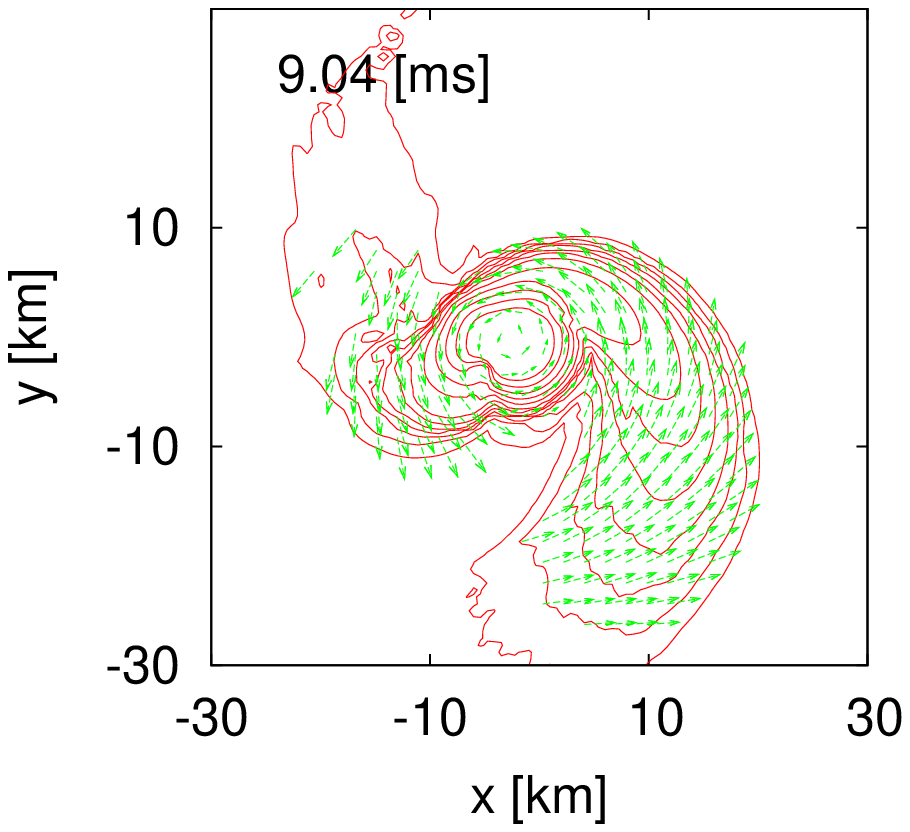}
      \end{minipage}
      \hspace{-5.24cm}
      \begin{minipage}{0.5\hsize}
      \includegraphics[width=8.0cm]{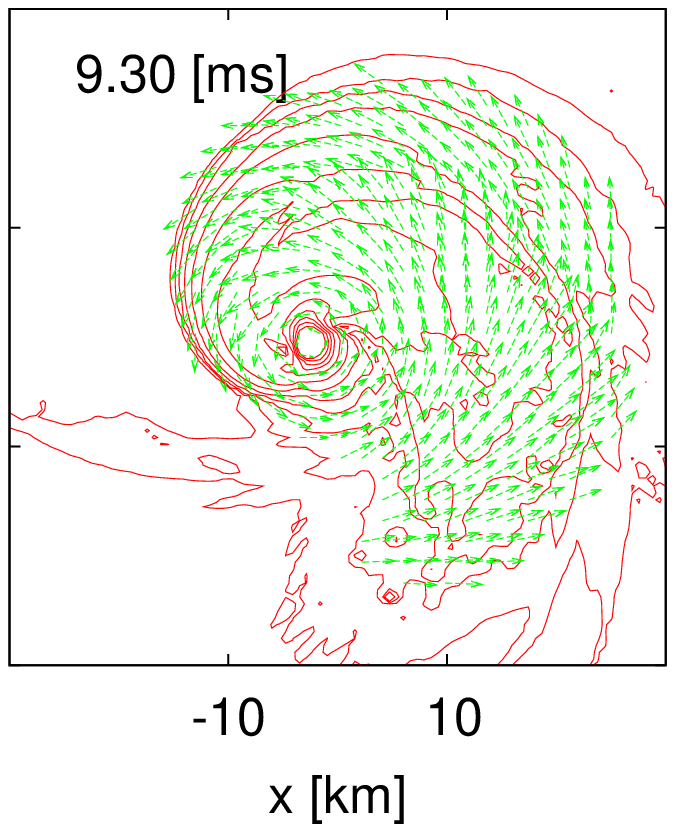}
      \end{minipage}
      \hspace{-4.94cm}
      \begin{minipage}{0.5\hsize}
      \includegraphics[width=8.0cm]{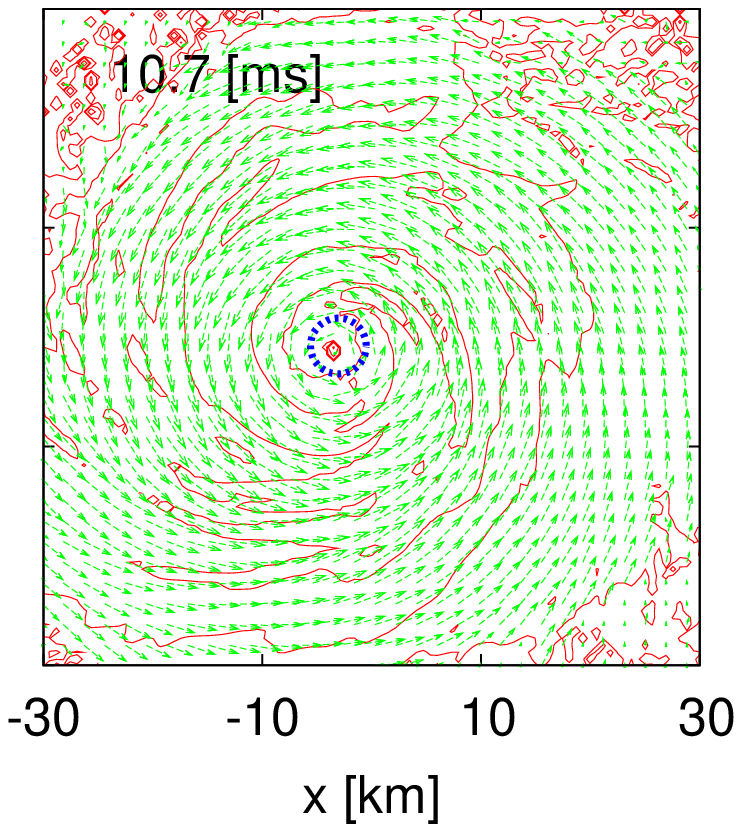}
      \end{minipage}
      \hspace{0.1cm}
      \\
    \end{tabular}
    \caption{\label{fig:1316-eq}
    The same as Fig. \ref{fig:15-eq} but for run APR1316H. 
    }
  \end{center}
\end{figure*}

\begin{figure*}
  \begin{center}
  \vspace*{40pt}
    \begin{tabular}{cc}
      \begin{minipage}{0.5\hsize}
      \includegraphics[width=8.0cm]{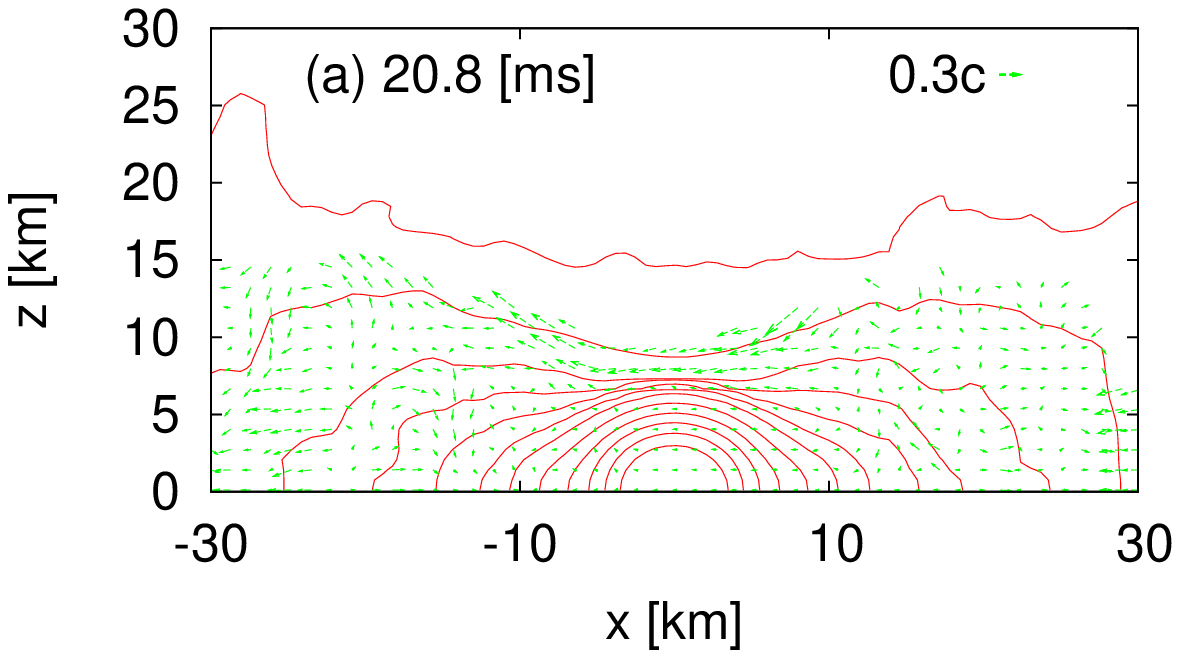}
      \end{minipage}
      \hspace{-1.0cm}
      \begin{minipage}{0.5\hsize}
      \includegraphics[width=8.0cm]{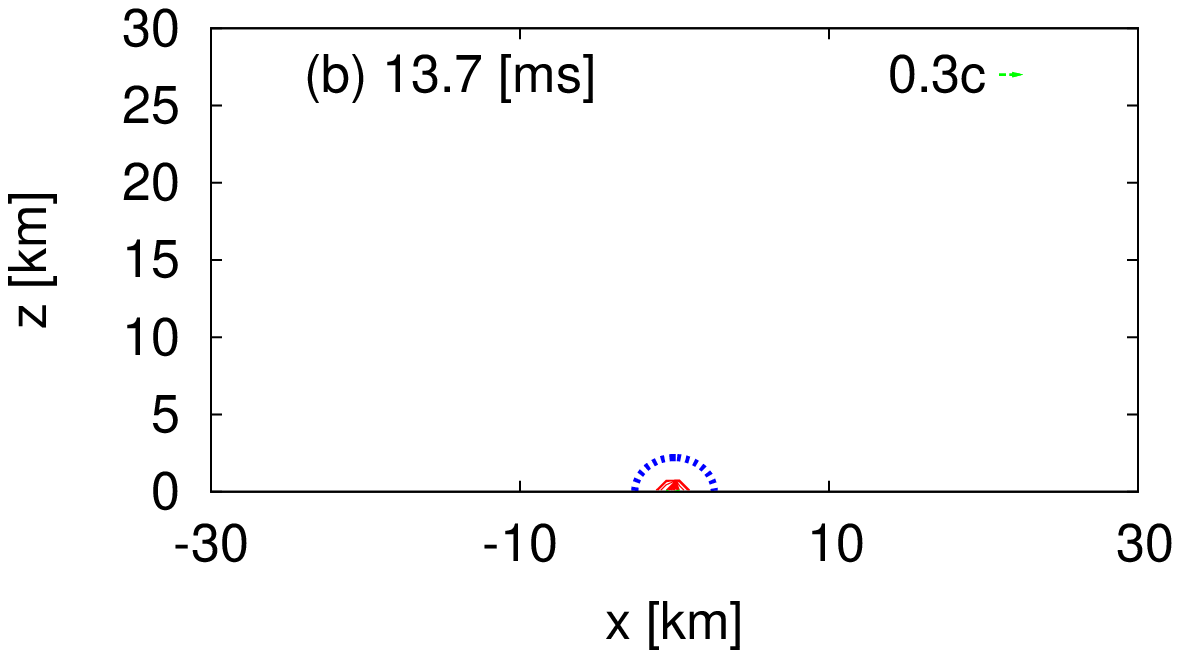}
      \end{minipage}
      \\
      \begin{minipage}{0.5\hsize}
      \includegraphics[width=8.0cm]{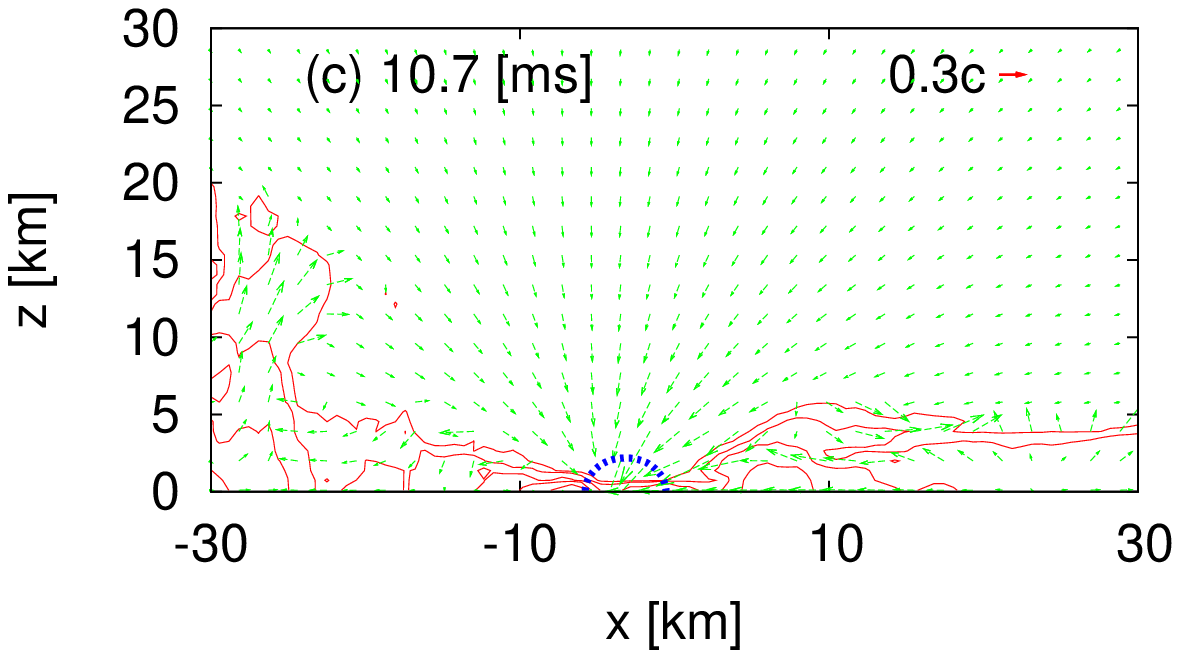}
      \end{minipage}
      \hspace{-1.0cm}
      \begin{minipage}{0.5\hsize}
      \includegraphics[width=8.0cm]{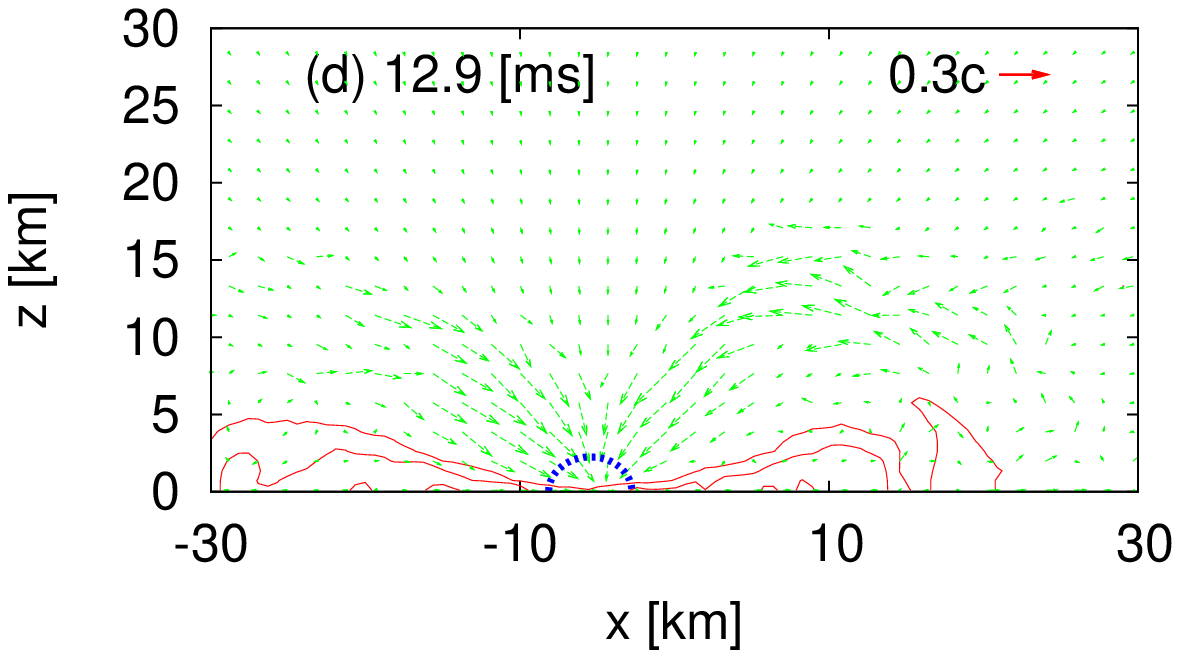}
      \end{minipage}
    \end{tabular}
    \caption{\label{fig:med} Snapshots of the density contour curves
      for $\rho$ and velocity field $(v^y,v^z)$ on the $x=0$ plane for
      runs (a) APR1414H, (b) APR145145H, (c) APR1316H, and (d)
      APR135165H.  The time shown in the upper-left side denotes the
      elapsed time from the beginning of the simulation.  The
      short-dashed half circles on the equatorial plane denotes the
      location of the apparent horizon.  }
  \end{center}
\end{figure*}

The total mass of this model is slightly smaller than the threshold
mass, i.e., $m_0 < M_{\rm thr}$, and hence, a compact hypermassive
neutron star is formed. Because the total mass is close to $M_{\rm
thr}$ in this model, the merged object first becomes very compact;
e.g., $\alpha_{\rm min}$ and $\rho_{\rm max}$ reach $\sim 0.2$ and
$\sim 2 \times 10^{15}~{\rm g/cm^3}$ soon after the first contact (see
Fig. \ref{fig:alpc-rhoc} (a)). Hence, the self-gravity is only
slightly insufficient for inducing collapse to a black hole. The
merged object subsequently bounces back to become an oscillating
hypermassive neutron star.  The quasiradial oscillation is repeated
for several times (see Fig. \ref{fig:alpc-rhoc}), and then the
oscillation is damped due to shock dissipation and the hypermassive
neutron star relaxes to a quasisteady ellipsoidal
configuration~\cite{Shibata:2006nm}, as found from the last panel of
Fig. \ref{fig:14-eq} and Fig. \ref{fig:med} (a). These figures show
that axial ratio of the two major axes on the equatorial plane is
$\sim 0.9$, and that of the polar coordinate radius $r_{\rm p}$ to the
equatorial one $r_{\rm e}$ is $r_{\rm p}/r_{\rm e} \sim 0.7$. Thus,
the ellipticity is not negligible. Such high ellipticity seems to
reflect the fact that the hypermassive neutron star is rapidly
rotating \cite{CH69}. 

\begin{figure*}
  \begin{center}
  \vspace*{40pt}
    \begin{tabular}{cc}
      \begin{minipage}{0.5\hsize}
      \includegraphics[width=8.0cm]{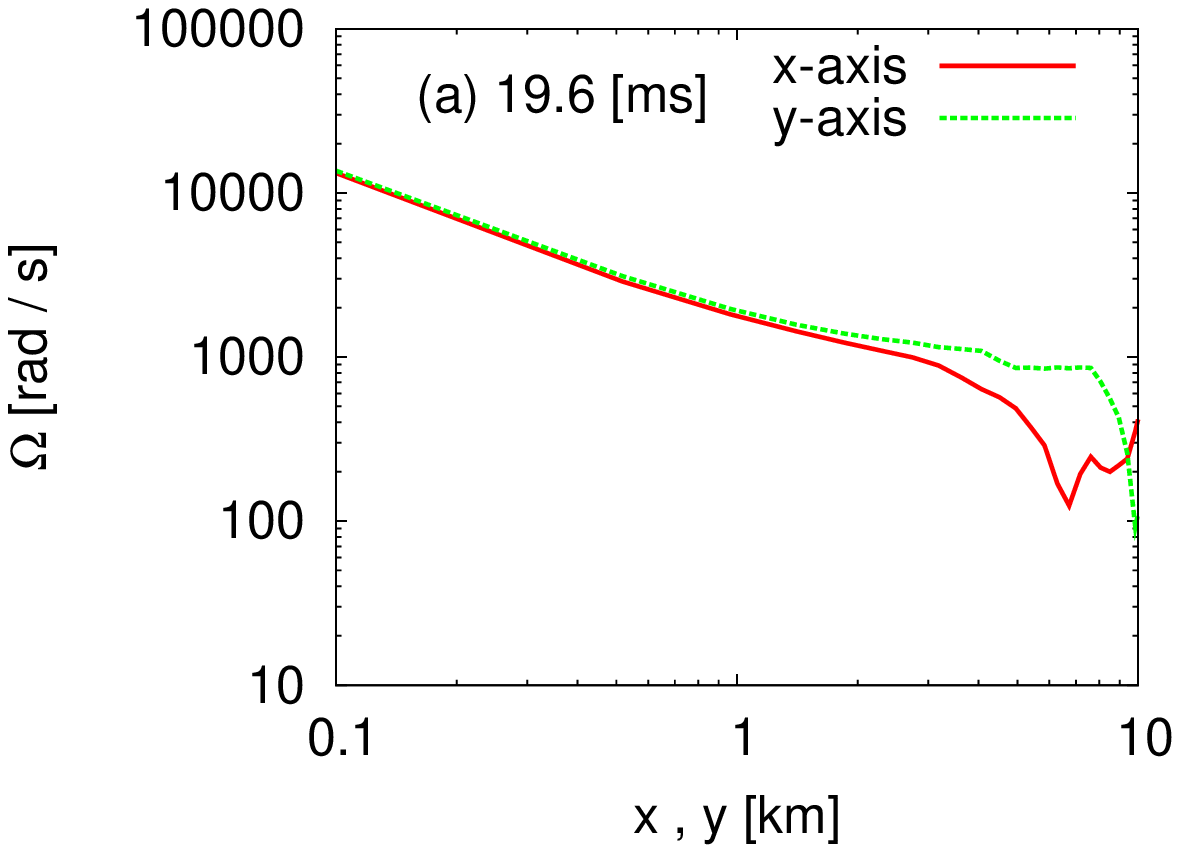}
      \end{minipage}
      \hspace{-1.0cm}
      \begin{minipage}{0.5\hsize}
      \includegraphics[width=8.0cm]{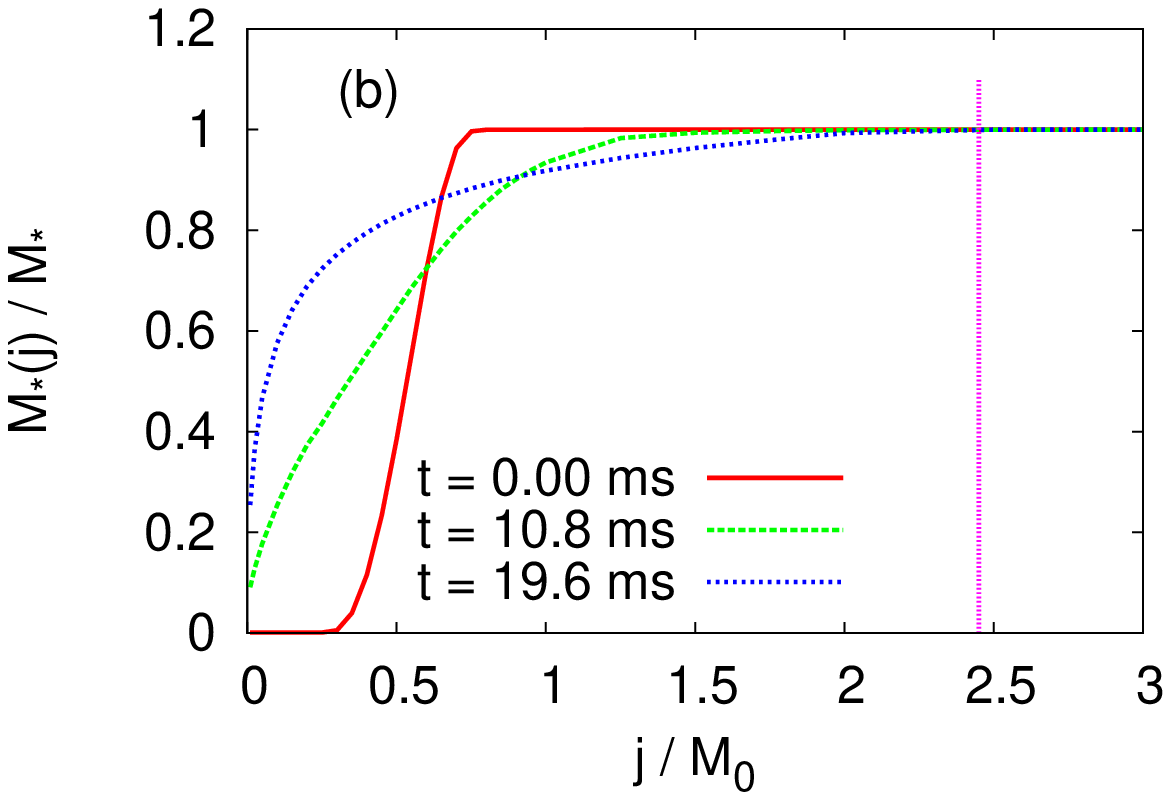}
      \end{minipage}
      \\
      \begin{minipage}{0.5\hsize}
      \includegraphics[width=8.0cm]{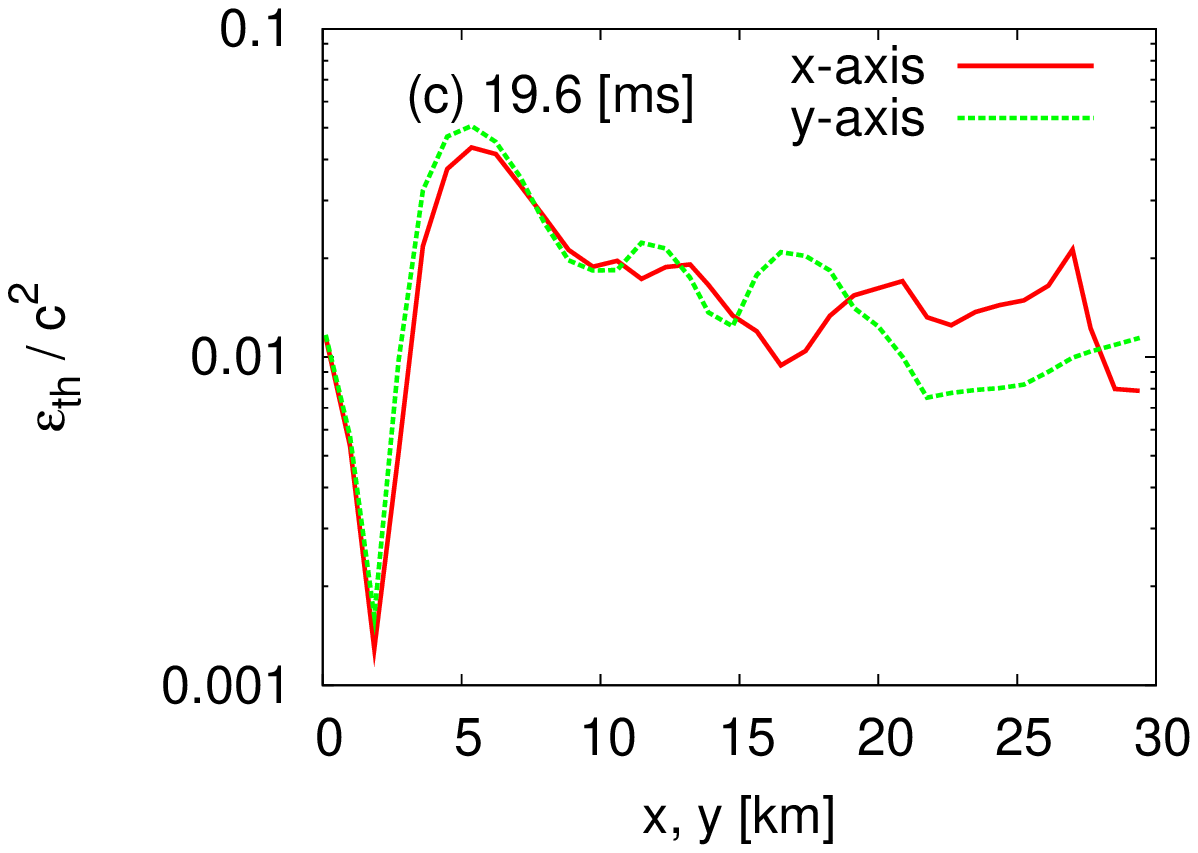}
      \end{minipage}
      \hspace{-1.0cm}
      \begin{minipage}{0.5\hsize}
      \includegraphics[width=8.0cm]{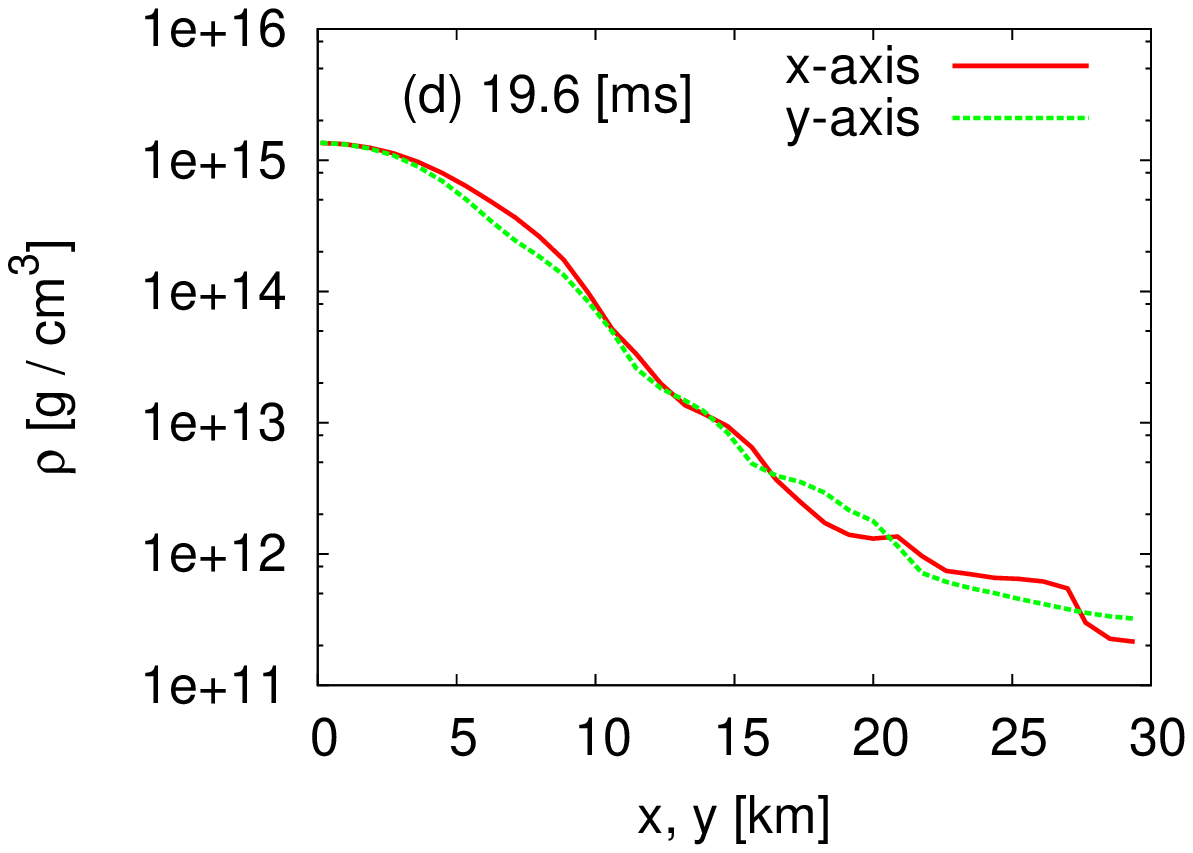}
      \end{minipage}
      \\
      \begin{minipage}{0.5\hsize}
      \includegraphics[width=8.0cm]{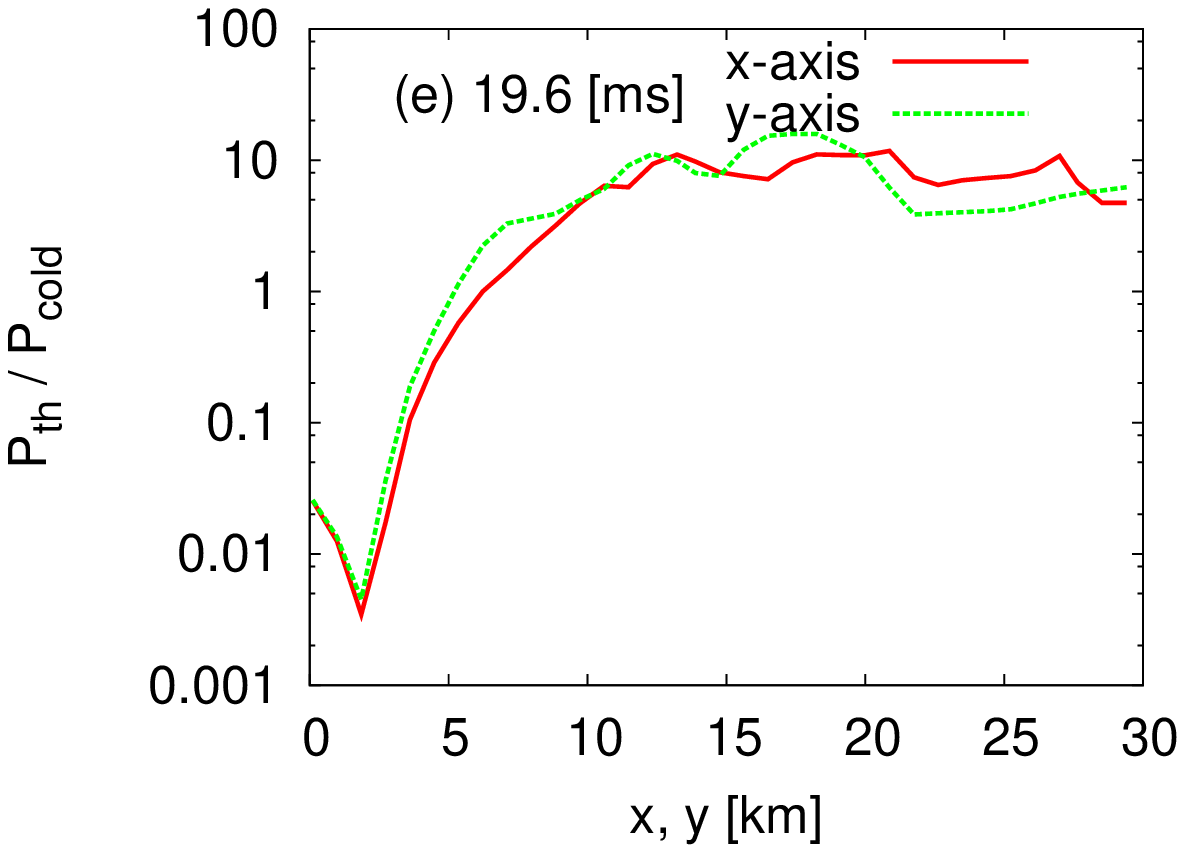}
      \end{minipage}
      \\
    \end{tabular}
    \caption{\label{fig:vel-prof} Various quantities of hypermassive
      neutron star formed for run APR1414H. (a) Profiles of
      angular velocity along $x$ and $y$ axes at $t=19.6$ ms; (b) Mass
      spectrum as a function of specific angular momentum at the
      selected time slices. The vertical dotted line denotes the specific
      angular momentum at the innermost stable circular orbit around
      the black hole of mass $0.97M_0$ and spin $a=0.77$; (c)--(e)
      Specific thermal energy, rest-mass density, and ratio of the
      thermal pressure to the nuclear-matter pressure, $P_{\rm
        th}/P_{\rm cold}$, along $x$ and $y$ axes at $t=19.6$ ms.  }
  \end{center}
\end{figure*}

Figure \ref{fig:vel-prof} (a) plots angular velocity profile of the
hypermassive neutron star in the late phase along $x$ and $y$
axes. This shows that the hypermassive neutron star rotates rapidly
and differentially. In the central region, the rotational period is
$\sim 0.5$ ms, comparable to the dynamical time scale.  This implies
that centrifugal force around the central region plays an important
role for supporting self-gravity. Note that the mass of the
hypermassive neutron star is by $\approx 30\%$ larger than the maximum
allowed mass of spherical neutron stars of the APR EOS (which is
$\approx 2.2M_{\odot}$), and hence in the absence of rotation, the
hypermassive neutron star would collapse to a black hole. Thus, the
rapid rotation near the central region seems to be an essential agent
for supporting its strong self-gravity. (We note that thermal energy
generated by shock heating during the merger phase in part plays a
role for supporting the self-gravity; see discussion below.)

As we will discuss in Sec.~\ref{sec:rad}, the hypermassive neutron
star continuously emits gravitational waves and loses angular
momentum because it has a nonaxisymmetric shape and rapid rotation.
This indicates that it may eventually collapse to a black hole after a
substantial fraction of the angular momentum is dissipated from the
central region.  The lifetime of the hypermassive neutron star may be
estimated by calculating the time scale for the angular momentum loss due
to the gravitational-wave emission.  At the end of the simulation ($t
\sim 20$ ms), the angular momentum and its dissipation rate are $J
\sim 0.74 J_0\sim 4.9\times 10^{49}{\rm g~cm^2~s^{-1}}$ and $dJ/dt
\sim 6.7 \times 10^{49}~{\rm g~cm^2s^{-2}}$, respectively.  The
lifetime of the hypermassive neutron star may be estimated by
$J/(dJ/dt) \sim 700~{\rm ms}$. $dJ/dt$ (namely the amplitude of
gravitational waves) gradually decreases with time, and hence, this
time scale should be regarded as the shortest one. However, the
decrease time scale of $dJ/dt$ is not as short as the dynamical time
scale, and thus, the estimated time scale is likely to be correct
within the factor of 2--3. If this estimation is correct, the
hypermassive neutron star would collapse to a black hole in a few
seconds (but see discussion below for other possibilities).  Other
physical processes, which are not taken into account in this work,
could also contribute to dissipating and/or transporting angular
momentum: Because the hypermassive neutron star rotates differentially
as shown in Fig. \ref{fig:vel-prof}, magnetic fields might be
amplified by the magnetorotational instability and/or magnetic
winding~\cite{Balbus:1991,Shibata:2006hr,Duez:2006qe}. As a result,
angular momentum may be transported efficiently, leading the
hypermassive neutron star to collapse to a black hole. If the time
scale for the magnetic processes is shorter than the emission time
scale of gravitational waves, they would be the main agent for
inducing gravitational collapse to a black hole.

As discussed above, the hypermassive neutron star eventually collapses
to a black hole in any scenario. Because the hypermassive neutron star
has a spread envelope (see the last three panels of Fig. 3), the final 
black hole formed after the gravitational collapse may be surrounded
by an accretion disk. To qualitatively estimate the outcome, we generate
Fig.~\ref{fig:vel-prof} (b) which shows the evolution of the mass
spectrum as a function of the specific angular momentum $M_*(j)$ where
$j=h u_\varphi$ is the specific angular momentum of each fluid
element. Here, $M_*(j)$ is defined as an integrated baryon mass of
fluid elements with $j > j'$;
\begin{eqnarray}
M_*(j) = \int_{j'<j} \rho_*(x') d^3x'.
\end{eqnarray}
Figure \ref{fig:vel-prof} (b) shows that the value of $j$ 
for most of the fluid elements is smaller than $2M_0$ for $t \alt 20$ ms. 
However, the fraction of the mass element of $j \geq M_0$ increases 
with time. This indicates that angular momentum is transported outward 
due to nonaxisymmetric hydrodynamic interaction. 

The ADM mass and angular momentum at the end of the simulation ($t
\approx 20$ ms) are $\approx 0.97M_0$ and $0.74J_0$, respectively, and
thus, the nondimensional spin parameter of the system is $\sim 0.77$.
Then, assume that the final mass and spin of the black hole would be
$M_{\rm BH} \sim 0.97M_0$ and $a \sim 0.77$. Specific angular momentum
at innermost stable circular orbit (ISCO), $j_{\rm ISCO}$, of such a
rotating black hole is given by $\approx 2.45M_0$~ (e.g.,
Ref. \cite{shapiro}).  This suggests that a fluid element of $j \agt
2.45M_0$ would be able to form a disk surrounding the black
hole.  Figure \ref{fig:vel-prof} (b) shows that any mass element does
not have specific angular momentum large enough to form the disk.
However, the profile of the mass spectrum quickly changes with time
by the hydrodynamic angular momentum transport, as mentioned above. 
The estimated lifetime of the hypermassive neutron star is a few seconds, 
and much longer than 20 ms. This suggests that a substantial fraction 
of the fluid elements may form an accretion disk. 


Finally, we touch on thermal effects on the evolution of the
hypermassive neutron star.  Because shocks are generated during the
merger and in the subsequent dynamical phase, the hypermassive neutron
star is heated up and as a result, the specific thermal energy,
$\varepsilon_{\rm th}$, becomes nonzero. To clarify the role of the
generated thermal energy, we plot profiles of $\varepsilon_{\rm th}$,
$\rho$, and $P_{\rm th}/P_{\rm cold}$ along $x$ and $y$ axes in
Fig. \ref{fig:vel-prof} (c)--(e). Figure \ref{fig:vel-prof} (c) shows
that the value of $\varepsilon_{\rm th}/c^2$ is 0.002--0.05 for a
region of $\rho \geq \rho_{\rm nuc} \approx 2 \times 10^{14}~{\rm
  g/cm^3}$, where $\rho_{\rm nuc}$ denotes the nuclear density, and
for the region of subnuclear density, it is 0.01--0.02. Assuming that
the matter field is composed of neutron gas and thermal radiation, the
temperature of the region with $\rho \agt 10^{11}~{\rm g/cm^3}$ (above
which the optical depth for neutrino transport would be larger than
unity and the cooling due to the neutrino emission would not be
efficient) is approximately calculated to give a high temperature as 
$7.2 \times 10^{10}(\varepsilon_{\rm th}/0.01c^2)~{\rm K}$. 
Nevertheless, the thermal pressure $P_{\rm th}$ in the central region
of $\rho \agt 10^{15}~{\rm g/cm^3}$ is only $\sim 1$--2\% of $P_{\rm
  cold}$ as shown in Fig. 7 (e), and hence, the thermal pressure plays
a minor role for supporting the self-gravity in the central region.
This indicates that centrifugal force plays a more important role for
supporting the self-gravity of the hypermassive neutron star in its
early evolution phase.  By contrast, for a region of $\rho \alt
\rho_{\rm nuc}$, the thermal pressure is larger than $P_{\rm
  cold}$. Namely, the effect of the thermal energy plays an important
role for determining the profile in the envelope of the hypermassive
neutron star.

Although the ratio $P_{\rm th}/P_{\rm cold}$ is small in the central
region, it is not a negligible value and the thermal pressure seems to
contribute in part to supporting the self-gravity. This suggests that
even after dissipation and/or transportation of angular momentum via
gravitational radiation or magnetic-field effects, the hypermassive
neutron star of relatively small mass may not collapse to a black hole
because of the presence of the thermal energy. If so, the collapse
will set in after the thermal energy is dissipated via neutrino
cooling. The cooling time scale by neutrino emission from the central
region of the hypermassive neutron star is likely to be of order
10--100 s (e.g., chapters 11 and 18 of Ref.~\cite{shapiro}), and thus,
the lifetime of the hypermassive neutron star of relatively low mass
may be rather long.

In all the above scenarios, the hypermassive neutron star eventually
collapses to a black hole after a substantial fraction of angular
momentum is dissipated.  This indicates that the resulting black hole
will not be rotating as rapidly as the black holes promptly formed
after the onset of merger (cf. Sec. \ref{sec:BHmass}). 

For the envelop of the hypermassive neutron star, $\varepsilon_{\rm
th} \sim 0.01$--0.02 and the temperature of this region is high, $\sim
10^{11}$ K.  Because of its high temperature and relatively low
density, such region is subject to a large amount of thermal neutrino
emission. In particular, for the region of $\rho \alt 10^{11}{\rm
g/cm^3}$, neutrinos are not trapped by the matter but escape freely,
so cooling will proceed rapidly. Thus, the thermal energy decreases on
a time scale, shorter than the dynamical time scale, until the cooling
time scale becomes as long as the dynamical one. This point should be
explored in the future, incorporating finite-temperature EOSs and the
neutrino emission, although such work is beyond scope of this paper.

\subsubsection{APR145145,APR1515} \label{subsec145}

In these equal-mass massive models, a black hole is formed after the
onset of merger in dynamical time scale $\sim 1$ ms, and most
of the material falls into the black hole (e.g., the last two panels
of Fig. \ref{fig:15-eq}).  Figure \ref{fig:AH-disk} (a) plots the evolution
of $M_{r>r_{\rm AH}}$, which denotes the rest mass of baryon located
outside the apparent horizon, for runs APR145145H and APR1515H. Here,
$M_{r>r_{\rm AH}}$ is defined by
\begin{eqnarray}
M_{r>r_{\rm AH}} \equiv \int_{r>r_{\rm AH}} \rho_* d^3x,
\label{eq:Mah}
\end{eqnarray}
where $r_{\rm AH}[=r_{\rm AH}(\theta,\varphi)]$ denotes the radius of
the apparent horizon. 

This figure shows that more than $99\%$ of the fluid elements are
swallowed by the black hole within $\sim 0.1$ ms after the formation
of the apparent horizon.  The primary reason for this rapid infalling
is that the specific angular momentum $j_{\rm ini}$ at the onset of 
merger is too small for all the fluid elements: The maximum value
of the specific angular momentum at the onset of merger, $j_{\rm
max}$, is about $1.25M_0$ for both the models.  As discussed in
Sec.~\ref{sec:BHmass}, the final value of the black-hole spin is
estimated to be $\approx 0.8$ for both the models. This predicts that
the specific angular momentum at ISCO around the formed black hole is
$\approx 2.4M_0$, which is much larger than $j_{\rm max}$. Moreover,
the angular momentum transport works inefficiently because of
equal-mass symmetry (see Fig. \ref{fig:15-eq}) and also the specific
angular momentum decreases due to the gravitational-wave emission during
the merger.  All these facts indicate that formation of massive disks
surrounding the black hole is unlikely. 

The disk mass for model APR1515 is smaller than that for model
APR145145.  The likely reason for this is that the object formed just after
the onset of merger for model APR1515 is more compact than that
for model APR145145. As a result, (i) the dynamical time scale becomes
shorter and the time duration for which the angular momentum transport
works does as well; (ii) dissipation of angular momentum by
gravitational waves is larger and disk formation becomes less likely. 
This dependence of the disk mass on the total mass is also found in
the merger of unequal-mass binary neutron stars (see Sec. \ref{subsec:1316}).

Accretion time scale for the disk and increase time scale of the area
of the apparent horizon are much longer than the dynamical time scale
(i.e., rotational time scale of the disk) for $t-t_{\rm AH} \agt 1.5$
ms, as shown in Fig. \ref{fig:AH-disk} (b).  Therefore, the final
state formed for these models is a rapidly rotating black hole surrounded by a
quasisteady disk of tiny mass (see Fig. \ref{fig:med} (b)). 

We compare the result for the disk mass in the present and 
previous simulations for model APR1515. 
In Ref.~\cite{Shibata:2006nm}, we reported that the final disk mass is
$\sim 4 \times 10^{-4}M_\odot$, whereas it is $\sim 7 \times
10^{-5}M_\odot$ for the present work. This difference seems to
originate simply from the fact that we evaluated the disk mass soon
after its formation ($t-t_{\rm AH}=0.5~{\rm ms}$) in
Ref.~\cite{Shibata:2006nm}; i.e., the disk mass is evaluated before
the disk relaxes to a quasisteady state. In the present work, we can
evaluate it at $t-t_{\rm AH}=3.0~{\rm ms}$, at which the disk is in a
quasisteady state, because we employ the moving puncture method which
enables us to perform a longterm simulation even after black hole
formation.

\subsubsection{APR1316,APR135165}\label{subsec:1316}

As shown in Ref.~\cite{Shibata:2006nm} and in Fig. 5, the less massive
neutron star is significantly tidally deformed in the final inspiral
phase of unequal-mass binary neutron stars, and then, the merger
occurs. At the onset of merger, the less massive star is highly
elongated and hence its outer part subsequently forms a large spiral
arm.  The merger in the central part simultaneously proceeds and, for
the case that $m_0 > M_{\rm thr}$, a black hole is formed around the
center in a dynamical time scale $\sim 1$ ms. One interesting feature
in the early stage of the merger (before gravitational collapse to a
black hole) is that material in the inner part of the less massive
star slips through the surface of the companion neutron star and forms
a small spiral arm. As a result, two asymmetric spiral arms are formed
around the central object which subsequently collapses to a black hole
(see the fourth panel of Fig. \ref{fig:1316-eq}).  Due to the
nonaxisymmetric structure, the angular momentum is transported
outwards in the spiral arms.  Because the rotation velocity of the
small spiral arm around the central object is faster than that of the
large spiral arm, they collide within one orbit. As a result, shocks
are formed and the material in the spiral arms is heated up.  In this
mechanism, kinetic energy of the material is converted into thermal
energy, and then a part of the fluid elements, which lose the kinetic
energy, are swallowed by the black hole (see
Fig. \ref{fig:AH-disk}). The fluid elements which escape from falling
into the black hole eventually form a disk around the black hole.  All
these features were already found in Ref.~\cite{Shibata:2006nm}, but
in the previous work, the simulation was not able to continue for a
time long enough to determine the final state because the simulation
crashed before a quasisteady state was reached.  As
Fig. \ref{fig:AH-disk} (b) indicates, the present simulations, by
contrast, have been done until the final state (composed of a black
hole and quasisteady accretion disk) is reached, and the conclusive
statement becomes possible.

Figure \ref{fig:med} (c) and (d) also show that the final outcomes for
models APR1316 and APR135165 are a rotating black hole surrounded by a
disk.  The disk mass for runs APR1316H and APR135165H evaluated at
$t-t_{\rm AH}=3.0~{\rm ms}$ is about $2.4 \times 10^{-2}$ and
$6.4\times10^{-3}M_\odot$, respectively (see Fig. \ref{fig:AH-disk}
(a) and Table~\ref{tab:BHmass}).  The likely reason for this 
significant difference is that for the less massive binary neutron
star, the system at the onset of merger is less compact and has a
longer dynamical time scale for transporting angular momentum outward
more efficiently, as already mentioned in Sec. \ref{subsec145}.

The disk mass computed for three grid resolutions are described in
Table \ref{tab:BHmass}. We find that the disk mass does not
systematically converge and magnitude of the error seems to be by
$\sim 50\%$.  The likely reason for the slow convergence is that the
grid resolution in the outer domain (where the fluid elements spread)
is not systematically improved as the grid resolution in the inner
domain is improved and disk mass depends sensitively on spurious
numerical transport process of angular momentum (see, e.g.,
Ref.~\cite{Shibata:2009cn} for discussion about slow convergence of
disk mass in black hole-neutron star merger).  However, all the
numerical results are derived in a fairly narrow range irrespective of
the grid resolution, and the order of magnitude of the disk mass does
not depend on the grid resolution.

In the previous paper \cite{Shibata:2006nm}, the merger of
unequal-mass binary neutron stars collapsing to a black hole was
studied only for $m_0=3M_{\odot}$. We reported that for the mass ratio
$Q_M \agt 0.8$, the rest mass of the disk surrounding the black hole was
likely to be smaller than $0.01 M_{\odot}$. The present results
indicate that the disk mass depends sensitively on the total mass of
the system, and even for the case that the mass ratio is not much
smaller than unity, a system composed of a black hole and massive disk
may be an outcome after the merger, if the total mass is close to
$M_{\rm thr}$. 

\begin{figure}
  \begin{center}
  \vspace*{40pt}
    \begin{tabular}{cc}
      \begin{minipage}{0.5\hsize}
      \includegraphics[width=8.0cm]{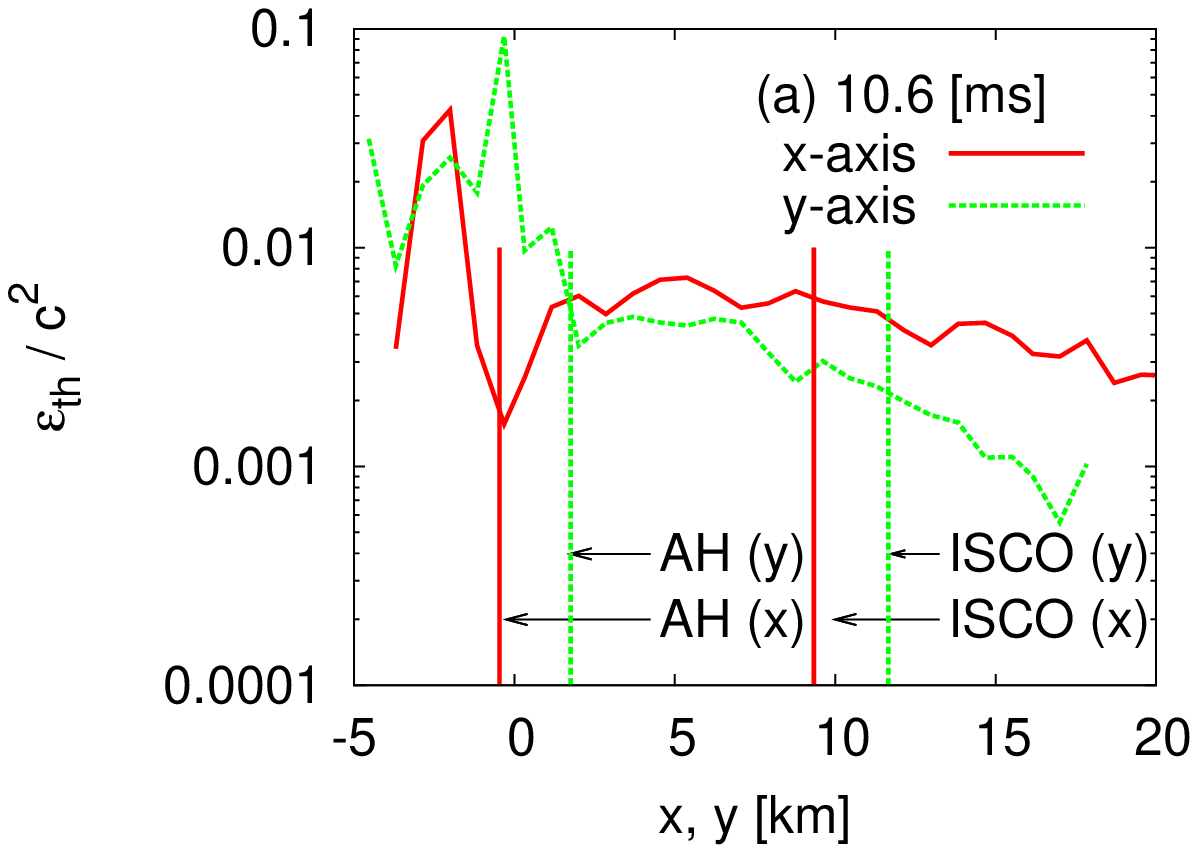}
      \end{minipage}
      \hspace{-1.0cm}
      \begin{minipage}{0.5\hsize}
      \includegraphics[width=8.0cm]{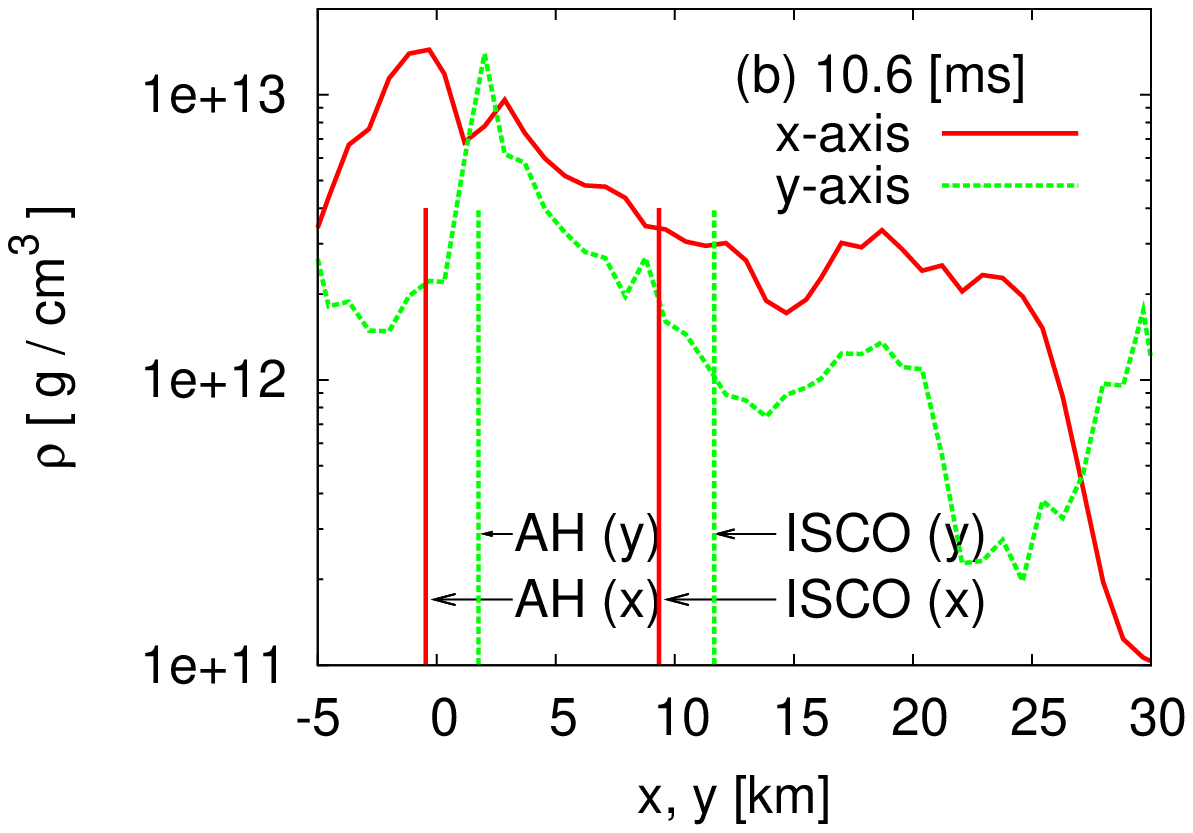}
      \end{minipage}
      \\
    \end{tabular}
    \caption{\label{fig:epsth-apr1316} Profiles of (a) specific
    thermal energy and (b) density for run APR1316H. 
    The vertical solid (dashed) lines denote the location of the AH 
    (ISCO). 
    }
  \end{center}
\end{figure}

\begin{figure}
  \begin{center}
  \vspace*{40pt}
    \begin{tabular}{cc}
      \begin{minipage}{0.5\hsize}
      \includegraphics[width=8.0cm]{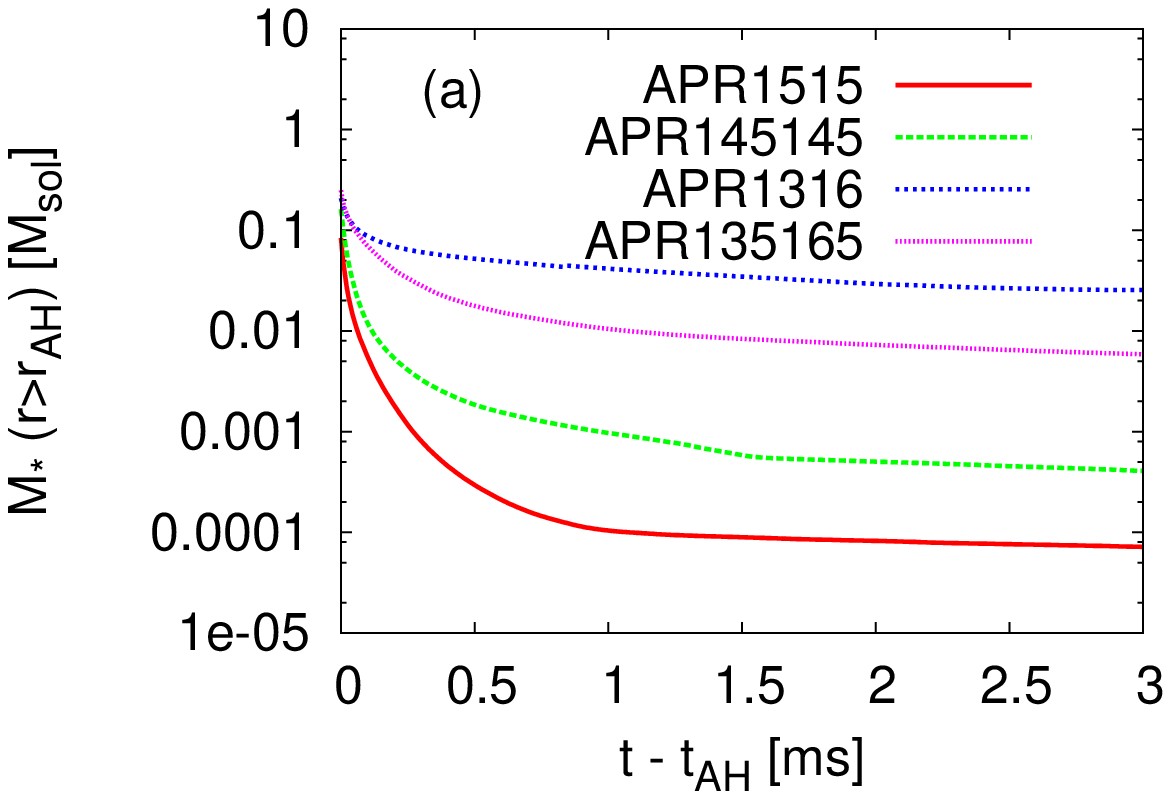}
      \end{minipage}
      \hspace{-1.0cm}
      \begin{minipage}{0.5\hsize}
      \includegraphics[width=8.0cm]{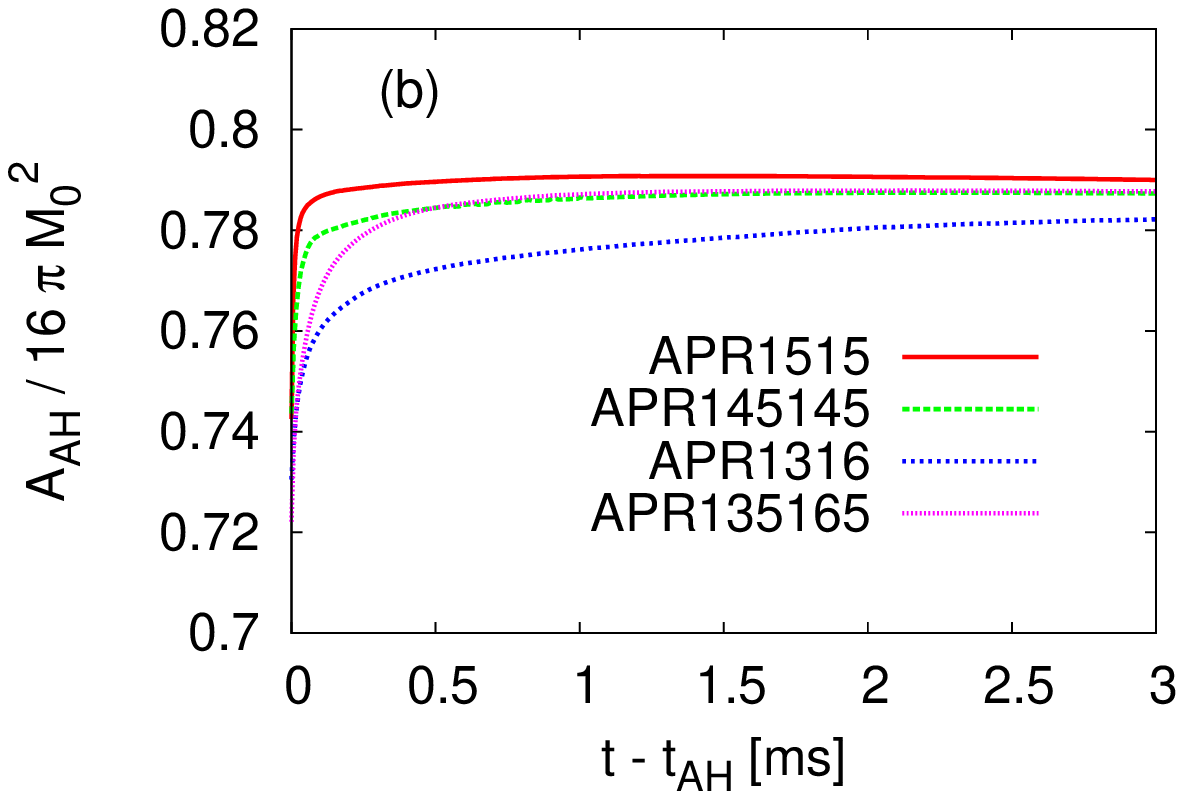}
      \end{minipage}
    \end{tabular}
    \caption{\label{fig:AH-disk} The evolution of (a) the rest mass of
      disks surrounding black holes and (b) the evolution of the area
      of apparent horizons for runs APR145145H, APR1515H, APR1316H, and
      APR135165H. $t_{\rm AH}$ denotes the time at which the apparent
      horizon is formed. }
  \end{center}
\end{figure}

For model APR1316, the disk mass is larger than $0.01M_{\odot}$ with a
high maximum rest-mass density $\sim 10^{12}~{\rm g/cm^3}$ in its
inner region, as shown in Fig. \ref{fig:epsth-apr1316}.  Also, the
thermal energy is generated by shock heating, resulting in a high
specific internal energy, $\varepsilon/c^2 \sim 0.01$ which implies a
high matter temperature as $7.2\times 10^{10}(\varepsilon_{\rm
  th}/0.01c^2)~{\rm K}$ (see Fig. \ref{fig:epsth-apr1316}).  All these
properties are favorable for producing a large amount of high-energy
neutrinos, from which electron-positron pairs and gamma-rays may be
generated \cite{GRBdiskN}. Thus, the final outcome is a candidate for
the central engine of short GRBs.

\subsection{Black hole mass and spin}\label{sec:BHmass}

As mentioned in the previous subsections, the final outcomes after the
merger for models APR145145, APR1515, APR1316, and APR135165 are a
rotating black hole.  We here determine the black-hole mass, $M_{\rm
BH,f}$, and spin, $a_{\rm f}$, using the same methods as those used in
Refs.~\cite{Shibata:2007zm,Yamamoto:2008js}.

There are at least two methods for (approximately) estimating the
black-hole mass and three methods for estimating the black-hole spin.
In the first method for estimating the black-hole mass and spin, we
use the conservation laws. The energy conservation law is
approximately written as
\begin{eqnarray}
M_{\rm BH,f} = M_0 - M_{r>r_{\rm AH}} - \Delta E, \label{enecon}
\end{eqnarray}
where $M_0$ is the initial ADM mass of the system. In Eq.~(\ref{enecon}), 
we neglect the binding energy between the black hole and surrounding
matter, but it is likely to be at most 10\% of $M_{r >r_{\rm AH}}$ and
thus a minor correction.

Conservation law of angular momentum is approximately written as
\begin{eqnarray}
J_{\rm BH,f} = J_0 - J_{r>r_{\rm AH}} - \Delta J,
\end{eqnarray}
where $J_{\rm BH,f}$ is angular momentum of the black hole and $J_0$
is total angular momentum at $t=0$.  $J_{r>r_{\rm AH}}$ approximately
denotes angular momentum of the material located outside apparent
horizon which is defined by
\begin{eqnarray}
J_{r>r_{\rm AH}} \equiv \int_{r>r_{\rm AH}}\rho_* h u_\varphi d^3x,
\end{eqnarray}
where $u_\varphi$ is the $\varphi$-component of the four velocity.  We
note that $J_{r>r_{\rm AH}}$ is strictly equal to the angular momentum
of material only for stationary axisymmetric systems. Thus, this 
should be adopted only for the case that the system approximately
relaxes to a stationary axisymmetric spacetime. 

From $M_{\rm BH,f}$ and $J_{\rm BH,f}$, a nondimensional spin
parameter is defined by 
\begin{eqnarray}
a = \frac{J_{\rm BH,f}}{M_{\rm BH,f}^2}.
\end{eqnarray}
Hereafter, we refer to this spin as $a_{\rm f1}$. 

In the second method, the mass and spin of the black holes are determined
using geometrical properties of apparent horizon.  When the system
relaxes to a quasisteady state, the black-hole mass may be
approximately estimated from equatorial circumferential length $C_e$
of apparent horizon, because $C_e/4\pi$ is equal to $M_{\rm BH,f}$ for
Kerr black holes. This value also gives an approximate value of the
black-hole mass even in the presence of surrounding torus \cite{bhdisk}. 

Next, we assume that area of apparent horizon, $A_{\rm AH}$, obeys the 
same relation as that of Kerr black holes; 
\begin{eqnarray}
\hat{A}_{\rm AH} \equiv \frac{\pi A_{\rm AH}}{C_e^2}
= \frac{1+\sqrt{1-a^2}}{2}. \label{aah}
\end{eqnarray}
Then, from $\hat A_{\rm AH}$ and $C_e$, a black-hole spin may be estimated. 
Hereafter, we refer to this spin as $a_{\rm f2}$.
Note that this method should be used only in the case that the system
relaxes to a quasisteady state, because Eq. (\ref{aah}) holds only for 
such spacetime. 

A black-hole spin is also estimated from polar and equatorial circumferential
radii of apparent horizon, $C_p$ and $C_e$.  For Kerr black holes, the
ratio $C_p/C_e$ is given by a known function composed only of the
black-hole spin as
\begin{eqnarray}
\frac{C_p}{C_e} = \frac{\sqrt{2\hat{r}_+}}{\pi}\int^{\pi/2}_0
\sqrt{1-\frac{a_{\rm f}^2}{2\hat{r}_+}\sin^2\theta}d\theta,
\end{eqnarray}
where $\hat{r}_+=1+\sqrt{1-a^2}$. The black-hole spin determined 
from $C_p/C_e$ is referred to as $a_{\rm f3}$ in the following. 

All the results for the black-hole mass and spin are summarized in
Table~\ref{tab:BHmass}. We find that $M_{\rm BH,f}$ and $C_e/4\pi$
agree within 0.5\% error for all the models.  This indicates that both
quantities at least approximately denote the black-hole mass and that
we obtain the black-hole mass within $\sim 0.5\%$ error.

The spin parameters $a_{\rm f2}$ and $a_{\rm f3}$ agree within 0.01
for all the models. However, the spin parameters $a_{\rm f1}$ does not
agree well with $a_{\rm f2}$ and $a_{\rm f3}$. The typical size of the
difference is 0.06--0.07, 0.05--0.07, and 0.04--0.06 for low-,
medium-, and high-grid resolutions, respectively.  Because the
magnitude of the difference decreases systematically with the
improvement of the grid resolution, this discrepancy originates
primarily from numerical error associated with the finite grid
resolution.  As discussed in Ref.~\cite{Yamamoto:2008js}, poor grid
resolution enhances spurious dissipation of angular momentum and
shortens the time duration of the inspiral phase. This leads to the
underestimation of $\Delta J$.  Indeed, the time duration of the
inspiral phase and $\Delta J$ increase with improvement of the grid
resolution, and as a result, the value of $a_{\rm f1}$ systematically
decreases (see Table III). By contrast, $C_p/C_e$ and $\hat{A}_{\rm
  AH}$ depend weakly on the grid resolution, and so do $a_{\rm f2}$
and $a_{\rm f3}$.  This suggests that the convergent value of the
black-hole spin is close to $a_{\rm f2}$ and $a_{\rm f3}$, and therefore, 
we conclude that for all the models, the black-hole spin 
is $a_{\rm f} \approx 0.78 \pm 0.02$.

In the merger of equal-mass, nonspinning binary black holes, $a_{\rm
f}$ is $\approx 0.69$~\cite{Boyle:2007ft,Buonanno:2006ui}, which is by
about 0.1 smaller than that for the merger of binary neutron stars.
This difference arises primarily from the magnitude of $dJ/dt$ in the
final phase of coalescence. In the merger of binary black holes, a
significant fraction of angular momentum is dissipated by
gravitational radiation during the last inspiral, merger, and ringdown
phases, because a highly nonaxisymmetric state is accompanied by a
highly compact state from the last orbit due to the high compactness
of the black holes. By contrast, compactness of the neutron stars is a
factor of $\sim 5$--7 smaller, and as a result, such a highly
nonaxisymmetric and compact state is not achieved for the binary
neutron stars. Indeed, the angular momentum loss rate by gravitational
waves during the last phases is much smaller than that of the binary
black holes.

Finally, we comment on dependence of the black hole mass and spin on
the approaching velocity.  As discussed in Sec.~\ref{sec:approach},
our prescription of the approaching velocity can reduce the initial
eccentricity.  By comparing the result for run APR1515L with/without
approaching velocity, we find that the spin and mass of the final
state of the black hole are insensitive to the initial orbital
eccentricity.

\begin{table*}
\centering
\begin{minipage}{140mm}
\caption{\label{tab:BHmass} Key numerical results for models
  APR145145, APR1515, APR1316, and APR135165. $\Delta E$, $\Delta J$,
  $M_{r>r_{\rm AH}}$, $\hat{A}_{\rm AH}$, $C_p$, $C_e$, and $f_{\rm
    QNM}$ denote energy and angular momentum carried by gravitational
  waves, rest mass of the material located outside apparent horizon,
  area of the apparent horizon in units of $16\pi M_{\rm BH,f}^2$,
  polar and equatorial circumferential radii of the apparent horizon,
  and quasinormal mode frequency, respectively.  $M_{r>r_{\rm AH}}$
  and $f_{\rm QNM}$ are given in units of $M_0$ and ${\rm kHz}$,
  respectively.  $f_{\rm QNM}$ is derived from Eq.~(\ref{eq:QNM})
  using $M_{\rm BH,f}$ and $a_{\rm f2}$.  For all the models, results
  for three grid resolutions are presented. All the quantities are
  evaluated when we stopped the simulations. }
\begin{tabular}{lccccccccccc}
\hline\hline
Model                        &
$\Delta E/M_0$               &
$\Delta J/J_0$               &
$M_{r>r_{\rm AH}}$           & 
$M_{\rm BH,f}/M_0$           &
$\hat{A}_{\rm AH}$           &
$C_e/4\pi M_0$               &
$C_p/C_e$                    &
$a_{\rm f1}$                 &
$a_{\rm f2}$                 &
$a_{\rm f3}$                 &
$f_{\rm QNM}$               \\
\hline
APR145145L& 1.15\% & 16.7\% & 0.025\%&0.988&0.8208&0.9880&0.8628&0.84&0.77&0.77&6.67\\
APR145145M& 1.16\% & 17.4\% & 0.010\%&0.987&0.8123&0.9863&0.8585&0.83&0.78&0.78&6.76\\
APR145145H& 1.19\% & 17.9\% & 0.017\%&0.988&0.8060&0.9843&0.8569&0.82&0.79&0.78&6.81\\
APR1515L  & 1.19\% & 16.9\% & 0.004\%&0.988&0.8092&0.9864&0.8576&0.84&0.79&0.78&6.56\\
APR1515M  & 1.21\% & 18.0\% & 0.002\%&0.988&0.8126&0.9862&0.8570&0.83&0.78&0.78&6.53\\
APR1515H  & 1.24\% & 18.9\% & 0.004\%&0.987&0.8101&0.9867&0.8569&0.81&0.78&0.78&6.56\\
APR1316L  & 1.11\% & 15.6\% & 0.72\% &0.982&0.8098&0.9730&0.8627&0.83&0.78&0.77&6.76\\
APR1316M  & 1.17\% & 16.9\% & 0.74\% &0.981&0.8181&0.9750&0.8662&0.83&0.77&0.76&6.70\\
APR1316H  & 1.18\% & 17.5\% & 0.85\% &0.980&0.8156&0.9752&0.8633&0.83&0.78&0.77&6.71\\
APR135165L& 1.10\% & 15.8\% & 0.23\% &0.987&0.8101&0.9804&0.8643&0.83&0.78&0.76&6.53\\
APR135165M& 1.15\% & 17.1\% & 0.14\% &0.987&0.8111&0.9801&0.8615&0.82&0.78&0.77&6.52\\
APR135165H& 1.17\% & 17.8\% & 0.21\% &0.986&0.8102&0.9818&0.8607&0.81&0.78&0.77&6.53\\
\hline\hline
\end{tabular}
\end{minipage}
\end{table*}

\subsection{Gravitational Waves}\label{sec:rad}

\subsubsection{General feature and convergence}\label{sec:gw0}

\begin{figure*}
  \begin{center}
  \vspace*{40pt}
    \begin{tabular}{c}
      \vspace{-4.0cm}
      \hspace{-6.0cm}
      \begin{minipage}{0.5\hsize}
      \includegraphics[width=16.0cm]{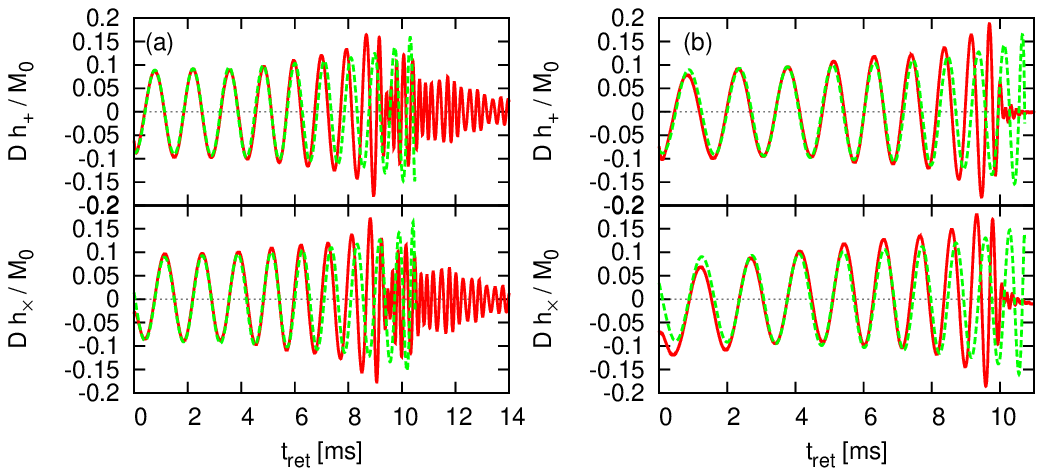}
      \end{minipage}  \\
      \hspace{-6.0cm}
      \begin{minipage}{0.5\hsize}
      \includegraphics[width=16.0cm]{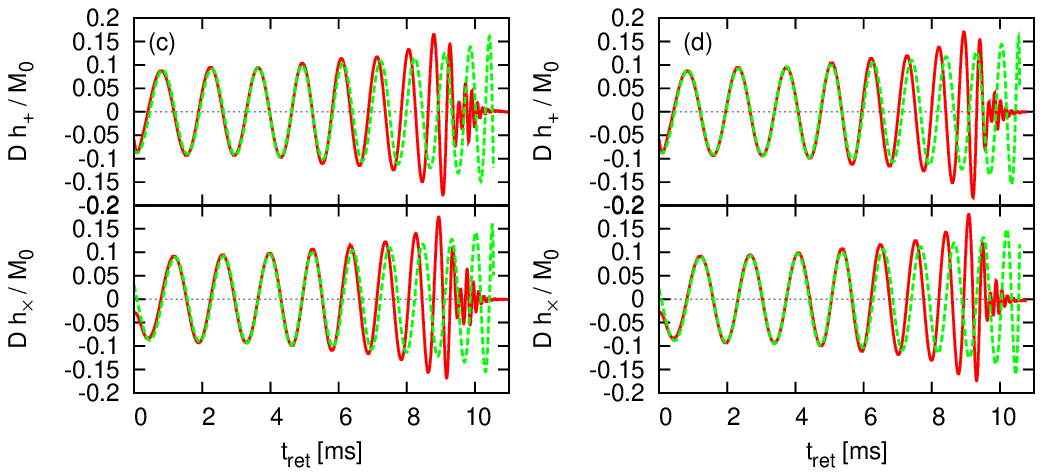}
      \end{minipage}
    \end{tabular}
    \caption{\label{fig:inspiral} $h_{+,\times}$ for runs
    (a) APR1414H, (b) APR1515H, (c) APR1316H, and (d) APR135165H. In each
    panel, the top (bottom) one is $h_+~(h_\times)$, and 
    the solid and dashed curves denote the waveforms
    calculated by the simulation and Taylor T4 formula, respectively.
    }
  \end{center}
\end{figure*}

Figure \ref{fig:inspiral} plots gravitational waveforms as a function
of a retarded time for runs APR1414H, APR1515H, APR1316H, and
APR135165H.  Here, the retarded time is defined by
\begin{equation}
t_{\rm ret} \equiv t - D - 2M_0 \ln (D / M_0),
\end{equation}
where $D$ is the distance between the source and an observer. For
comparison, inspiral waveforms calculated by the Taylor T4 formula are
plotted together by the dashed curves. 


For the case that a hypermassive neutron star is the outcome after the
merger (for model APR1414), gravitational waves are composed of the
inspiral, merger, and quasiperiodic waveforms. Here, the merger
waveform denotes short-term burst-type waves emitted after the
inspiral phase and before the hypermassive neutron star relaxes to a
quasisteady state. The quasiperiodic waveforms seen for model APR1414
are emitted by rapidly rotational motion of the ellipsoidal
hypermassive neutron star and their amplitude is comparable to that of 
the late inspiral phase. Gravitational waves of nearly identical
frequency ($f \sim 3.8$ kHz for this model) are emitted for many
cycles in the quasiperiodic phase, because the dissipation time scale
of the rotational kinetic energy by gravitational radiation reaction
is much longer than the rotational time scale. Consequently, the
effective (time-integrated) amplitude of gravitational waves is much
larger than that for the late inspiral phase of frequency $f \sim 1$
kHz (cf. Sec. \ref{sec:fourier}). This property is essentially the
same as that found in our previous works
\cite{Shibata:2005ss,Shibata:2006nm}.

For the case that a black hole is formed in a dynamical time scale 
after the onset of merger (for models APR145145, APR1515, APR1316,
and APR135165), gravitational waves are composed of the inspiral,
merger, and ringdown waveforms. Because a rotating black hole in a
quasisteady state is the final outcome after the merger, the ringdown
waveform is characterized by the fundamental quasinormal mode of the
formed black hole. This is a universal qualitative feature that holds
irrespective of the total mass and mass ratio of the binary neutron
stars, but the amplitude and characteristic frequency of gravitational
waves depend strongly on these parameters (see
Sec. \ref{sec:qnm}). All these properties were not clarified in the
previous work \cite{Shibata:2006nm}.  The merger waveform denotes
short-term burst-type waves emitted after the inspiral phase and before
the ringdown phase. The frequency of gravitational waves in the merger
phase is slightly lower than the frequency of the black-hole
quasinormal mode, but the amplitude is much larger than that of the
ringdown waveform. The frequency and amplitude of the 
merger waveform are seen in the Fourier space in a better manner 
(see discussion of Secs.~\ref{sec:qnm} and \ref{sec:fourier}). 

\subsubsection{Inspiral gravitational waves}\label{sec:inspiral}

\begin{figure*}
  \begin{center}
  \vspace*{40pt}
    \begin{tabular}{cc}
      \hspace{-2.0cm}
      \begin{minipage}{0.5\hsize}
      \includegraphics[width=8.0cm]{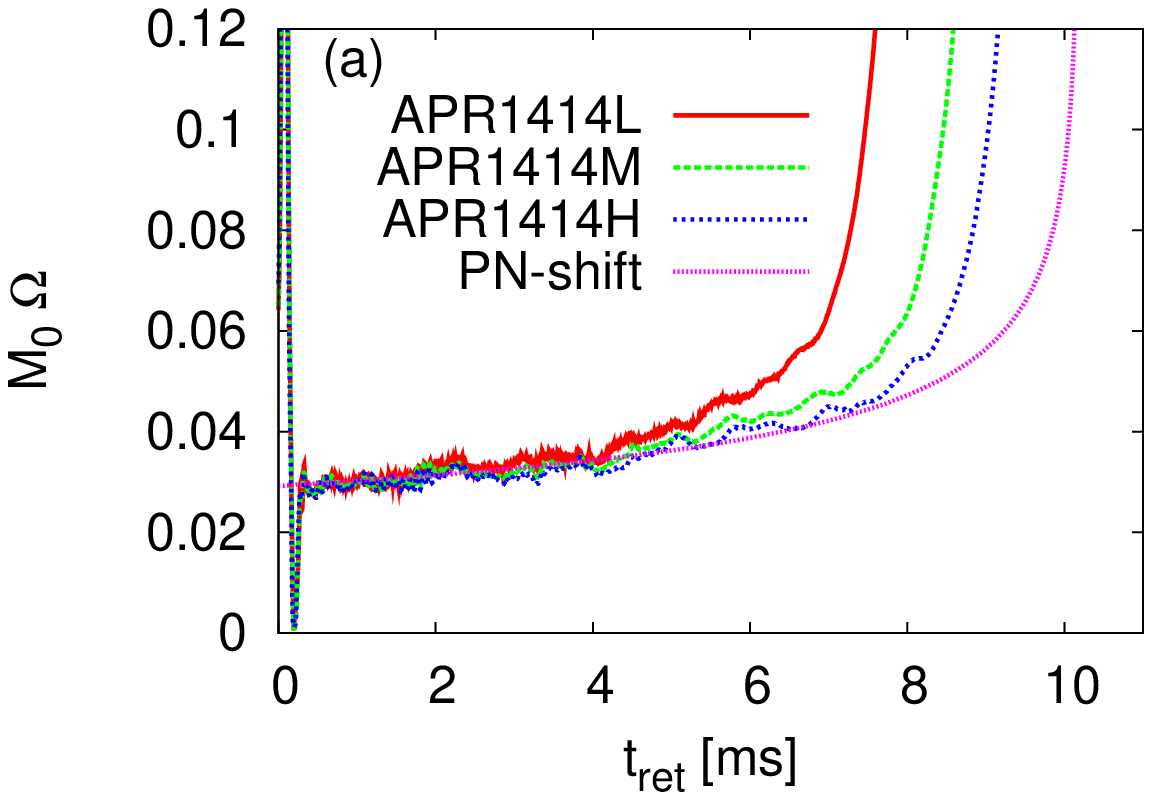}
      \end{minipage}
      \hspace{-1.0cm}
      \begin{minipage}{0.5\hsize}
      \includegraphics[width=8.0cm]{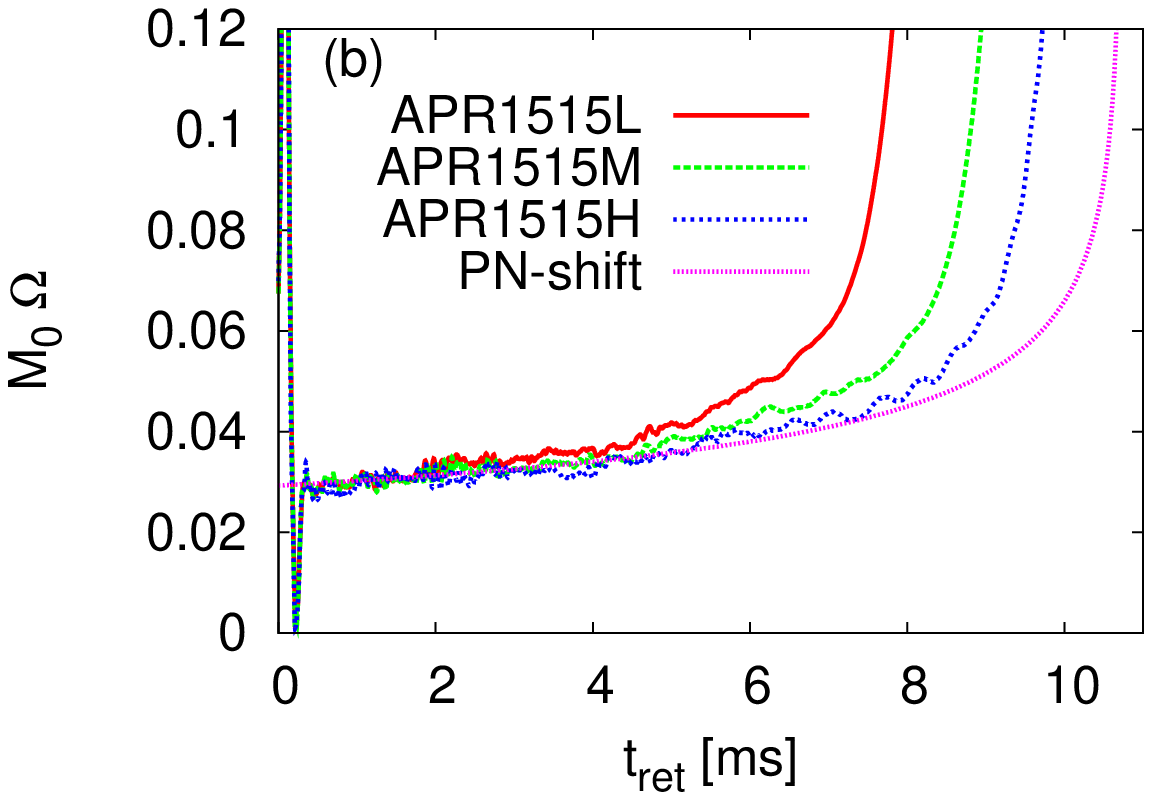}
      \end{minipage}
      \\
      \hspace{-2.0cm}
      \begin{minipage}{0.5\hsize}
      \includegraphics[width=8.0cm]{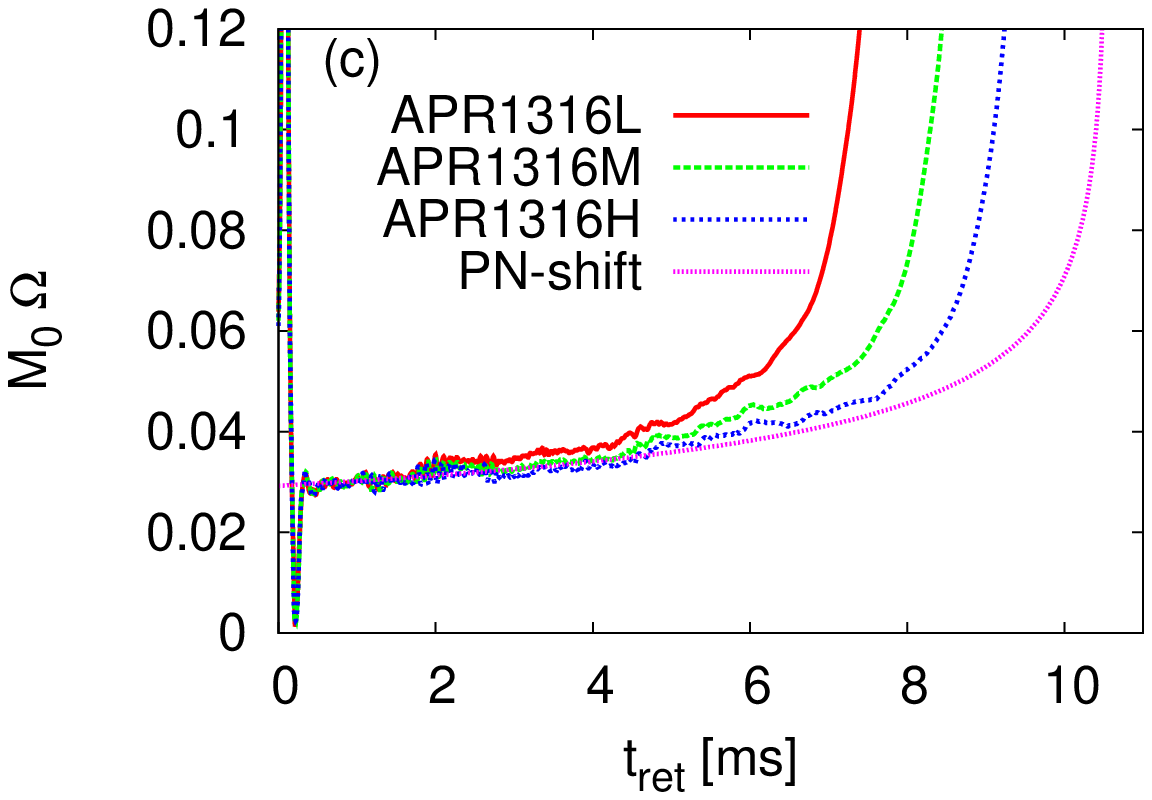}
      \end{minipage}
      \hspace{-1.0cm}
      \begin{minipage}{0.5\hsize}
      \includegraphics[width=8.0cm]{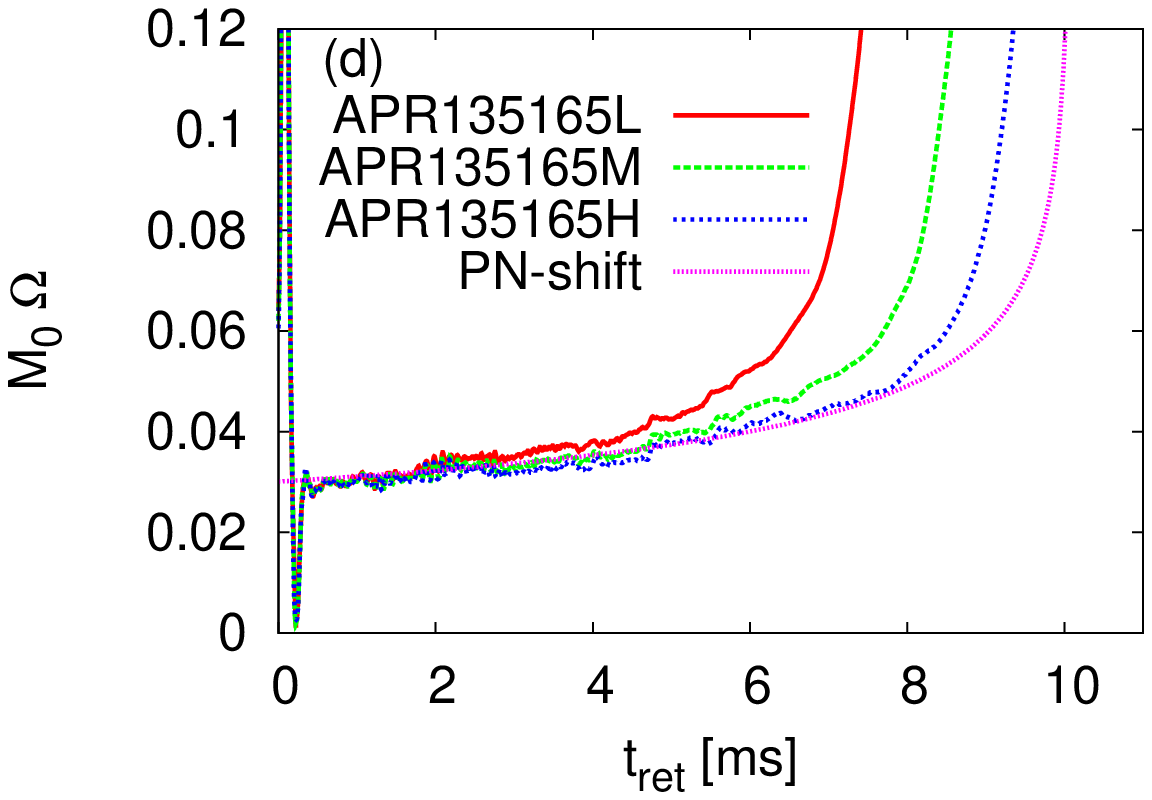}
      \end{minipage}
    \end{tabular}
      \hspace{-2.0cm}
    \caption{\label{fig:MW} The evolution of the orbital angular
      velocity $M_0\Omega$ derived from $\Psi_4$ for models (a)
      APR1414, (b) APR1515, (c) APR1316, and (d) APR135165. The solid,
      dashed, short-dashed, and dotted curves denote the angular
      velocity derived by runs of different grid resolutions and
      Taylor T4 formula, respectively.  }
  \end{center}
\end{figure*}

Figure \ref{fig:inspiral} shows that the inspiral waveforms are fitted
in part well with those derived by the Taylor T4 formula.  For
comparing the frequency of gravitational waves as a function of time,
the orbital angular velocity derived from gravitational waves is
plotted in Fig. \ref{fig:MW} for models APR1414, APR1515, APR1316, and
APR135165. Here, the angular velocity, $\Omega$, is computed from 
$\Psi_4$ by
\begin{eqnarray}
\Omega = \frac{|\Psi_4(l=m=2)|}{2|\int dt\Psi_4 (l=m=2)|}. \label{eq:MW}
\end{eqnarray}
In Fig. \ref{fig:MW}, the solid, dashed, short-dashed, and dotted
curves are numerical results with three grid resolutions and results
by the Taylor T4 formula, respectively.  
In comparison, the time axis for the results by the Taylor T4 formula 
is shifted  according to the prescription described in Sec.~\ref{sec:PN}. 

\begin{figure}
  \begin{center}
  \vspace*{40pt}
    \begin{tabular}{cc}
      \hspace{-8.0cm}
      \begin{minipage}{0.5\hsize}
      \includegraphics[width=16.0cm]{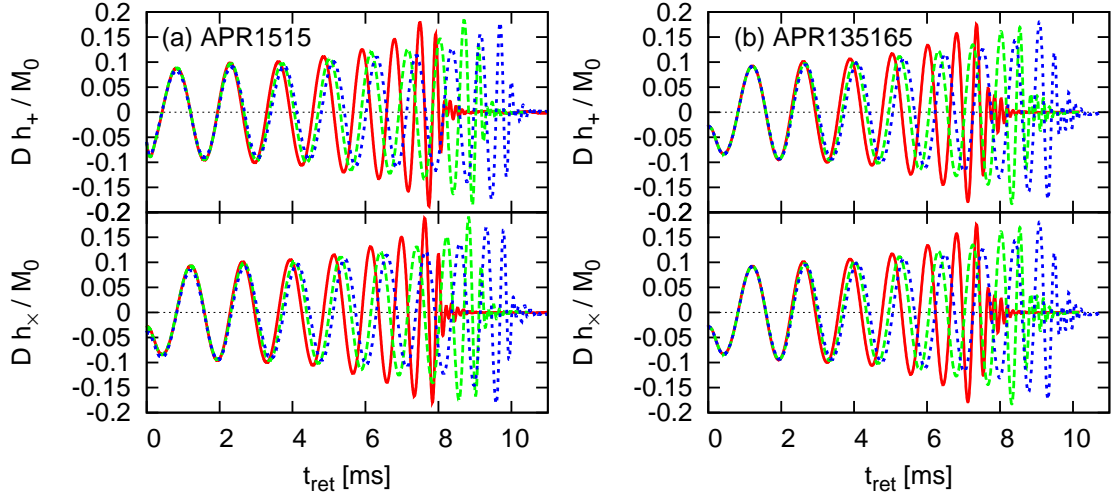}
      \end{minipage}
    \end{tabular}
      \vspace{-1.0cm}
      \vspace{-1.0cm}
    \caption{\label{fig:inspiralH} The same as
      Fig. \ref{fig:inspiral}, but for models APR1515 and APR135165.
      In each panel, the solid, dashed, and short-dashed curves denote
      the results for low, medium, and high-resolution runs,
      respectively. }
  \end{center}
\end{figure}

Figure~\ref{fig:MW} shows that with the lowest-resolution run, the
numerical results agree approximately with that of the Taylor T4
formula only for a short time (2 ms $\alt t_{\rm ret} \alt 4$ ms).
The time span, during which the numerical results are fitted well by
the Taylor T4 formula, is longer for the better grid resolutions.
Thus, disagreement of the numerical waveforms with those by the Taylor
T4 formula is simply due to the fact that the grid resolution is
insufficient. The more specific reason is that angular momentum is
numerically dissipated more for the lower grid resolution, as
mentioned in Sec.~\ref{sec:BHmass}. As a result, merger time (defined
as the time when the merger sets in, and approximately equal to the
time spent in the inspiral phase) is spuriously shortened and the number of
the wave cycles in the inspiral phase is spuriously reduced, resulting
in a disagreement of the orbital phase with the Taylor-T4's result.
These properties are observed in all the models (see also
Fig. \ref{fig:inspiralH}).

\begin{figure}
  \begin{center}
  \vspace*{40pt}
    \begin{tabular}{cc}
      \hspace{-1.0cm}
      \begin{minipage}{0.5\hsize}
      \includegraphics[width=8.0cm]{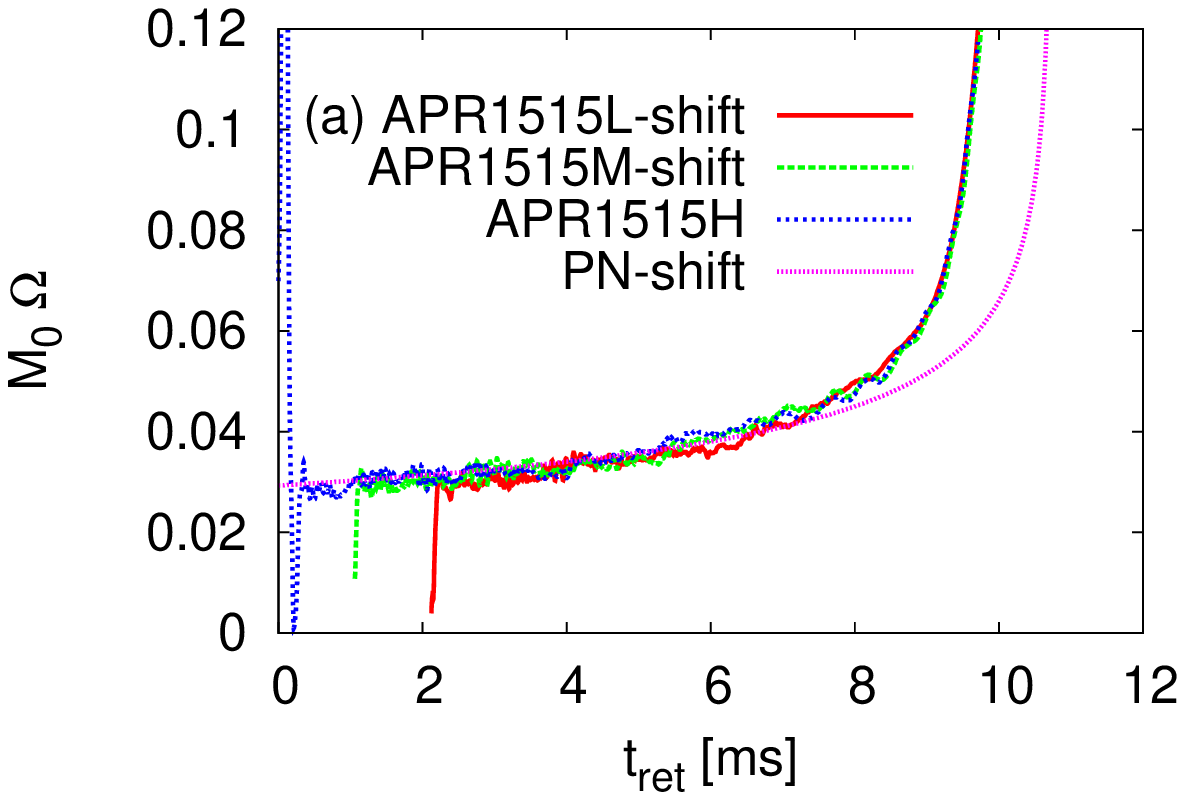}
      \end{minipage}
      \hspace{-1.0cm}
      \begin{minipage}{0.5\hsize}
      \includegraphics[width=8.0cm]{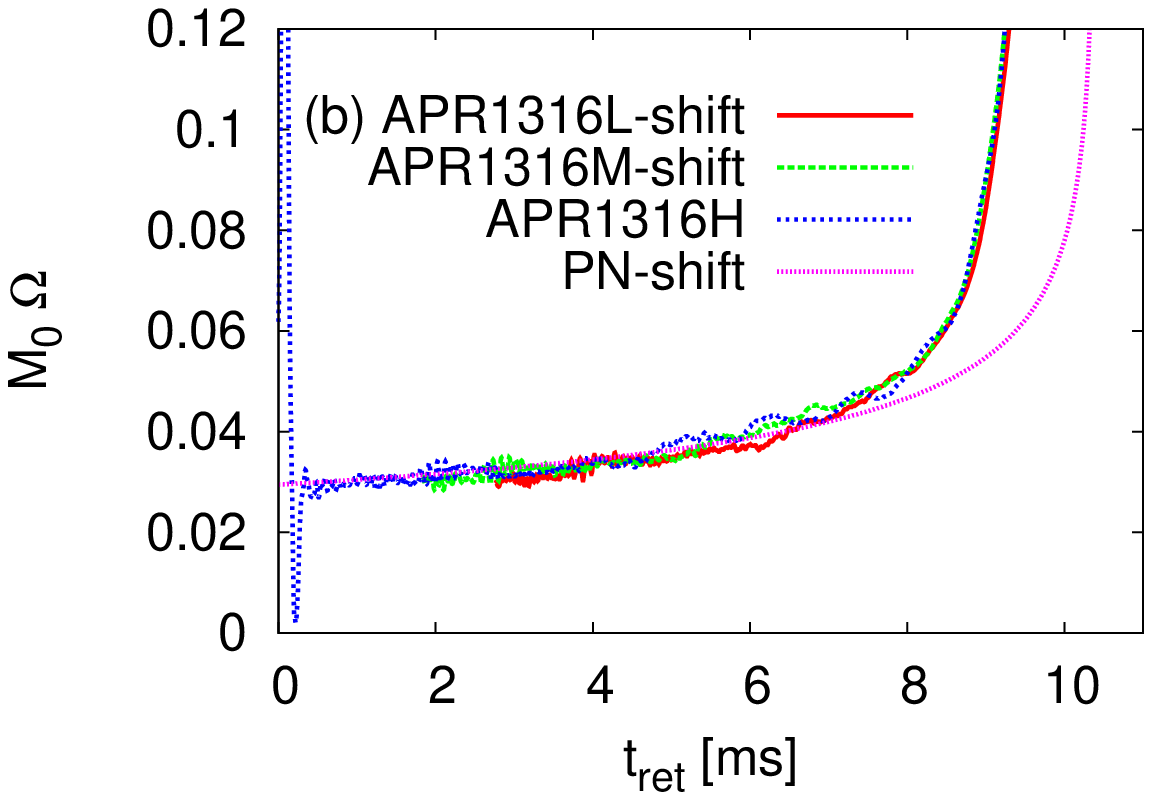}
      \end{minipage}
    \end{tabular}
    \caption{\label{fig:MWH} (a) The same as Fig.~\ref{fig:MW}(b) but
      the numerical results for runs APR1515L and APR1515M are plotted
      by shifting the time coordinates to align the merger time with
      run APR1515H. (b) The same as (b) but for model APR1316.}
  \end{center}
\end{figure}

The numerical results for the angular velocity curve do not agree with
the curve derived by the Taylor T4 formula even with the highest
resolution run.  To clarify the reason for this disagreement, we
generate Fig. \ref{fig:MWH}, in which the results (a) for runs
APR1515L and APR1515M and (b) for runs APR1316L and APR1316M are 
replotted by shifting the time axis to align the merger time for the
results of three grid resolutions.  Then, we also replot the result by
the Taylor T4 formula so that it fits well with the overlapped
numerical results. This figure shows that the results of three grid
resolutions approximately agree each other and also with the result
obtained by the Taylor T4 formula for 4 ms $\alt t_{\rm ret}\alt 8$
ms, whereas for $t_{\rm ret} \gtrsim $ 8 ms three numerical results
deviate from the result of the Taylor T4 formula.  This suggests that
(i) for the late inspiral phase (for 4 ms $\alt t_{\rm ret}\alt 10$ ms
in this figure), the numerical results are approximately convergent
and reliable and (ii) the disagreement with the results by the Taylor
T4 formula comes from a reason different from the numerical error. The
most plausible reason is that for such a late inspiral phase, a finite
size effect of neutron stars, which is not taken into account in the
Taylor T4 formula, plays an important role for accelerating the inward
motion \cite{LRS} because tidal deformation of two
neutron stars and its effect are not negligible for the last orbits.

In binary BH system, the comparison of the PN and numerical-relativity
waveforms have been extensively done, varying the PN order, changing
the expression of the PN formula (e.g., Taylor T1, T4, and Et), and
including spins of black holes
~\cite{Boyle:2007ft,Gopakumar:2007vh,Hannam:2007wf,Scheel:2008rj,Boyle:2009dg}.
The consensus is that the Taylor T4 formula is valid for $m_0\Omega
\lesssim 0.1$.  However, this is not the case for binary neutron stars
because the finite size effect plays an important role for them, as
mentioned above.  We infer that the criterion will depend on EOS of
neutron stars, because it determines compactness of neutron stars and
thus degree of tidal deformation.  For proving this point more 
strictly, however, it is necessary to compute waveforms with a higher 
precision in which a sufficient convergence can be achieved.  This is 
the issue left for the future. 

\begin{figure}
  \begin{center}
  \vspace*{40pt}
    \begin{tabular}{c}
      \begin{minipage}{0.5\hsize}
      \includegraphics[width=8.0cm]{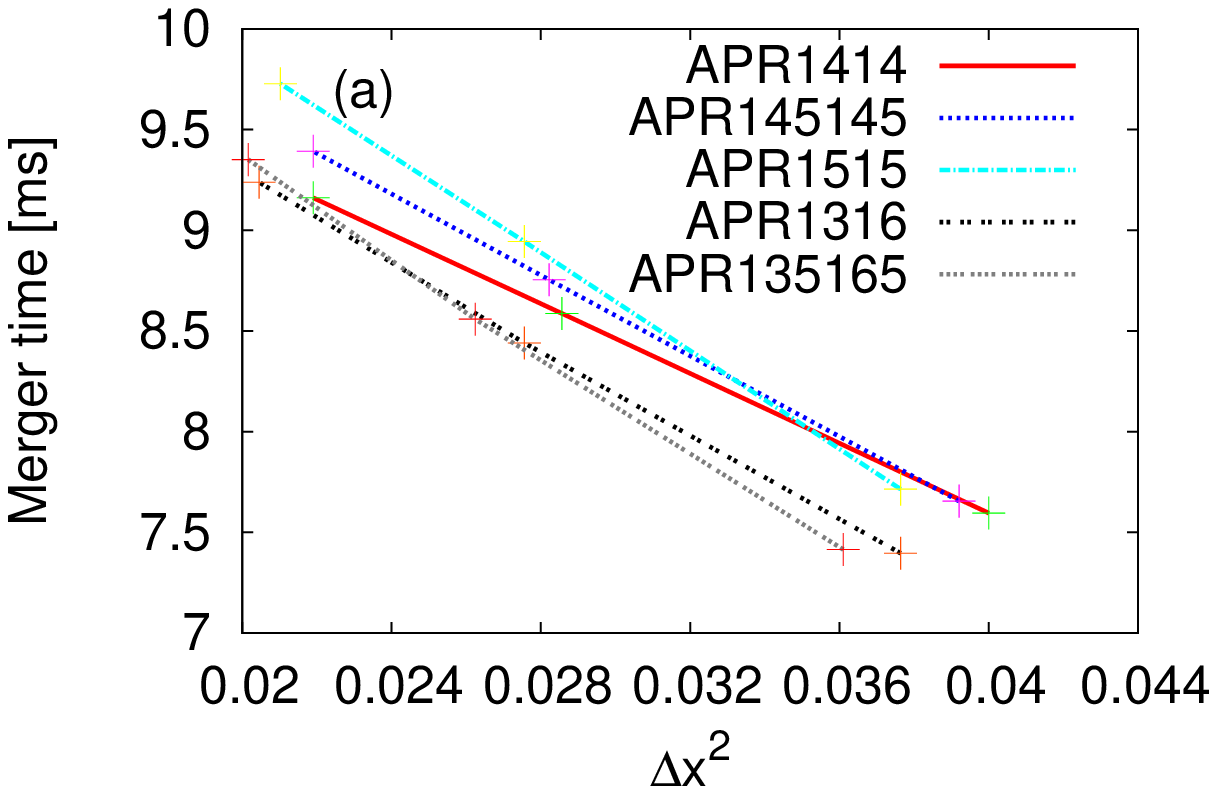}
      \end{minipage}
      \hspace{-1.0cm}
      \begin{minipage}{0.5\hsize}
      \includegraphics[width=8.0cm]{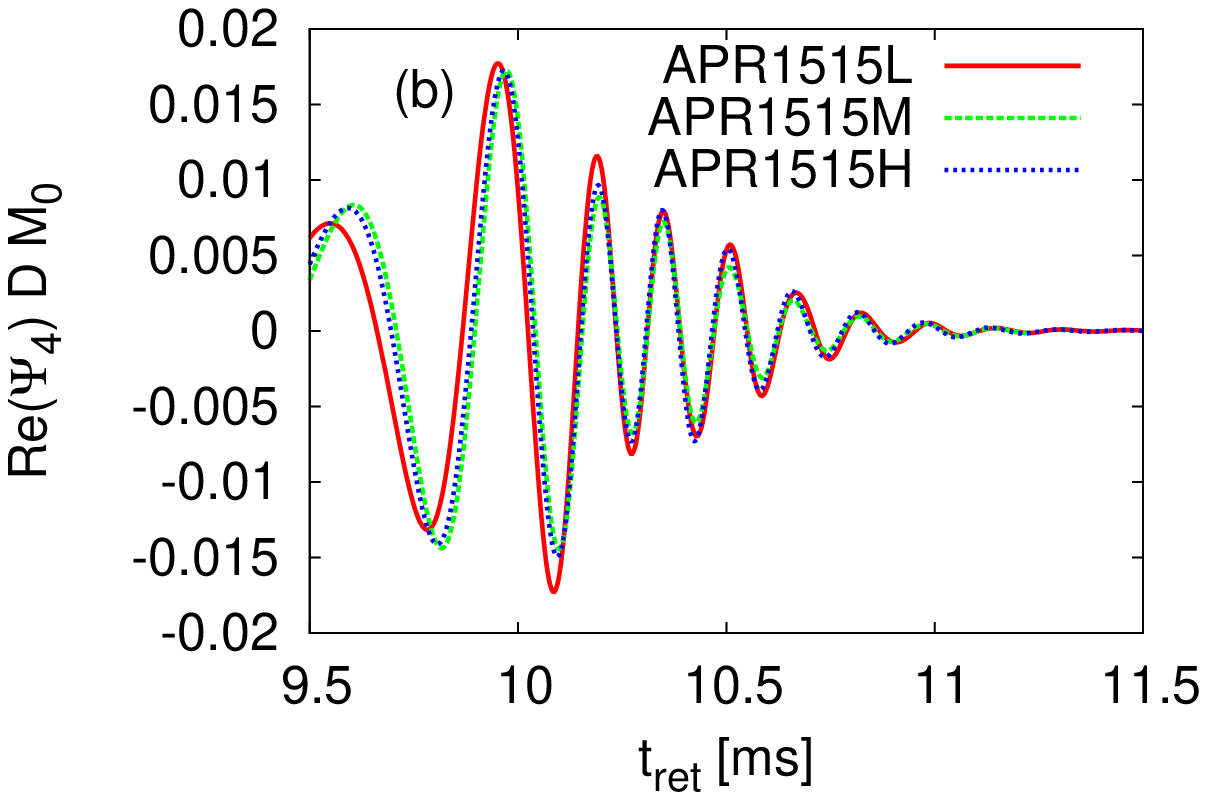}
      \end{minipage}
    \end{tabular}
    \caption{\label{fig:conv-gw} (a) The numerical results for merger
      time as a function of $\Delta x^2$ for all the models and (b)
      $\Psi_4$ as a function of retarded time for runs APR1515L,
      APR1515M, and APR1515H.  To align the phase, the time coordinates for
      runs APR1515L and APR1515M are shifted.  }
  \end{center}
\end{figure}

Before closing this subsection, we touch on the convergence of the
inspiral waveforms. As mentioned above, the wave cycle is spuriously
shorten for the lower grid-resolution simulations. Figure
\ref{fig:conv-gw} (a) plots the merger time as a function of $\Delta
x^2$, where we define the merger time as the time when $M_0 \Omega$
reaches $0.12$ (see Fig.~\ref{fig:MW}). We find that the merger time
converges approximately at second order.  Comparing the merger time
derived by the highest-resolution run with that derived by the
extrapolated result, we find that the merger time is by $\sim 2$ ms
underestimated irrespective of models. This suggests that $\sim 2$
wave cycles are spuriously lost: Figure 13 (b) suggests that the wave
cycles in an early phase of the numerical simulation seem to be lost.
To derive more accurate gravitational waveforms for the inspiral
phase, say, those with the error of half cycle, a higher grid
resolution is required. For this purpose, the present simulation with
uni-grid domain is computationally expensive (although it is in
principle possible to perform the simulation), and hence, an adaptive
mesh refinement (AMR) algorithm would be required
\cite{Yamamoto:2008js,Baiotti:2008ra,Anderson:2007kz}.  Because the
purpose of this paper is not to derive highly precise inspiral
gravitational waveforms but to qualitatively explore gravitational
waves in the merger and subsequent phases, we do not pay attention to 
the inspiral waveforms in more detail. 

\subsubsection{Merger and ringdown gravitational waves}\label{sec:qnm}

In contrast to the results for the inspiral waveforms, the numerical
results for the merger and ringdown waveforms have a good convergence,
and the grid resolution chosen in this work appears to be acceptable.
Figure \ref{fig:conv-gw} (b) plots $\Psi_4$ as a function of the
retard time for runs APR1515L, APR1515M, and APR1515H. To focus on the
ringdown waveforms in comparison among three runs, the time is shifted
for runs APR1515L and APR1515M, to align the phase in the ringdown
waveforms.  This figure shows that the phase of the waveforms agrees
well for $10~{\rm ms} \lesssim t_{\rm ret} \lesssim 11.5~{\rm ms}$.
The amplitude of gravitational waves slightly disagree, but the error
is at most $\sim 20\%$ for the lowest resolution run and for the
highest resolution run, it appears to be less than 5\% (note that this
does not originate from the finite extraction radius as shown in
Fig.~\ref{fig:ext}).  This indicates that the grid resolution for model
APR1515H and resulting spurious short merger time do not seriously
affect the merger and ringdown waveforms. This feature does not depend
on the total mass or mass ratio of the model. Thus for a quantitative
study of the merger and ringdown waveform, the highest grid resolution
adopted in this paper is acceptable. 

\begin{figure*}
  \begin{center}
  \vspace*{40pt}
    \begin{tabular}{cc}
      \begin{minipage}{0.5\hsize}
      \includegraphics[width=8.0cm]{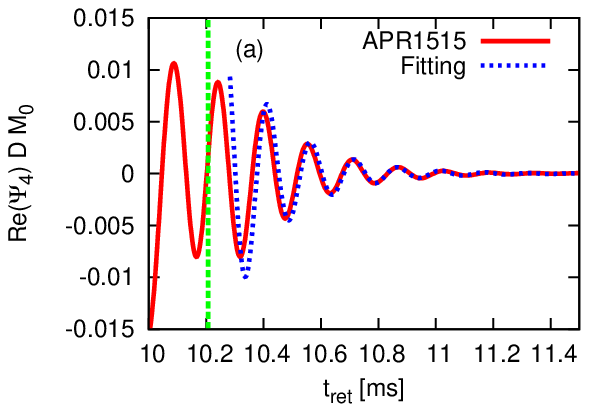}
      \end{minipage}
      \hspace{-1.0cm}
      \begin{minipage}{0.5\hsize}
      \includegraphics[width=8.0cm]{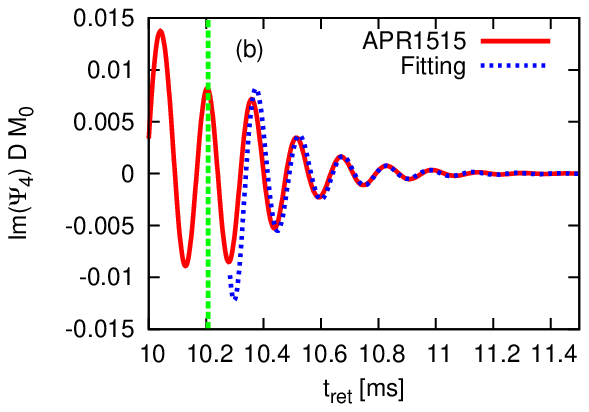}
      \end{minipage}
      \\
      \begin{minipage}{0.5\hsize}
      \includegraphics[width=8.0cm]{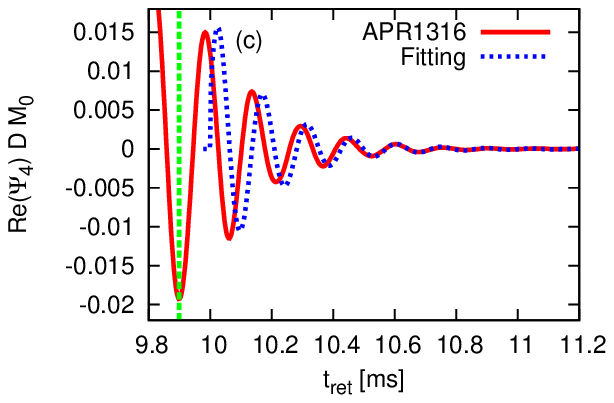}
      \end{minipage}
      \hspace{-1.0cm}
      \begin{minipage}{0.5\hsize}
      \includegraphics[width=8.0cm]{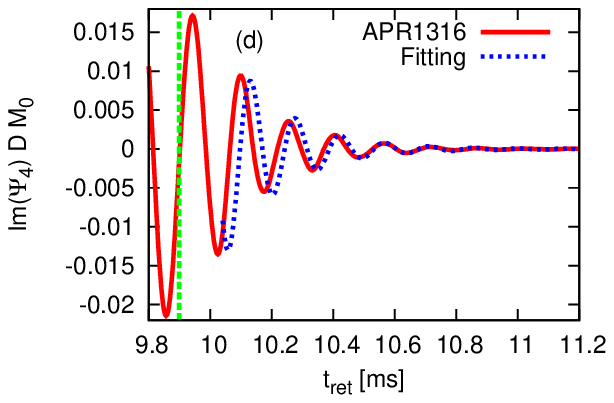}
      \end{minipage}
    \end{tabular}
    \caption{\label{fig:ring} Ringdown waveforms associated primarily
    with the fundamental quasinormal mode (a), (b) for run APR1515H 
    and (c), (d) for run APR1316H. The panels (a) and (c) plot the
    real part of $\Psi_4$ and (b) and (d) the imaginary part.  For all
    the panels, the short-dashed curves denote the fitting curves calculated
    by Eq.~(\ref{eq:QNM}). The vertical dashed line in each panel denotes 
    the formation time of the apparent horizon.}
  \end{center}
\end{figure*}

As mentioned in Sec.~\ref{sec:gw0}, the waveforms in the ringdown
phase are primarily characterized by the fundamental quasinormal mode
of the formed black holes for models APR145145, APR1515, APR1316, and
APR135165. To clarify this fact, Fig. \ref{fig:ring} plots $\Psi_4$
together with a fitting formula in the form
\begin{eqnarray}
A {\rm e}^{-t/t_d}\sin(2\pi f_{\rm QNM} t + \delta), \label{eq:QNM}
\end{eqnarray}
where $A$ and $\delta$ are constants, and the frequency and damping
time scale are predicted by a linear perturbation analysis as
\cite{Leaver:1985ax,Echeverria:1989hg}
\begin{eqnarray}
&& f_{\rm QNM} \approx 10.7\left(\frac{M_{\rm BH,f}}{3.0
M_\odot}\right)^{-1} [1-0.63(1-a_{\rm f})^{0.3}]~{\rm kHz},\\ 
&& t_d \approx \frac{2(1-a_{\rm f})^{-0.45}}{\pi f_{\rm QNM}}~{\rm ms}. \label{eq:decay}
\end{eqnarray}
Figure \ref{fig:ring} (a) and (b) show that gravitational waves for
model APR1515 are well fitted by the hypothetical waveforms given by
Eq.~(\ref{eq:QNM}) for $t_{\rm ret} \agt 10.5$ ms. [Here, we set $a_{\rm
f}=a_{\rm f 2}$ for the fitting (cf. Sec.~\ref{sec:BHmass})].  This is
reasonable because the final outcome for model APR1515 is a stationary
rotating black hole with negligible disk mass, and hence, the
black-hole perturbation theory (i.e., Eq.~(\ref{eq:QNM})) should work
well.  This is also the case for the gravitational waveforms for model
APR145145, for which the merger proceeds in essentially the same
manner as that for model APR1515.  

We note that gravitational waves emitted for $t_{\rm ret} \alt 10.5$ ms 
are not well fitted by the hypothetical fitting formula. This implies that 
they are not the ringdown waves but merger waves. Namely, 
these gravitational waves are not emitted by the ringdown oscillation of 
the black hole but probably by a motion of the material moving around 
the black hole. Figure \ref{fig:ring} (a) and (b) show that the 
gravitational-wave amplitude in the merger phase is much larger than 
that in the ringdown phase. 

The fitting between the numerical and analytic waveforms for models
APR1316 and APR135165 works fairly well, but is not as good as that
for models APR1515 and APR145145 (see Fig. \ref{fig:ring} (c) and (d)
for the results of run APR1316H); the numerical waveform may be fitted
by the analytic one for $t_{\rm ret} \agt 10.4~{\rm ms}$, but the
damping time appears to modulate.  This disagreement is reasonable
because for models APR1316 and APR135165, a fraction of material is
located outside the black hole horizon even after formation of the
black hole, and it subsequently falls into the black hole.
Thus, the system is not completely in vacuum nor in a stationary
state, and hence, the numerical waveforms may not be well fitted by
the analytic results derived in an ideal assumption.

\begin{figure*}
  \begin{center}
  \vspace*{40pt}
    \begin{tabular}{cc}
      \begin{minipage}{0.5\hsize}
      \includegraphics[width=8.0cm]{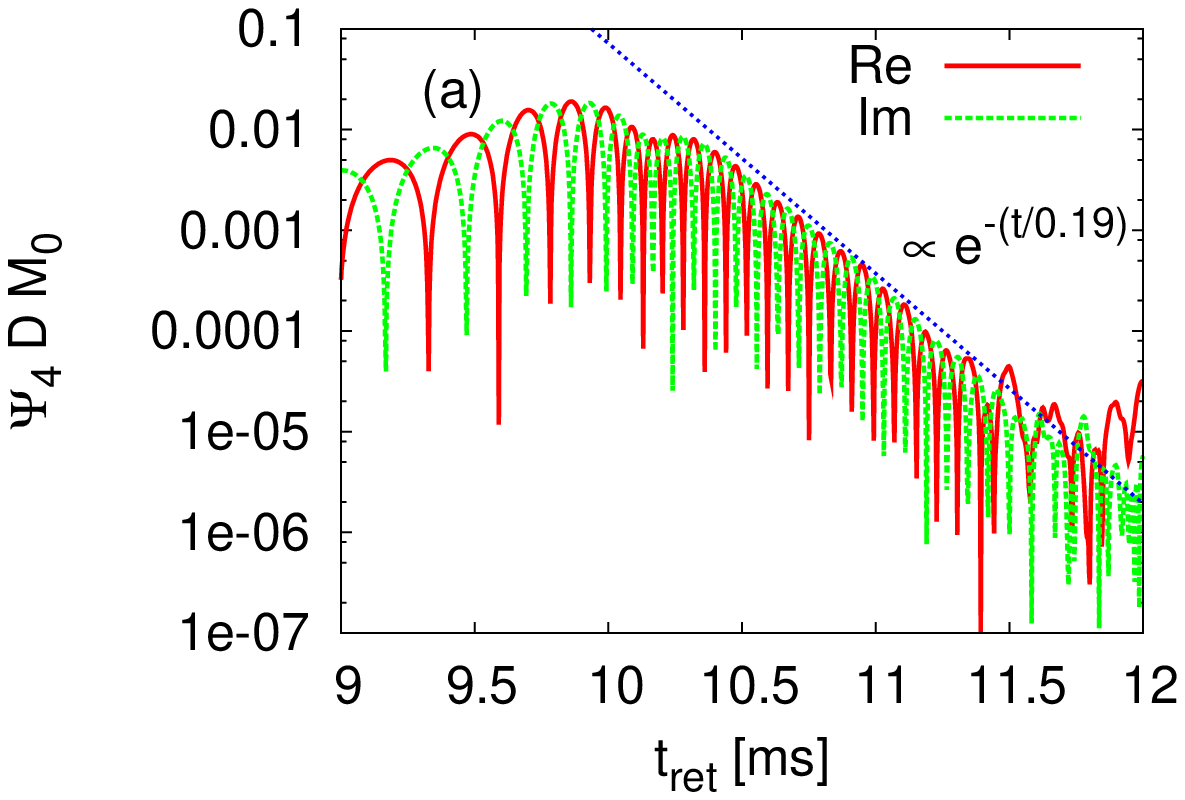}
      \end{minipage}
      \hspace{-1.0cm}
      \begin{minipage}{0.5\hsize}
      \includegraphics[width=8.0cm]{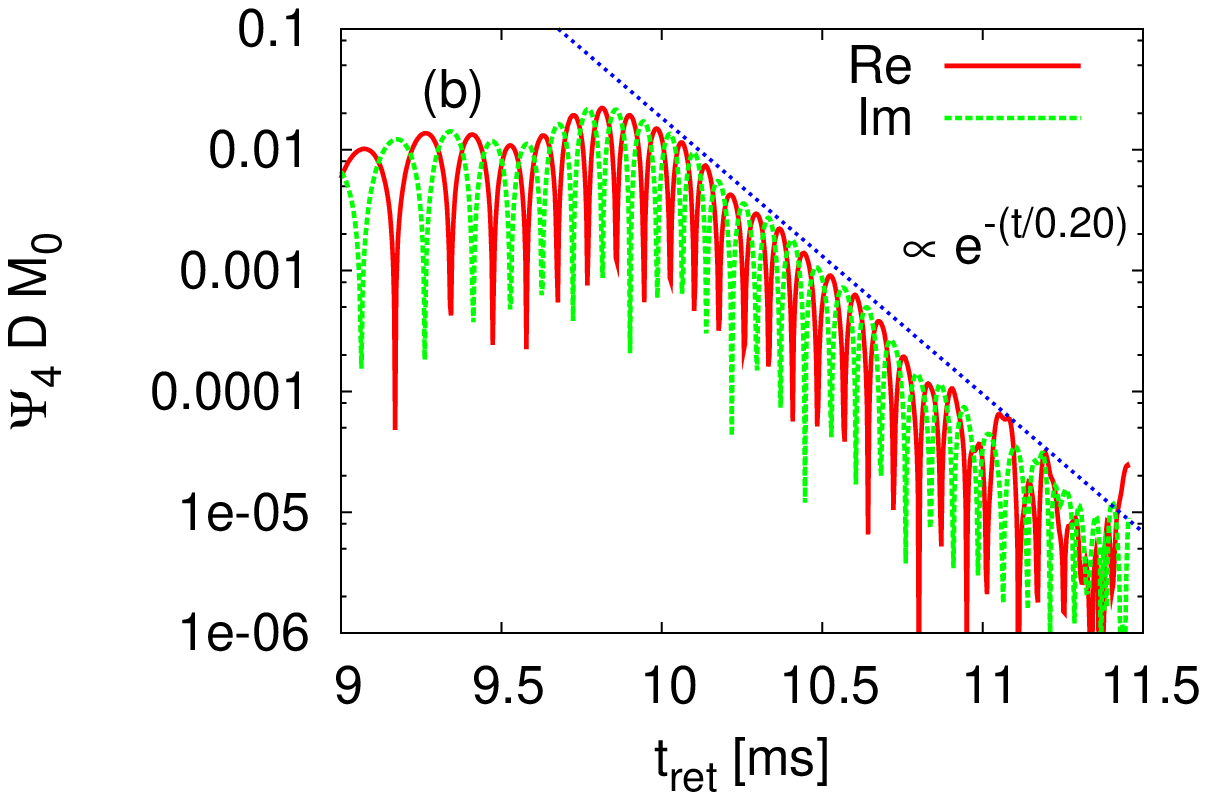}
      \end{minipage}
    \end{tabular}
    \caption{\label{fig:decay} Decay of $\Psi_4$ for (a) run APR1515H 
    and (b) run APR1316H. Solid, dashed, and short-dashed curves denote the real, imaginary part 
    of $\Psi_4$, and exponential decay with $t_d=0.19$ ms for run APR1515H and 0.20 ms for run 
    APR1316H.}
  \end{center}
\end{figure*}

Damping time scale $t_d$ in Eq.~(\ref{eq:decay}) allows us to evaluate
black hole spin. By this method, we can check the accuracy of the spin
parameters estimated by the methods described in Sec~\ref{sec:BHmass}.
Figure~\ref{fig:decay} depicts the evolution of $|\Psi_4|$ for runs
APR1515 and APR1316, which shows a clear exponential decay in
$10.5~{\rm ms}\lesssim t \lesssim 11.5~{\rm ms}$ for APR1515 and
$10~{\rm ms}\lesssim t \lesssim 10.5~{\rm ms}$ for APR1316.  We find
that the damping time scale is estimated from this figure as
$0.19\pm0.01$ ms for run APR1515H and $0.20\pm 0.02$ ms for run
APR1316H. The larger uncertainty for run APR1316H reflects the fact
that surrounding matter falls into black hole even after the black
hole formation for a relatively long time scale and ringdown
gravitational waves are not simply characterized by the fundamental
quasinormal modes.  Putting the values of the damping time scale into
Eq.~(\ref{eq:decay}), the black hole spin is estimated as $0.77\pm0.04$
for run APR1515H and $0.84 \pm 0.06$ for run APR1316H.  As expected,
these values agree with the results in Sec.~\ref{sec:BHmass} within
the error bar. 

\begin{figure}
  \begin{center}
  \vspace*{40pt}
    \begin{tabular}{c}
      \begin{minipage}{0.5\hsize}
      \includegraphics[width=8.0cm]{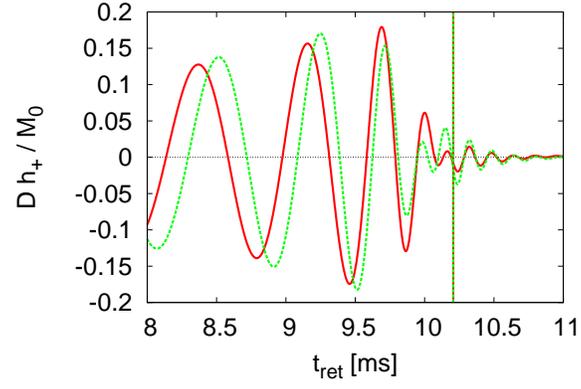}
      \end{minipage}
    \end{tabular}
    \vspace{-0.5cm}
    \caption{\label{fig:h1515-135165} The plus mode of gravitational
      waves $(h_+)$ for runs APR1515H (solid curve) and APR135165H
      (dashed curve).  The vertical dashed line denotes the formation
      time of apparent horizon for run APR135165H. The waveform for
      run APR1515H is shifted to align the formation time of the apparent
      horizon.  }
  \end{center}
\end{figure}

The merger waveforms also depend on the mass ratio.  We plot the plus
mode of gravitational waves ($h_+$) for runs APR1515H and APR135165H
in Fig. \ref{fig:h1515-135165}.  To focus on waveforms just
before/after BH formation, waveforms are shifted to align the
formation time of apparent horizon (vertical dashed line). The peak at
$t \sim 10$ ms for model APR1515 is higher than that for model
APR135165. This reflects the fact that the less massive neutron star
for model APR135165 is tidally deformed from the late inspiral phase,
and at the merger, the material of it starts expanding. Consequently,
its compactness decreases quickly, so does the degree of
nonaxisymmetry of the system, reducing the amplitude of gravitational
waves.

The amplitude of the first peak after the black hole formation is, by
contrast, higher for model APR135165 than for model APR1515. This
seems to reflect the fact that more material is located outside the
black-hole horizon at the black hole formation for model APR135165,
because such material subsequently falls into the black hole to excite
gravitational waves. 

\subsubsection{Fourier spectrum}\label{sec:fourier}


We define a Fourier power spectrum of gravitational waves by
\begin{eqnarray}
h(f)\equiv \sqrt{\frac{|h_+(f)|^2+|h_\times(f)|^2}{2}},
\end{eqnarray}
where
\begin{eqnarray}
&&
h_+(f)=\int e^{2i\pi ft}h_+(t)dt,\\
&&
h_\times(f)=\int e^{2i\pi ft}h_\times(t)dt, 
\end{eqnarray}
and $h_+$ and $h_{\times}$ denote the $+$ and $\times$ modes of
gravitational waves of $l=|m|=2$. Then, from $h(f)$, we define a
nondimensional spectrum (or effective amplitude) as
\begin{eqnarray}
h_{\rm eff}(f) \equiv  h(f)f.
\end{eqnarray}

\begin{figure*}
  \begin{center}
  \vspace*{40pt}
    \begin{tabular}{cc}
      \begin{minipage}{0.5\hsize}
      \includegraphics[width=9.0cm]{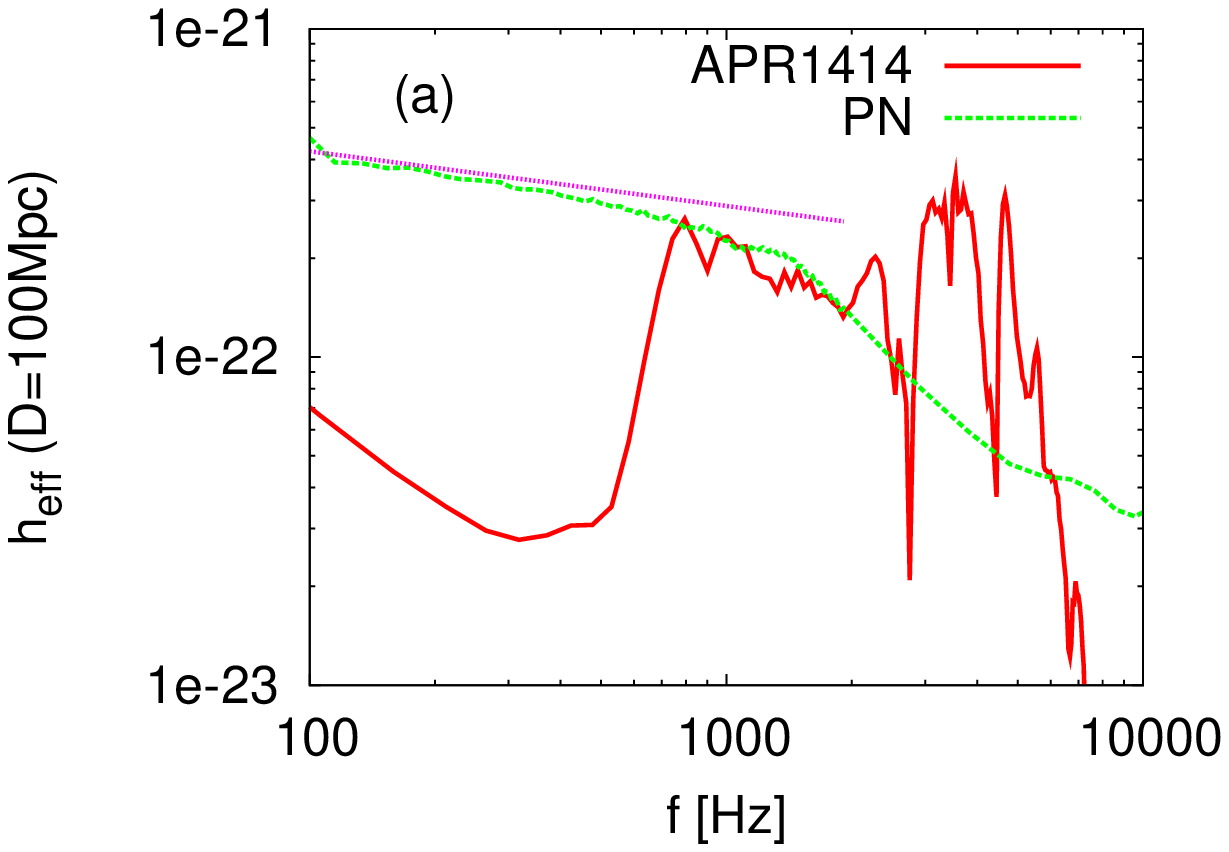}
      \end{minipage}
      \hspace{-0.5cm}
      \begin{minipage}{0.5\hsize}
      \includegraphics[width=9.0cm]{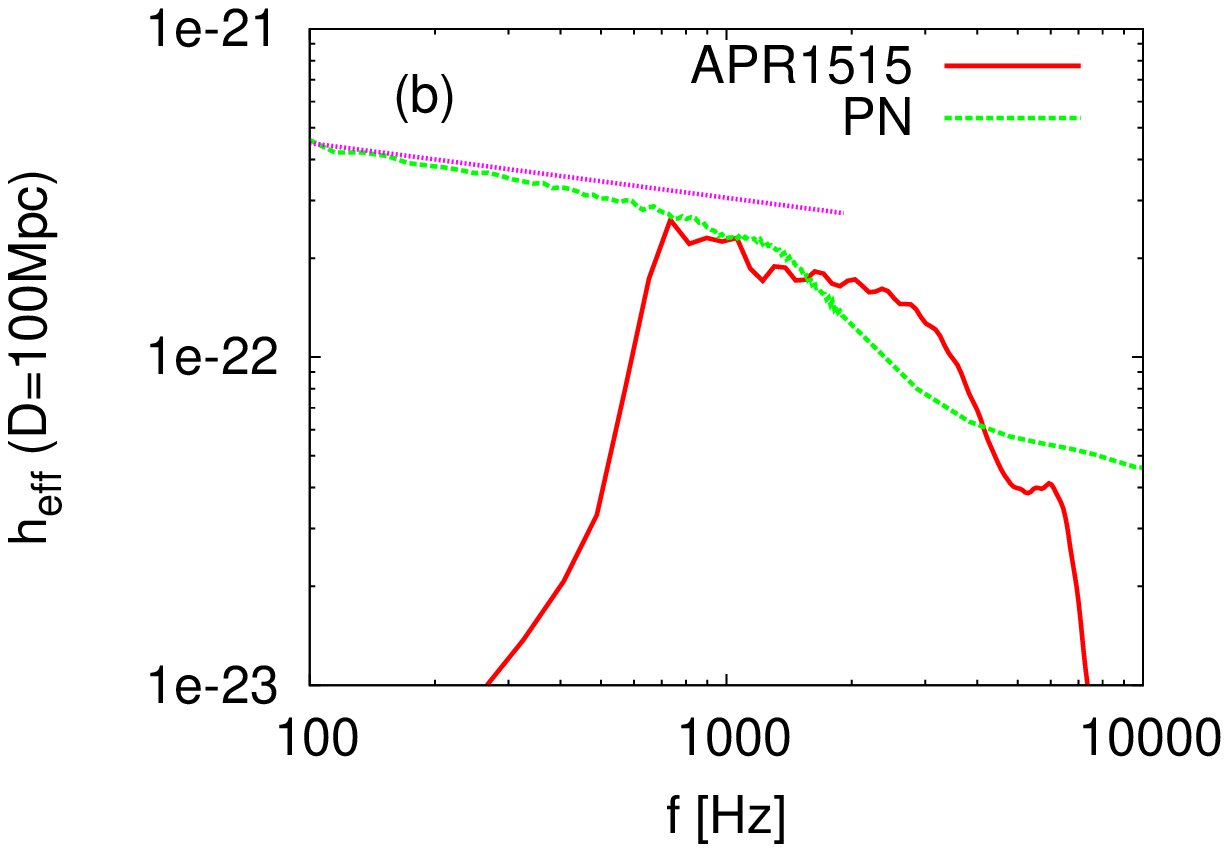}
      \end{minipage}
      \\
      \begin{minipage}{0.5\hsize}
      \includegraphics[width=9.0cm]{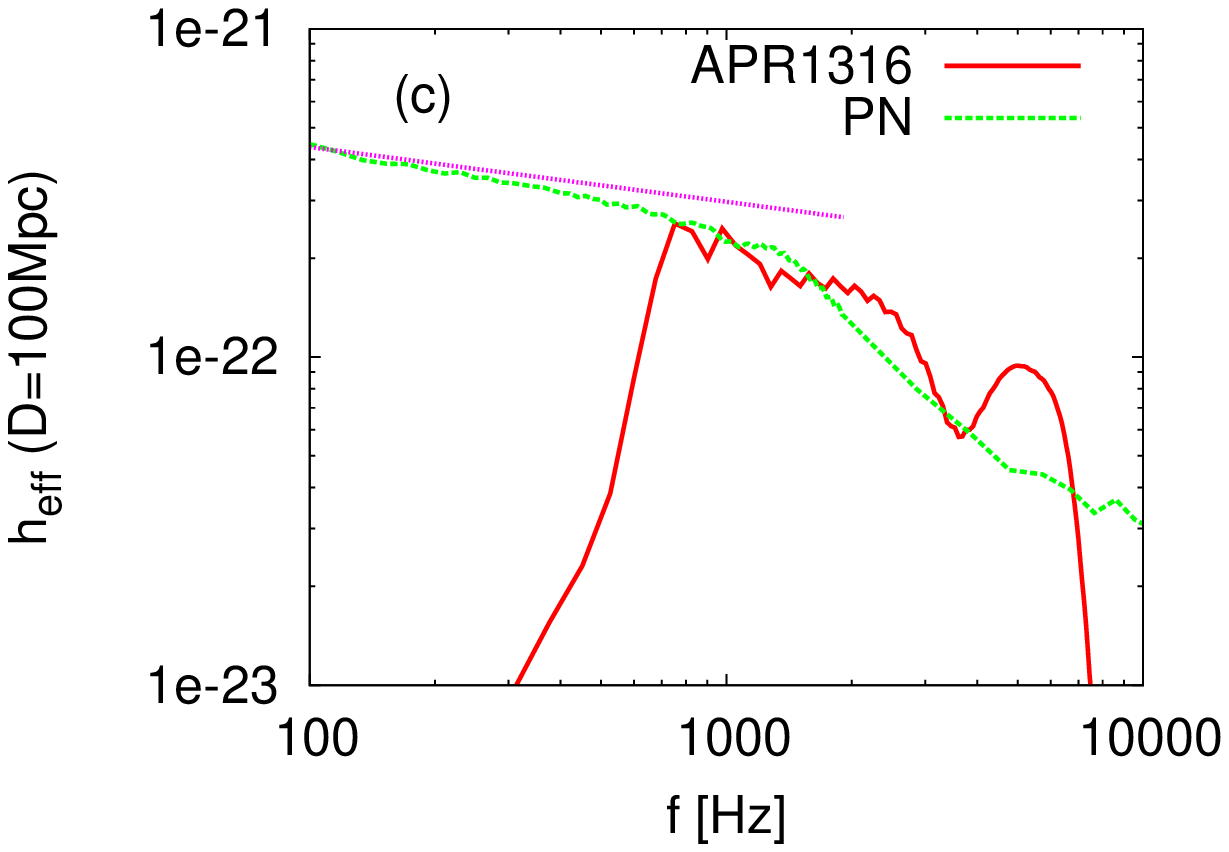}
      \end{minipage}
      \hspace{-0.5cm}
      \begin{minipage}{0.5\hsize}
      \includegraphics[width=9.0cm]{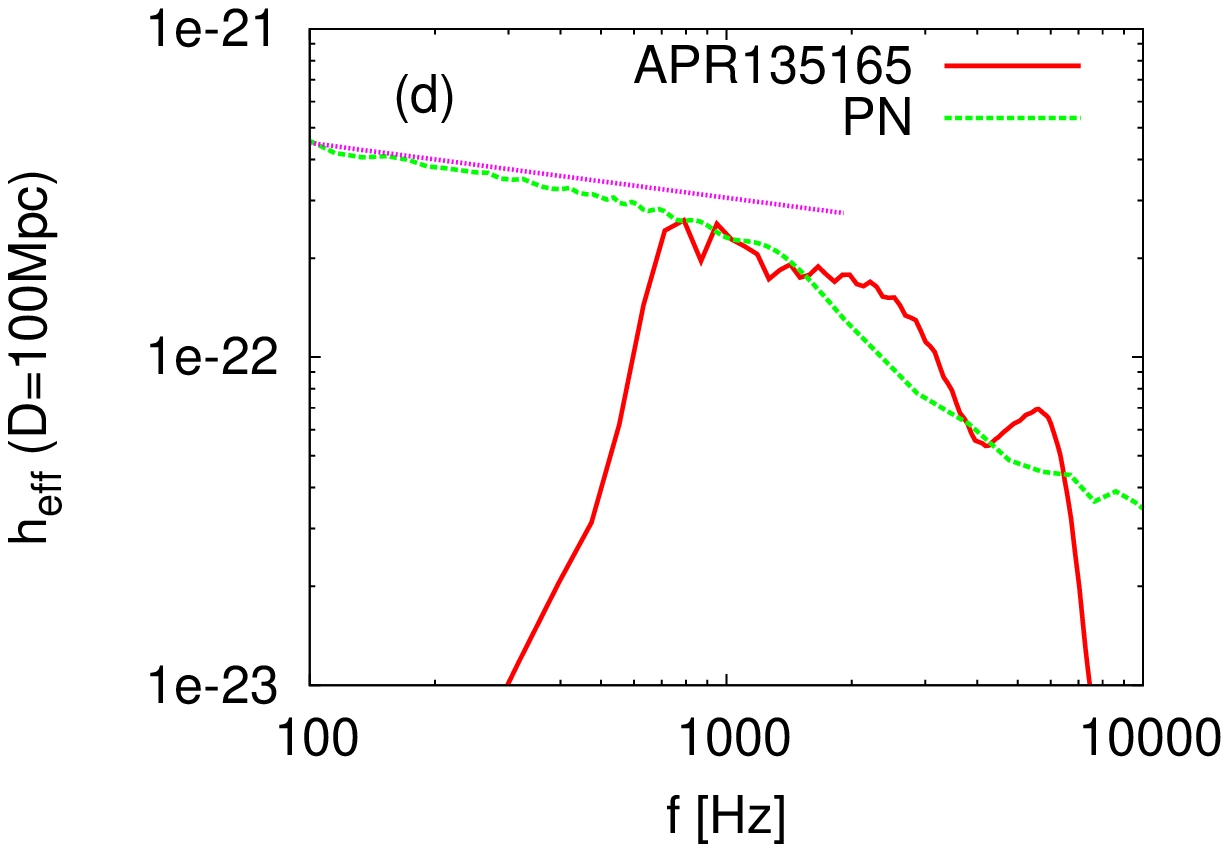}
      \end{minipage}
    \end{tabular}
    \caption{\label{fig:Fouri} Spectrum of gravitational waves as a
      function of frequency for runs (a) APR1414H, (b) APR1515H, (c)
      APR1316H, and (d) APR135165H for a hypothetical distance of 100
      Mpc. For comparison, the spectrum calculated by the Taylor
      T4 (Newtonian quadrupole, e.g., $\propto f^{-1/6}$) formula is
      shown by the dashed curve (dotted line).  }
  \end{center}
\end{figure*}

Figure \ref{fig:Fouri} shows the spectrum ($h_{\rm eff}$) of
gravitational waves for runs APR1414H, APR1515H, APR1316H, and
APR13516H. To plot Fig. \ref{fig:Fouri}, we assume $D=100$
Mpc. Because the simulations are started with an orbit of a finite
value of $\Omega_0$ ($f \equiv \Omega_0/\pi \sim 700$ Hz), the
spectrum amplitude is not realistic for a low-frequency side $f \alt
800$ Hz. To compensate for this drawback, we plot the spectrum of
gravitational waves derived by the Taylor T4 formula by the dotted
curve, which approximately behaves as $\propto f^{-n}$ where $n$
depends weakly on $f$ and is slightly larger than 1/6 around $f \alt
1$ kHz (note that for $f \ll 1$ kHz, $n \rightarrow 1/6$
\cite{Cutler:2002me}).  The figures show that the spectrum computed 
numerically smoothly connects to that of the Taylor T4 formula at $\sim
0.8$--1 kHz, indicating an acceptable accuracy of the numerical results.

The first noteworthy feature found from Fig. \ref{fig:Fouri} is that
the spectrum amplitude does not steeply damp even at $f \sim
1.2$--1.3~kHz~\cite{Bejger:2004zx} where the binary neutron stars
reach the ISCO, although this is predicted by the results derived from
the Taylor T4 formula.  This indicates that gravitational waves are
emitted by a high-velocity motion of a merging object even after the
binary neutron stars reach the ISCO. Rather, the spectrum amplitude
steeply damps at a fairly high frequency $f=f_{\rm cut} \sim 2.5$--3
kHz for the case that a black hole is formed. This indicates that by
an inspiral-type motion, gravitational waves are emitted even up to
such a high frequency: Even after the onset of merger, two
high-density peaks remain in the merging object (see the fourth panel
of Fig.~4), and the system emits gravitational waves for which the
waveform is similar to the inspiral one. 

\begin{figure*}
  \begin{center}
  \vspace*{40pt}
    \begin{tabular}{cc}
      \begin{minipage}{0.5\hsize}
      \includegraphics[width=9.0cm]{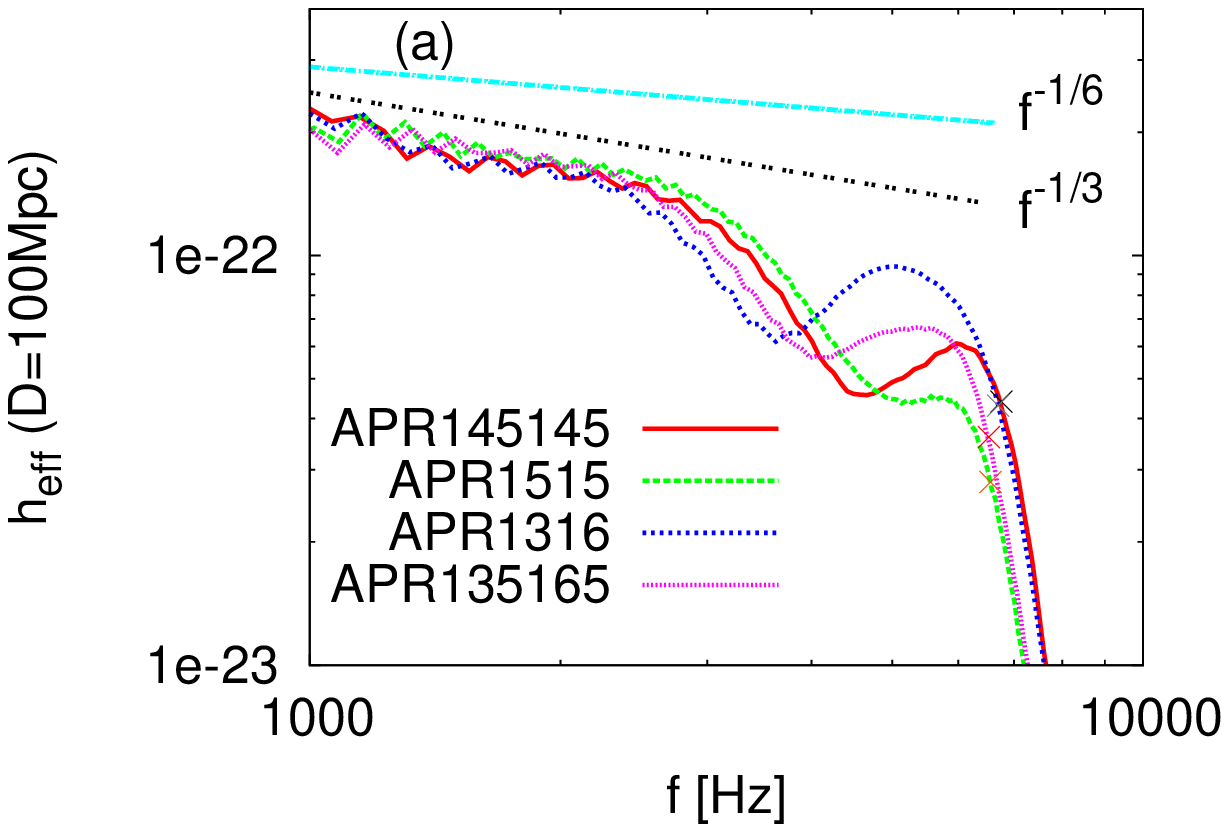}
      \end{minipage}
      \begin{minipage}{0.5\hsize}
      \includegraphics[width=9.0cm]{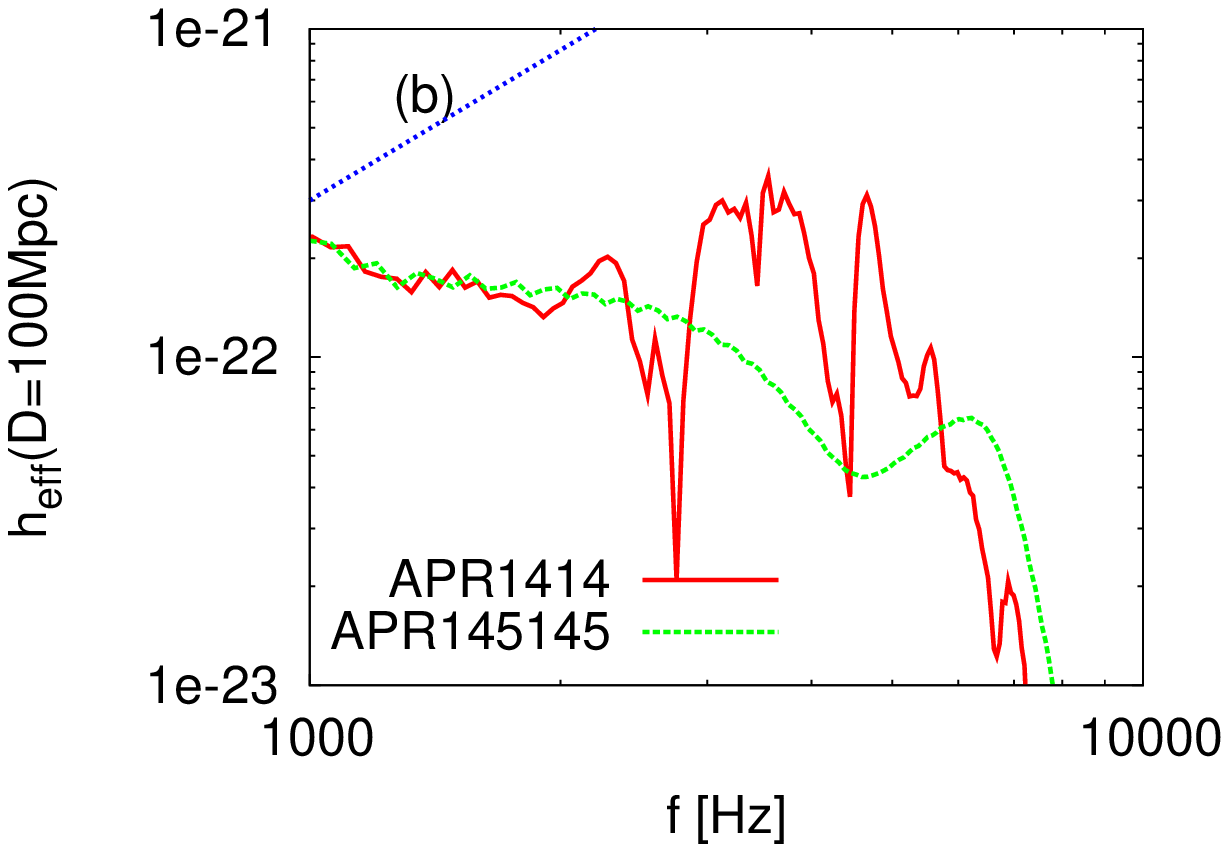}
      \end{minipage} \\
      \begin{minipage}{0.5\hsize}
      \includegraphics[width=9.0cm]{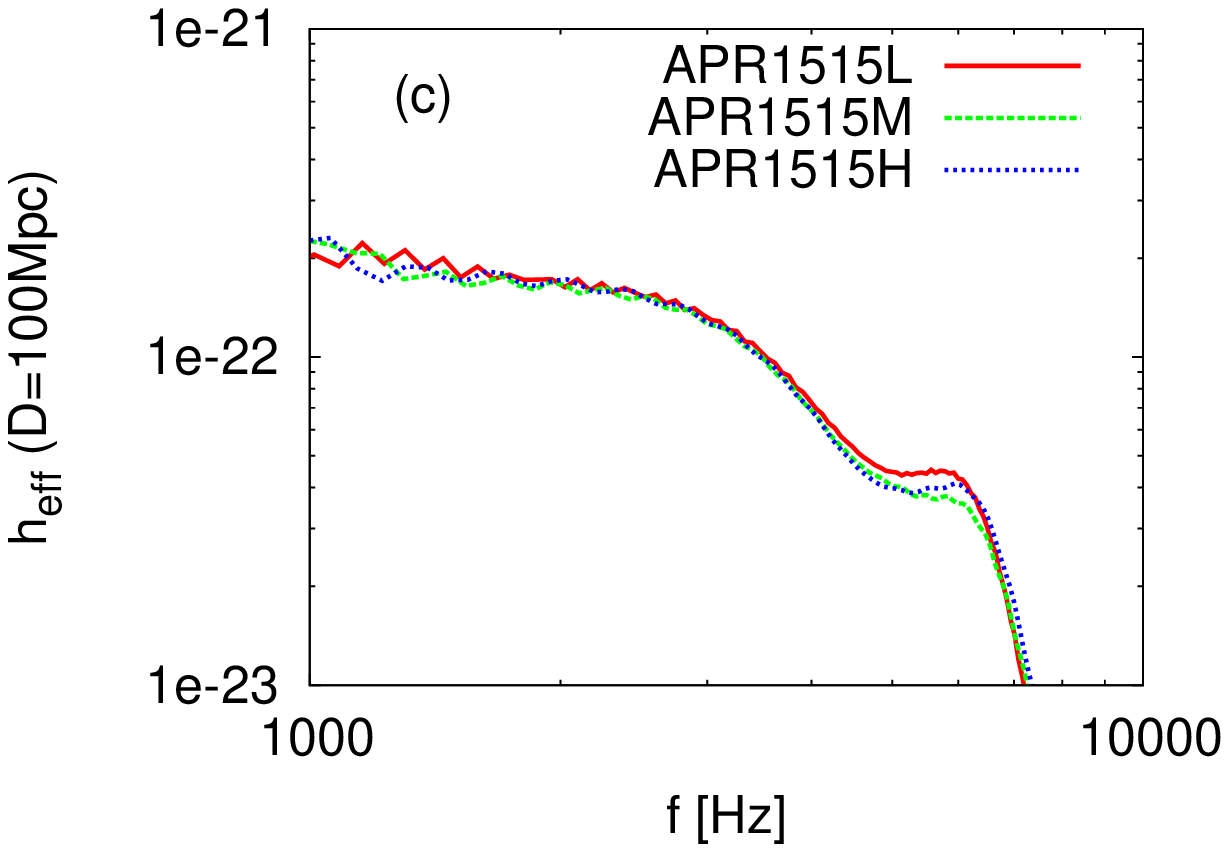}
      \end{minipage}
      \begin{minipage}{0.5\hsize}
      \includegraphics[width=9.0cm]{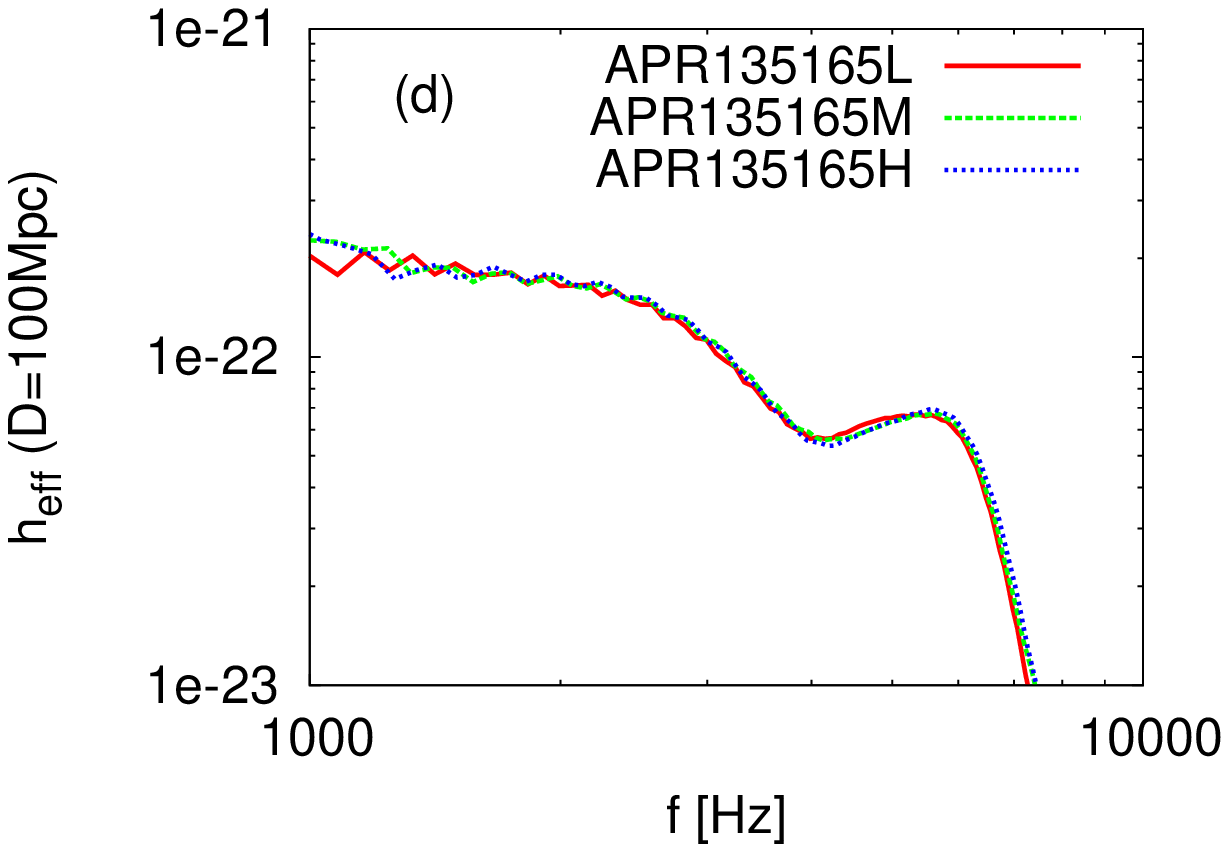}
      \end{minipage}
    \end{tabular}
    \caption{\label{fig:Fouri2} (a) Spectrum of gravitational waves
      for runs APR145145H, APR1515H, APR1316H, and APR135165H, (b) spectrum
      for runs APR1414H and APR145145H, (c) the spectrum for runs
      APR1515L, APR1515M, and APR1515H, and (d) the spectrum for runs
      APR135165L, APR135165M, and APR135165H.  Lines and cross symbols in
      the panel (a) denote $f^{-1/6}$, $f^{-1/3}$, and QNM frequencies
      for all the models.  The dotted line in the panel (b) shows 
      the planned noise level of the advanced LIGO.}
  \end{center}
\end{figure*}

For the case that a hypermassive neutron star is formed (for model
APR1414), multiple characteristic peaks for $\sim 2$--5 kHz are
seen. These peaks are related to quasiperiodic gravitational waves
emitted by the hypermassive neutron star. As reported in
Ref.~\cite{Shibata:2006nm}, the peak of the highest amplitude appears
at $f \sim 3.8$ kHz for model APR1414, which is associated with
gravitational waves emitted by the quasiperiodic rotation of the
hypermassive neutron star. The side-band peaks are generated by the
coupling modulation between the modes of the quasiperiodic rotation
and of a quasiradial oscillation of the hypermassive neutron star for
which the oscillation frequency is $\sim 1.2$ kHz. We note that the
simulation was artificially stopped at $t \sim 20$ ms for this model,
but the hypermassive neutron star is likely to survive for a longer
time $\agt 1$ s. Hence, the peak amplitude of the spectrum may be a
factor of $\sim 7 (\sim \sqrt{1000/20})$ larger in reality.  In
Fig.~\ref{fig:Fouri2} (b), we plot $h(f)$ together with the planned
noise level of the advanced LIGO, to clarify detectability of
quasiperiodic gravitational waves emitted by the hypermassive neutron
star.  Based on the hypothesis that the hypermassive neutron star 
survives for $\sim 1$ s, the peak amplitudes around $3$--4 kHz may 
be raised up by a factor of several. This suggests that such 
gravitational waves may be detected for an event of $D \alt 20$--30
Mpc, as mentioned in Refs. \cite{Shibata:2005ss,Shibata:2006nm}.

For models APR1515, APR145145, APR1316, and APR135165, the spectrum
shape has qualitatively an universal feature (see
Fig. \ref{fig:Fouri2} (a)): (i) for $f \leq f_{\rm cut} \approx 2.5$--3
kHz, the spectrum amplitude gradually decreases with $f$ according to
the relation $\propto f^{-n}$ where $n$ takes values between 1/6 and
$\sim 1/3$; (ii) for $f \geq f_{\rm cut}$, the amplitude steeply
decreases. This is likely to reflect the fact that two high-density
peaks in the merged object disappear during the collapse to a black
hole; (iii) for $f = f_{\rm peak} \approx 5$--6 kHz, which is slightly
smaller than $f_{\rm QNM} \approx 6.5$--6.9 kHz, a broad peak
appears. Because the frequency is always smaller than $f_{\rm QNM}$,
this peak is not associated with the ringdown gravitational waveform
but with the merger waveform; it is likely to be emitted by matter
moving around the black hole; (iv) for $f > f_{\rm fin} \approx f_{\rm
QNM}$, the amplitude damps in an exponential manner. 

We note that the feature of the spectrum shape is qualitatively
similar to that for the merger of binary black holes
\cite{Buonanno:2006ui}, except for the following differences: One of
the most noticeable differences is found in the phase (ii).  For the
case of binary neutron stars, the spectrum amplitude steeply decreases
for $f \geq f_{\rm cut}$.  By contrast for binary black
holes such a steep decrease is not found.  The other remarkable
difference is found in the peak amplitude associated with the phase
(iii) and (iv): For the binary neutron stars, this is at most half of
the amplitude at the ISCO ($f \sim 1$ kHz for the binary neutron
stars), whereas for binary black holes, this peak amplitude is as
large as or even larger than the amplitude at the ISCO 
(e.g., Ref.~\cite{Buonanno:2006ui}). 

The spectrum shape in the case of black hole formation is
significantly different from that of hypermassive neutron star
formation for $f \agt 2$ kHz. Figure \ref{fig:Fouri2} (b) compares the
spectra for models APR1414 and APR145145. For model APR1414, there are
multiple spiky peaks in the spectrum for $2~{\rm kHz} \alt f \alt 6$
kHz. By contrast, the spectrum shape is fairly smooth for model
APR145145. This suggests that if gravitational waves of frequency
between about 2 kHz and 6 kHz are detected, the outcome (hypermassive
neutron star or black hole) can be distinguished.

The most remarkable difference among the spectrum shapes of the four
black-hole formation models shown in Fig. \ref{fig:Fouri2} (a) is
found in the amplitude and width of peak associated with the merger
waveform: The peak amplitude for models APR1515 and APR145145 is much
less prominent than that for models APR135165 and APR1316 (see
Fig. \ref{fig:Fouri2} (a) and compare the same-mass models),
reflecting the fact that the amplitude of the merger gravitational
waveform is smaller. This difference seems to reflect the difference
in the disk-formation process, as mentioned in Sec. \ref{sec:qnm}: For
the equal-mass models, nearly all the material simultaneously
collapse to a black hole, whereas for the unequal-mass models, disks
surrounding the black hole are formed and subsequently the
material of the disk fall into the black hole, enhancing the
amplitude of gravitational waves.  This difference in the accretion
process is also reflected in the width of the peak: For the
unequal-mass models, the width of the peak is broader. This indicates
that a variety of the matter motion induces those gravitational waves.

The merger process is also reflected in the value of $f_{\rm cut}$.
For equal-mass runs APR145145H and APR1515H, $f_{\rm cut}\sim 3.0~{\rm
  kHz}$, while for unequal-mass runs APR1316H and APR135165H, $f_{\rm
  cut}\sim 2.5~{\rm kHz}$ ($f_{\rm cut}$ does not depend strongly on
the grid resolution).  Thus, $f_{\rm cut}$ is smaller for the
unequal-mass models for a given value of the total mass. The reason
for this is that tidal elongation and disruption of the less massive
neutron star occur for the unequal-mass binary neutron stars in the
late inspiral phase.  The tidal elongation sets in at a relatively low
frequency just after the binary neutron stars reach the ISCO. This is
reflected in a steep decrease of the spectrum amplitude at a smaller
frequency.

The total mass of the system is reflected in the value of $f_{\rm
  fin}$.  Figure \ref{fig:Fouri2} (a) shows that the frequency at
which $h_{\rm eff}$ reaches $2 \times 10^{-23}$ (for $D=100$ Mpc) is
approximately inversely proportional to the total mass; e.g., $f_{\rm
  fin} \sim 7.2~{\rm kHz}$ for models APR1515 and APR135165 and
$f_{\rm fin} \sim 7.6~{\rm kHz}$ for models APR145145 and
APR1316. This is reasonable because the value of $f_{\rm fin}$ is
determined primarily by $f_{\rm QNM}$ which is inversely proportional
to the black-hole mass for a given spin, and hence, approximately 
to the total mass.

Before closing this section, we comment on the convergence.  Figure
\ref{fig:Fouri2} (c) and (d) plots the spectrum for runs APR1515L,
APR1515M, and APR1515H and that for runs APR135165L, APR135165M, and
APR135165H, respectively. This shows that these do not overlap
completely, but the amplitude for a given frequency agrees among three
results within $\sim 1 \times 10^{-23}$ error for both models. 
Such error does not affect the findings and conclusions in this 
section.

\section{Summary}\label{sec:sum}

This paper reports new numerical results of general relativistic
simulations for binary neutron stars, focusing in particular on the
case that a black hole is formed in dynamical time scale after the
onset of merger. Following Ref.~\cite{Shibata:2006nm}, the APR EOS and
irrotational velocity field are employed for modeling the binary
neutron stars in the inspiral phase.  We prepare initial conditions
with an orbital separation which is much larger than that in the
previous work \cite{Shibata:2006nm}, and hence, unphysical effects
associated with initial nonzero eccentricity and incorrect approaching
velocity are excluded with a much better manner.  We adopt the moving
puncture approach, which enables to simulate black hole formation and
subsequent longterm evolution of the black hole spacetime.
Furthermore, the simulations are performed with a much better grid
resolution than that in the previous work \cite{Shibata:2006nm} to
obtain more reliable numerical results.

As a result of the improvements summarized above, we have obtained
the following new results for the outcome formed after the merger in
the present work: (i) For the binary neutron stars modeled by the APR
EOS, a black hole is formed in the dynamical time scale after the
onset of merger, if the total mass of the system $m_0$ is larger
than a threshold mass $M_{\rm thr}=2.8$--$2.9M_\odot$. This holds
irrespective of the mass ratio as long as $0.8 \alt Q_M \leq 1$. In
the present work, the value of $M_{\rm thr}$ is determined with less
uncertainty than that in Ref. \cite{Shibata:2006nm}. (ii) For the case
that the black hole is formed in the dynamical time scale after the
onset of merger, the resulting black-hole spin $a_{\rm f}$ is
$\approx 0.78 \pm 0.02$. This value depends weakly on the total mass
and mass ratio of the binary neutron stars. (iii) The mass of the
formed black hole, $M_{\rm BH,f}$, is calculated from the approximate 
energy conservation relation (\ref{enecon}) and the circumferential
equatorial radius of apparent horizon within $\approx 0.5\%$
error. (iv) The mass and spin of the formed black hole are consistent
with those determined by the frequency and damping time of the
quasinormal mode of gravitational waves, even when a disk of mass
$\alt 0.03M_{\odot}$ is formed around the black hole. (v) A
quasisteady disk of mass $\agt 0.01M_{\odot}$ is formed around the
black hole for the mass ratio $Q_M \approx 0.8$.  The disk mass
depends not only on the mass ratio but also on the total mass of the
binary neutron stars. We find that for the total mass ($m_0$) closer
to $M_{\rm thr}$, the resulting disk mass is larger.

Gravitational waves emitted during the merger and ringdown phases are
also analyzed. For the case that a hypermassive neutron star is
formed, gravitational waves are composed of the inspiral, merger, and
quasiperiodic waveforms. The quasiperiodic waves are emitted due to
quasiperiodic rotation of the formed ellipsoidal hypermassive neutron
star. For model APR1414, its characteristic frequency is $\approx 3.8$
kHz, which agrees with that found in the previous paper
\cite{Shibata:2006nm}.  As mentioned in the previous works
\cite{Shibata:2005ss,Shibata:2006nm}, the effective amplitude for the
quasiperiodic waves could be larger than $\sim 10^{-21}$ at a distance
of $D \leq 100$ Mpc. Although the frequency is rather high and
slightly outside the best-sensitive band of ground-based
gravitational-wave detectors, such gravitational waves are an
interesting target for the next-generation interferometric detectors.

For the case that a black hole is formed in the dynamical time scale
after the onset of merger, gravitational waves are composed of the
inspiral, merger, and ringdown waveforms. The feature is qualitatively
universal, irrespective of the total mass and mass ratio of the binary
neutron stars. The feature is clearly seen in the Fourier spectrum,
and quantitatively summarized as follows: (i) For $f \leq f_{\rm cut}
\approx 2.5$--3~kHz, the spectrum amplitude gradually decreases
according to the relation $\propto f^{-n}$ where $n$ is a slowly
varying function of $f$; $n=1/6$ for $f \ll 1$ kHz, and $n \sim
1/3$ for $f \rightarrow f_{\rm cut}$. $f_{\rm cut}$ is much larger
than the frequency at the ISCO. This is due to the fact that even
after the onset of merger, the merging object has two high-density
peaks, and emits gravitational waves for which the waveform is similar
to the inspiral one. (ii) For $f \geq f_{\rm cut}$, the spectrum
amplitude steeply decreases. This seems to reflect the fact that at
such frequency, two density peaks disappear during the collapse to a
black hole. (iii) For $f =f_{\rm peak} \approx 5$--6 kHz, which is
slightly smaller than $f_{\rm QNM} \sim 6.5$--6.9 kHz, a broad peak
appears. Because the frequency is always smaller than $f_{\rm QNM}$,
this peak is not associated with the ringdown gravitational waveform 
but with the merger waveform. (iv) For $f > f_{\rm fin} \approx f_{\rm
QNM}$, the amplitude damps in an exponential manner. 

Although the features (i)--(iv) are qualitatively universal, the
values of $f_{\rm cut}$, $f_{\rm peak}$, $f_{\rm QNM}$, and $f_{\rm
fin}$, and the height and width of the peak at $f=f_{\rm peak}$ depend
on the total mass and mass ratio of binary neutron stars, i.e., merger
and black hole formation processes. This implies that if gravitational
waves of high frequency $f=2$--8 kHz are detected, we will be able to
get information about merger and black hole formation precesses.

As mentioned in Sec. I, the hybrid EOS adopted in this work seems to
be appropriate for studying black hole formation in the merger of the
binary neutron stars.  However, for studying hypermassive neutron star
formation or evolution of accretion disk around the formed black hole,
the present choice of the EOS is not very appropriate, because for
such cases, effects associated with the thermal energy (finite
temperature) and neutrino cooling are likely to play an important
role~(e.g., Ref.~\cite{Ruffert:2001gf}).  Magnetic fields will also
play an important role for longterm evolution of the hypermassive
neutron star and black hole accretion disks if they are amplified in a
short time scale (say in 10 dynamical time scales of the system).  In
the future paper, we would like to report our new studies in which
these effects are taken into account.

In the present work, we are not able to compute the inspiral waveforms
with a high precision because the grid resolution is not high enough.
Even with the highest-grid resolution (e.g., in run APR1515H), the
phase error may amount to 2 wave cycles (see
Fig. \ref{fig:inspiralH}).  To suppress the phase error within, e.g.,
a half wave cycle in the inspiral phase, the grid resolution would
have to be by 30--40\% as high as that for model APR1515H. To perform
such high-resolution simulation, the present choice of the grid
structure requires a high computational cost, 
in particular for a systematic survey of gravitational
waveforms for a variety of parameters. We have to adopt an AMR
algorithm \cite{Yamamoto:2008js,Baiotti:2008ra,Anderson:2007kz} to
save the computational cost. Currently, we are studying the inspiral
waveforms in the simulation using a code with the AMR
algorithm~\cite{Yamamoto:2008js}.  We hope to present such results in
the near future.

\section*{Acknowledgments}

We thank J. Hansen and L. Baiotti for revising English grammar.
Numerical computations were in part carried on XT4 and general common
use computer system at the Center for Computational Astrophysics 
in the National Astronomical Observatory of Japan and on NEC-SX8
at Yukawa Institute for Theoretical Physics in Kyoto University.  This
work was supported by the Grant-in-Aid for Scientific Research
(No. 21340051), by the Grant-in-Aid for Scientific Research on
Innovative Areas (No. 20105004) of the Japanese Ministry of Education,
Culture, Sports, Science and Technology, and by NSF Grant PHY-0503366. 

\clearpage

\end{document}